\documentclass[onecolumn]{aastex631}
\pdfoutput=1
\usepackage{amsmath}
\usepackage{color}
\usepackage{amssymb}
\usepackage{amsfonts}
\usepackage{stackengine}
\usepackage{changepage}
\usepackage{array} 
\usepackage{subfigure}
\usepackage{natbib}
\usepackage{parskip}
\usepackage{multirow,booktabs}
\usepackage{graphicx}
\usepackage{float}
\usepackage{xspace}
\newcommand{\beq}{\begin{equation}}
\newcommand{\eeq}{\end{equation}}
\newcommand{\g}{\gamma}
\def\ba{\begin{eqnarray}}
\def\ea{\end{eqnarray}}

\newcommand{\fermi}{{\em Fermi}\xspace}

\newcommand{\gbm}{{\em GBM}\xspace}

 \usepackage{amsmath}
\usepackage{pifont}
\usepackage[normalem]{ulem}
\usepackage{tikz}
\shortauthors{Liu et al.}
\begin{document}

\title{Spectral Lag Transition of 32 Fermi Gamma-ray Bursts and their Application on Constraining Lorentz Invariance Violation }

\correspondingauthor{Bin-Bin Zhang}
\email{bbzhang@nju.edu.cn}
\author[0000-0002-5550-4017]{Zi-Ke Liu}
\affiliation{School of Astronomy and Space Science, Nanjing University, Nanjing 210093, China}
\affiliation{Key Laboratory of Modern Astronomy and Astrophysics (Nanjing University), Ministry of Education, China}
\author[0000-0003-4111-5958]{Bin-Bin Zhang}
\affiliation{School of Astronomy and Space Science, Nanjing
University, Nanjing 210093, China}
\affiliation{Key Laboratory of Modern Astronomy and Astrophysics (Nanjing University), Ministry of Education, China}

\author[0000-0002-1122-1146]{Yan-Zhi Meng}
\affiliation{School of Astronomy and Space Science, Nanjing University, Nanjing 210093, China}
\affiliation{Key Laboratory of Modern Astronomy and Astrophysics (Nanjing University), Ministry of Education, China}

\begin{abstract}
\noindent

The positive-to-negative transition of spectral lag is an uncommon feature reported in a small number of GRBs. An application of such a feature has been made to constrain the critical quantum gravity energy ($E_{\rm QG}$) of the light photons under the hypothesis that the Lorentz invariance might be violated. Motivated by previous case studies, this paper systematically examined the up-to-date {\it Fermi}/GBM GRB sample for the lag transition feature to establish a comprehensive physical limit on the Lorentz Invariance Violation (LIV). This search resulted in 32 GRBs with redshift available, which exhibit the lag-transition phenomenon. We first fit each of the lag-E relations of the 32 GRBs with an empirical smoothly broken power law function, and found that the lag transition occurs typically at about 400 keV. We then implemented the LIV effect into the fit, which enabled us to constrain the lower limit of the linear and quadratic values of $E_{\rm QG}$, which are typically distributed at $1.5\times 10^{14}$ GeV and $8\times 10^{5}$ GeV, respectively.

\end{abstract}

\section{Introduction}

 The spectral lag of Gamma-ray Bursts (GRBs) was first introduced by \cite{1996ApJ...459..393N} to describe the phenomenon that GRB light curves in higher energy bands peak earlier than those in lower energy bands. Subsequent studies \citep[e.g.,][]{2000ApJ...534..248N,2006MNRAS.367.1751Y} showed that long GRBs are always characterized by significant lags, whereas short GRBs always present zero, sometimes negative, lags. On the other hand, the exact physical mechanism that causes spectral lags remains incompletely resolved to date. Considering the relativistic beaming nature of a GRB jet, one may naturally expect that the so-called ``high-latitude" effect can cause photons at higher latitudes to arrive at the observer later and with softened observed energy. Such effect was used by \cite{2000ApJ...544L.115S,2001ApJ...554L.163I, 2006ApJ...643..266N} to explain the overall statistical properties of the observed lags as well as the luminosity-lag correlation \citep{2004ApJ...602..306S}. By assuming an intrinsic spectral shape and a temporal profile, \citet{2005MNRAS.362...59S} found that curvature effect marginally interpreted the observed lags, yet extreme physical parameter values are required. Furthermore, \cite{2016ApJ...825...97U} demonstrated that the high-latitude curvature effect alone was not sufficient to account for the spectral lags. Rather, one must consider the intrinsic curved spectral shape, the evolution of the magnetic field strength, and the rapid bulk acceleration of the emission zone to interpret some observed spectral lag features. The aforementioned theories successfully explain the positive lags, however the rarely observed negative lags remain a more complex matter that can be used to infer the different radiation origins \citep{2011ApJ...730..141Z}, radiation mechanisms \citep{2010ApJ...709..525L,2011ApJ...730..141Z}, or emission regions \citep{2009ApJ...707.1404T} of low and high energy photons. 
 
 In the context the fundamental physics, the delay of high energy photons, formulated as negative lag in this study, can be used to test the violations of Lorentz invariance, a hypothesis that is widely pursued in quantum gravity (QG). LIV occurs at the Planck energy scale ($E_{\mathrm{P} 1}=\sqrt{\hbar c^{5} / G} \simeq 1.22 \times 10^{19} \mathrm{GeV}$) in QG theories \citep{2005LRR.....8....5M,2013LRR....16....5A}. The LIV effect can be manifested through vacuum dispersion, which leads photons with higher energy to travel at lower speeds \citep{1998Natur.393..763A}. GRBs are one of the ideal probes for testing LIV due to their large cosmological distance, small variability time scale, and very high energy photons. Using the delay time of individual photons reaching a few GeV, several studies \citep[e.g.,][]{2009Natur.462..331A,2009Sci...323.1688A,2020PhRvL.125b1301A} have shown that the linear QG energy scale, $E_{\mathrm{QG}}$, can be constrained at \textcolor{black}{$\gtrsim 10^{18-19}$} GeV. In order to further constrain the LIV effect, the most optimal use of lag to date has been to fit the keV-MeV multi-wavelength measurements of spectral lags, including their positive-to-negative transitions, with a model incorporating LIV information. A fit of this type yields some deeper limits on the QG energy \citep[see e.g.,][]{2017ApJ...834L..13W,2021ApJ...906....8D}. This approach has, however, only been applied in a few GRBs. A systematic study is required to determine whether the positive-to-negative transition is common in a large sample of GRBs and, if possible, to identify some additional constraints on the LIV effect according to the large sample lag data.

 In this {\it Letter}, we utilize the {\it Fermi}/GBM GRB catalog to analyze all z-known GRBs with positive-to-negative lag transitions and use these results to place some further constraints on the LIV effect. Data selection and reduction are described in \S 2. Our model of the spectral lag is presented in \S \ref{sec3}. The constraints on LIV and fitting results with our model are presented in section \S \ref{sec4}, followed by a brief summary and discussion in \S 5.

\section{Data}

\begin{figure*}
\vspace{-0.7cm}
\gridline{\hspace{-1.0cm}\fig{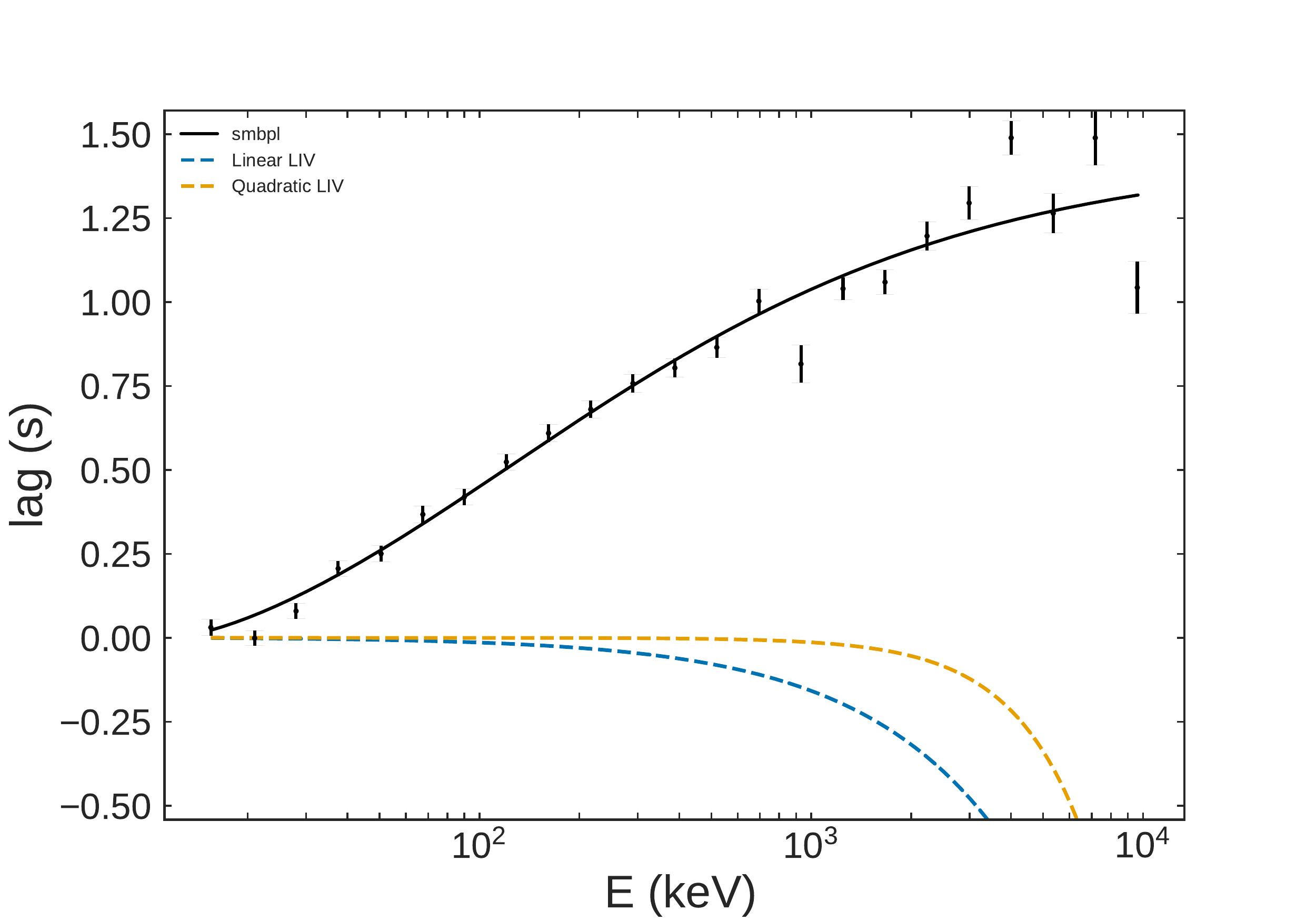}{0.29\textwidth}{GRB 210619B} \hspace{-0.4cm}\fig{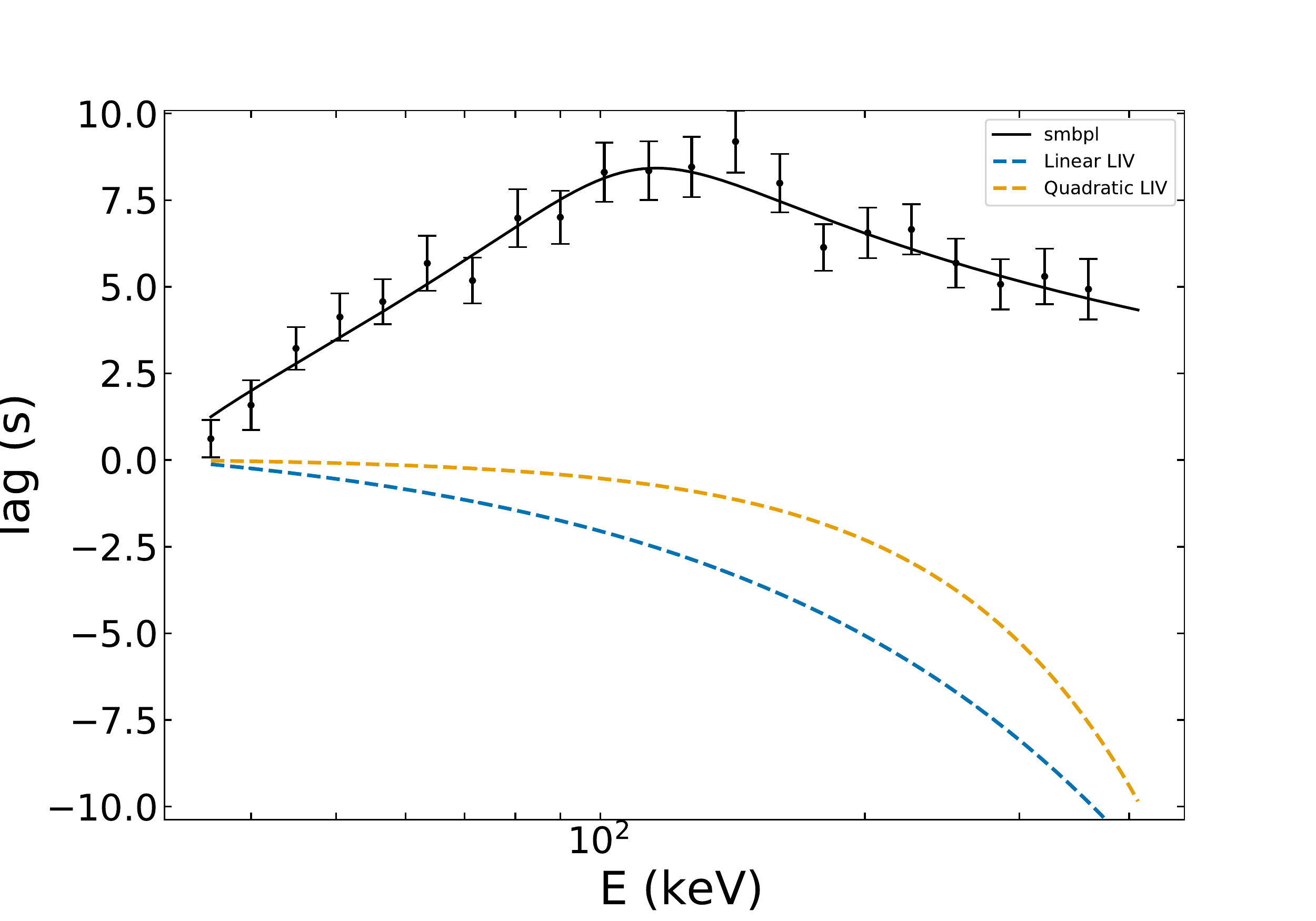}{0.29\textwidth}{GRB 210610B}
 \hspace{-0.4cm}\fig{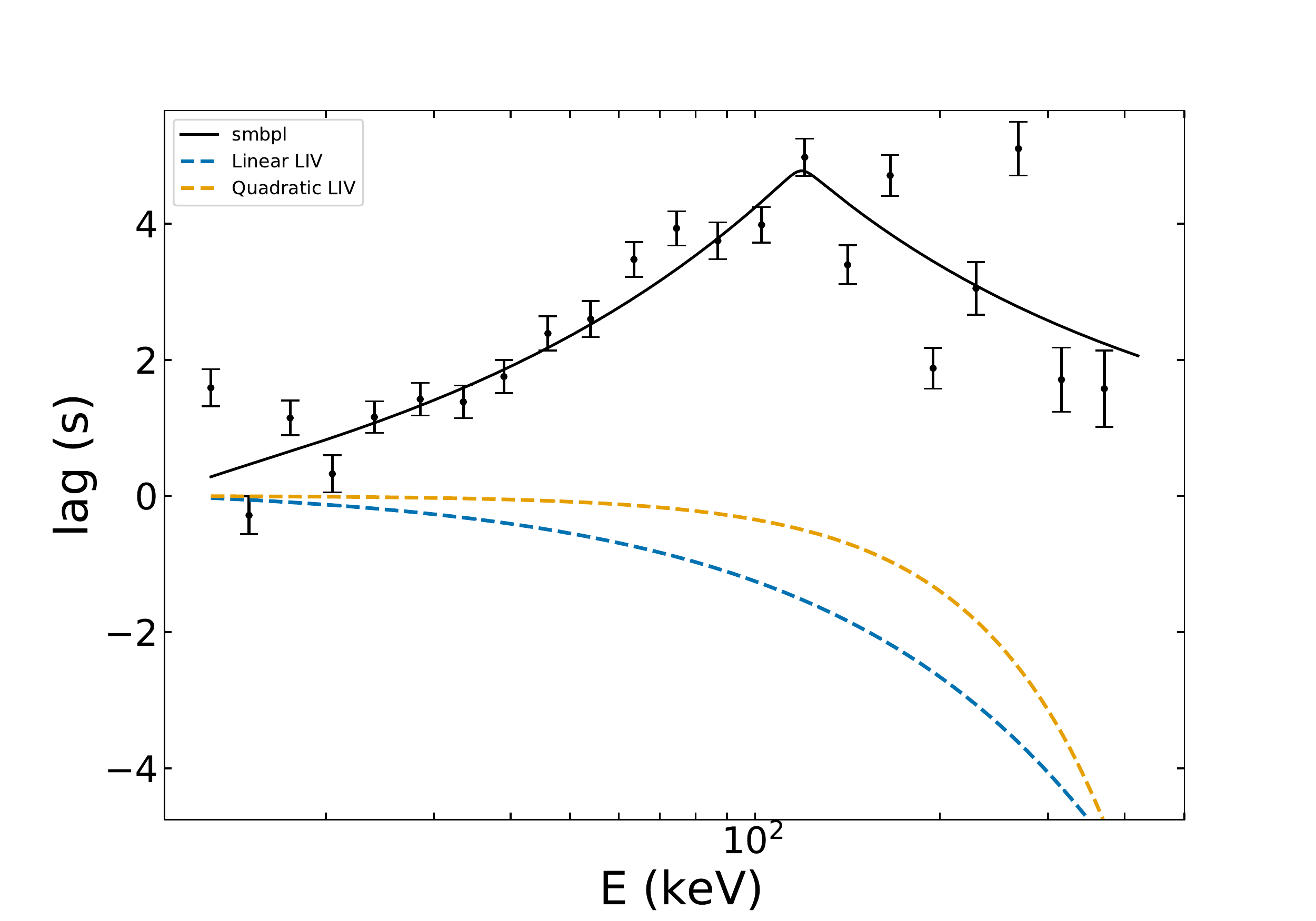}{0.29\textwidth}{GRB 210204A}
 \hspace{-0.4cm}\fig{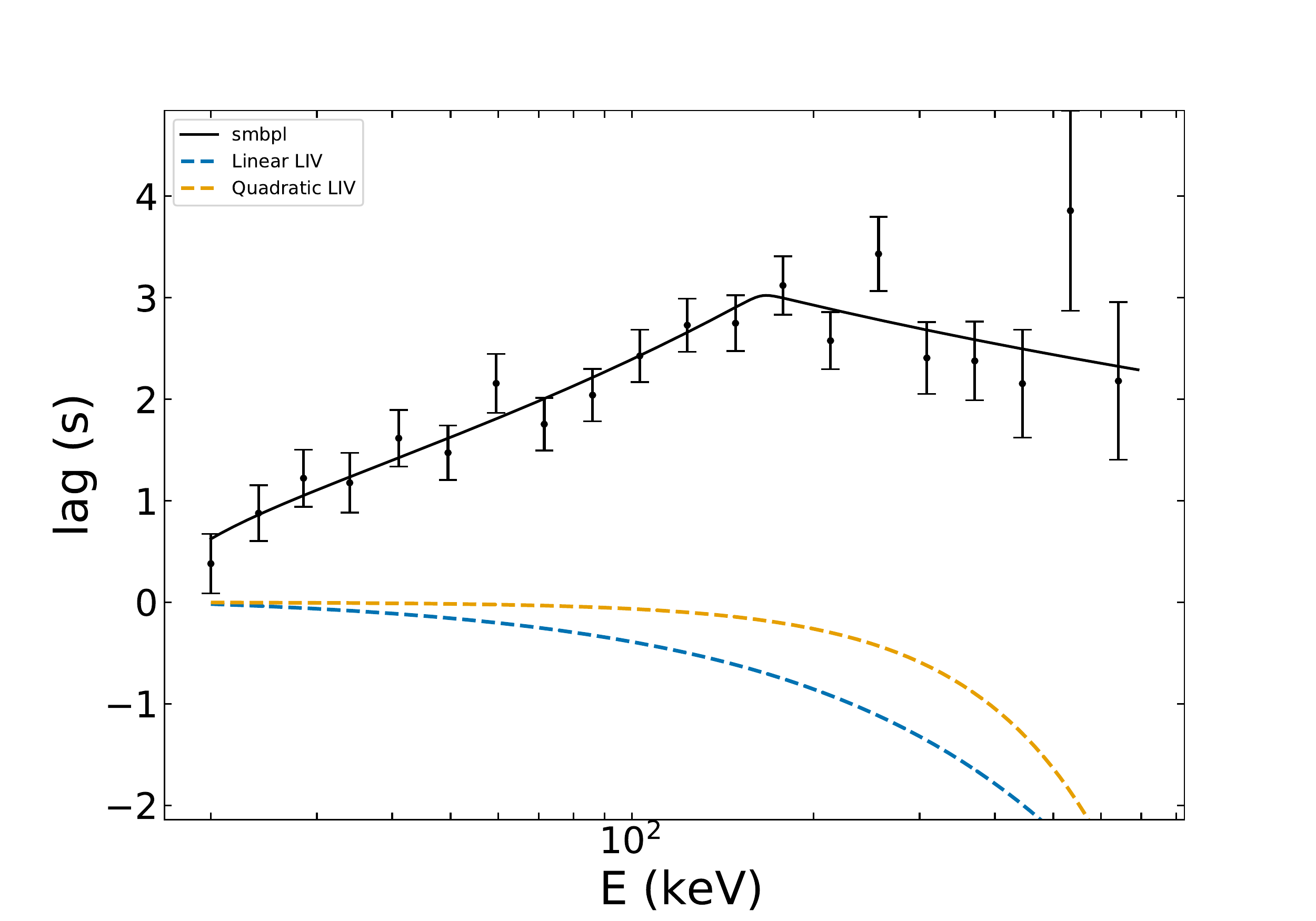}{0.29\textwidth}{GRB 201216C}
 }
\vspace{-0.5cm}
\gridline{\hspace{-1.0cm}\fig{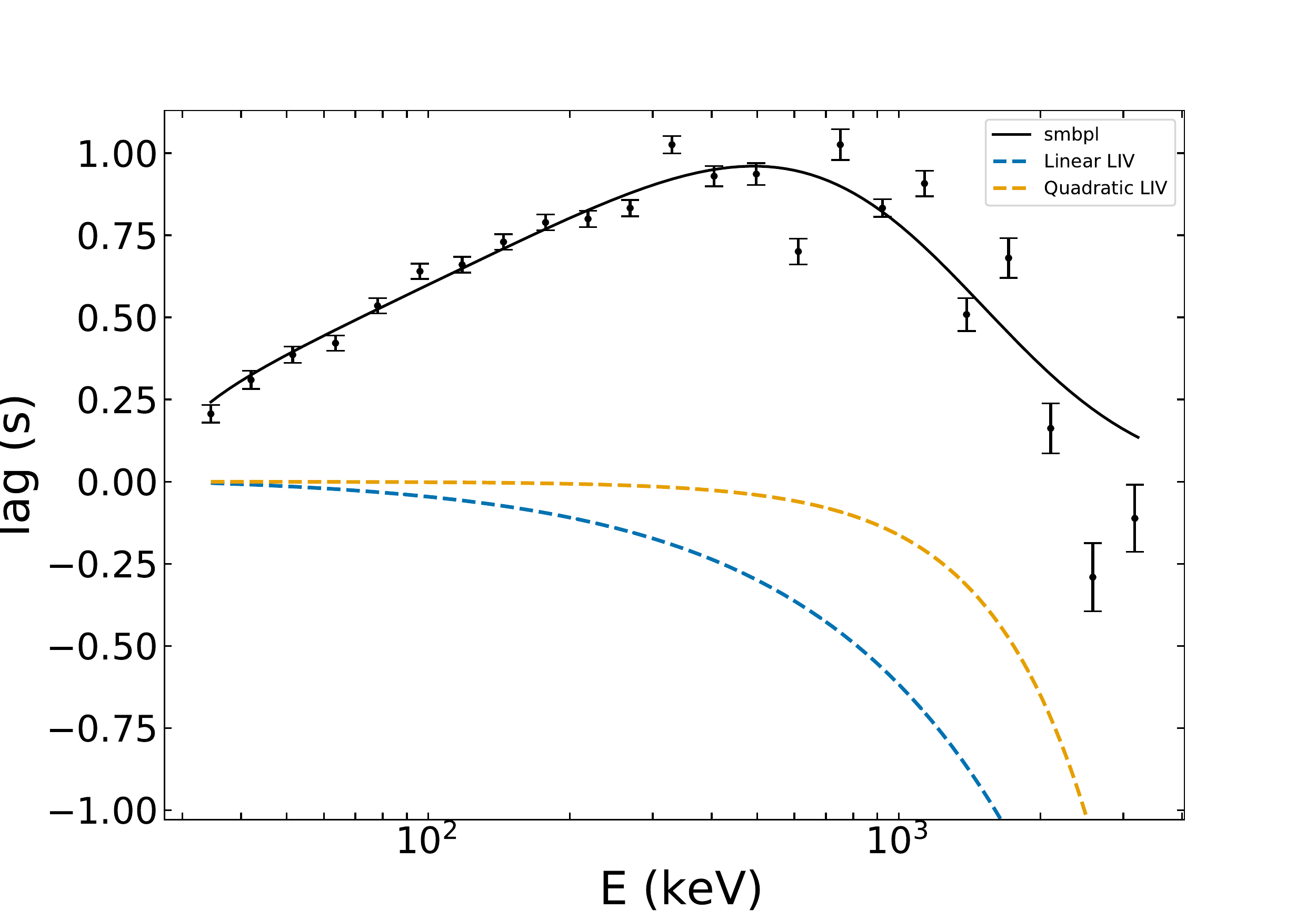}{0.29\textwidth}{GRB 200829A} \hspace{-0.4cm}\fig{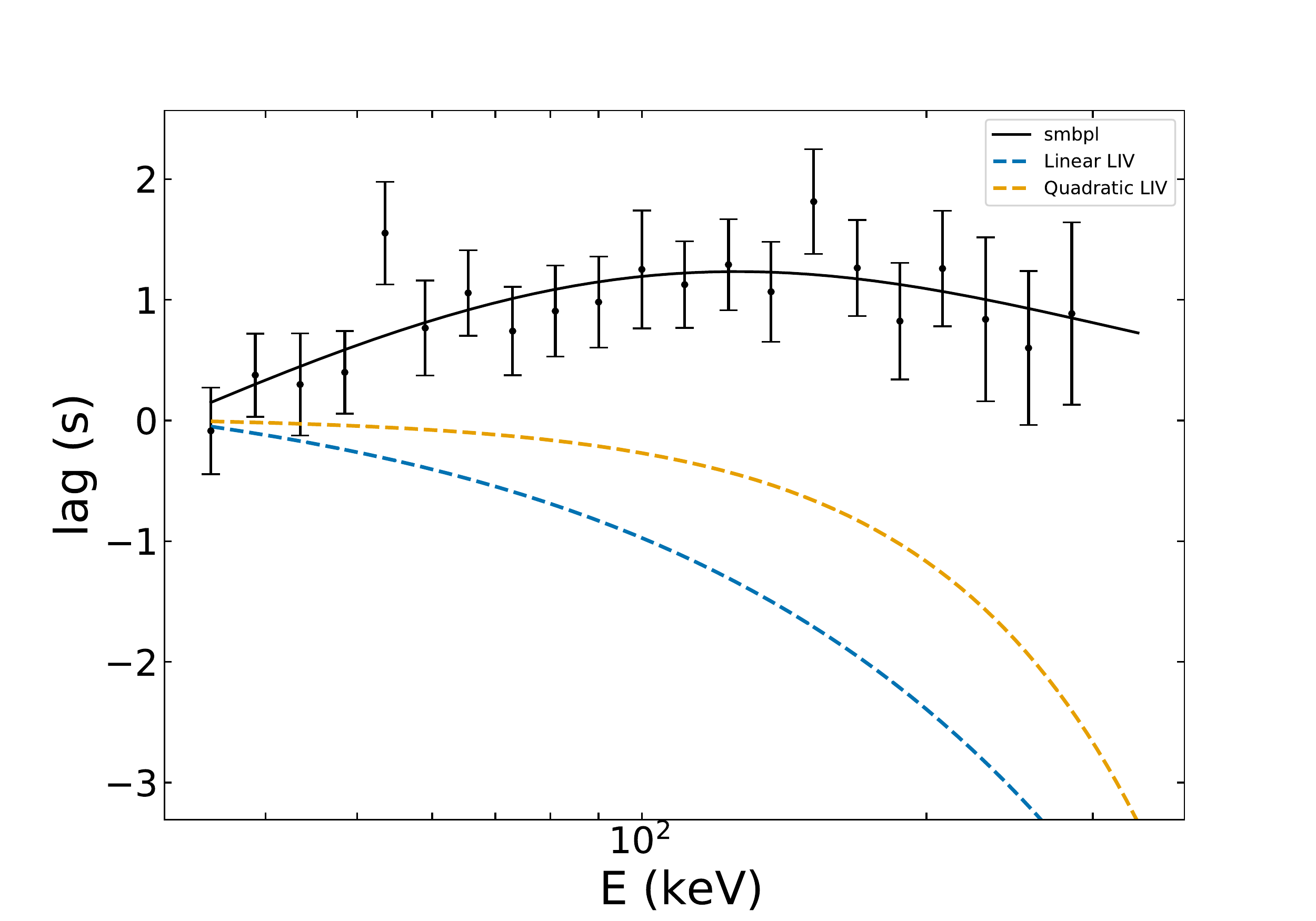}{0.29\textwidth}{GRB 200613A}
 \hspace{-0.4cm} \fig{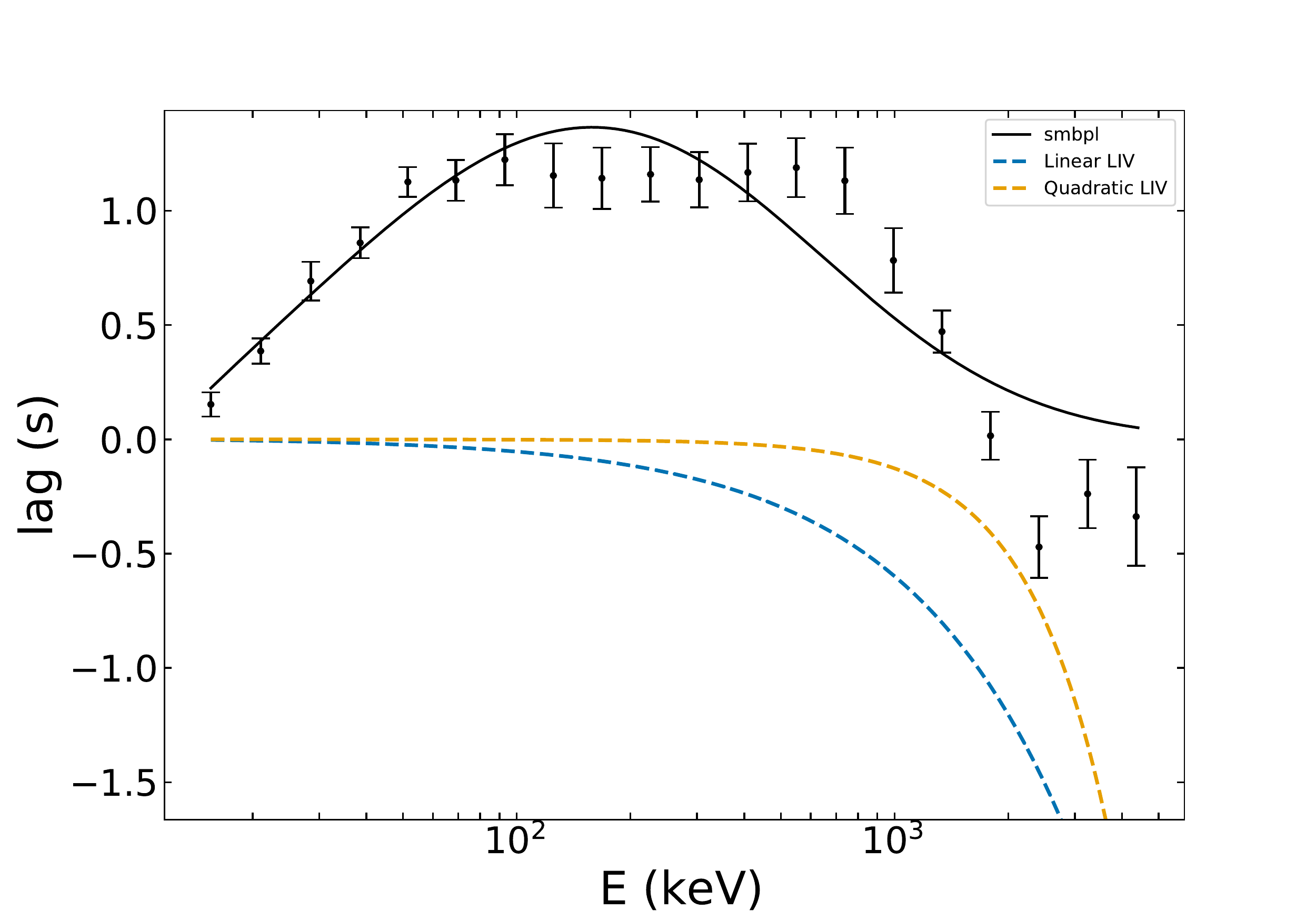}{0.29\textwidth}{GRB 190114C}
 \hspace{-0.4cm} \fig{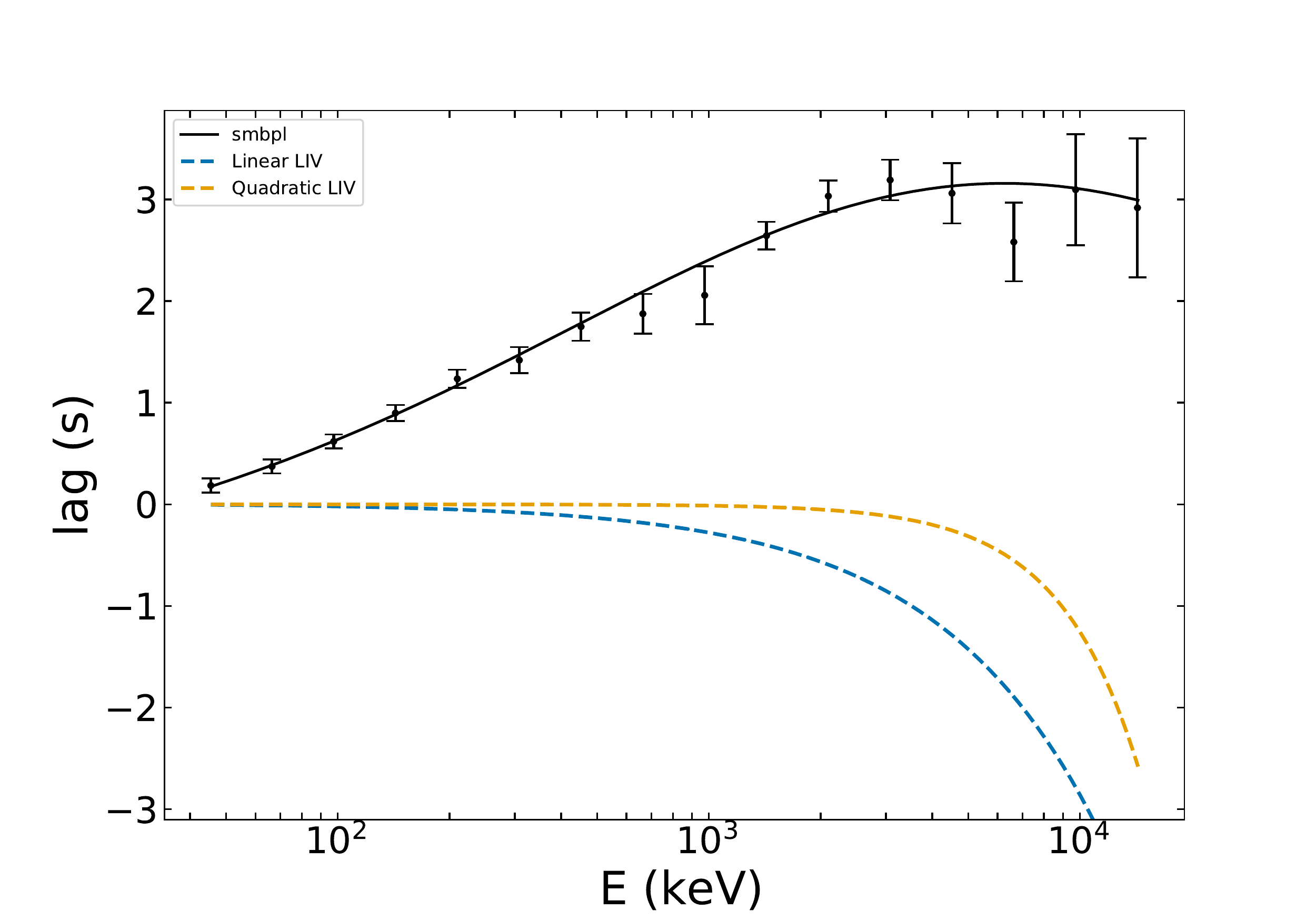}{0.29\textwidth}{GRB 180720B}
 } 
\vspace{-0.5cm}
\gridline{\hspace{-1.0cm}\fig{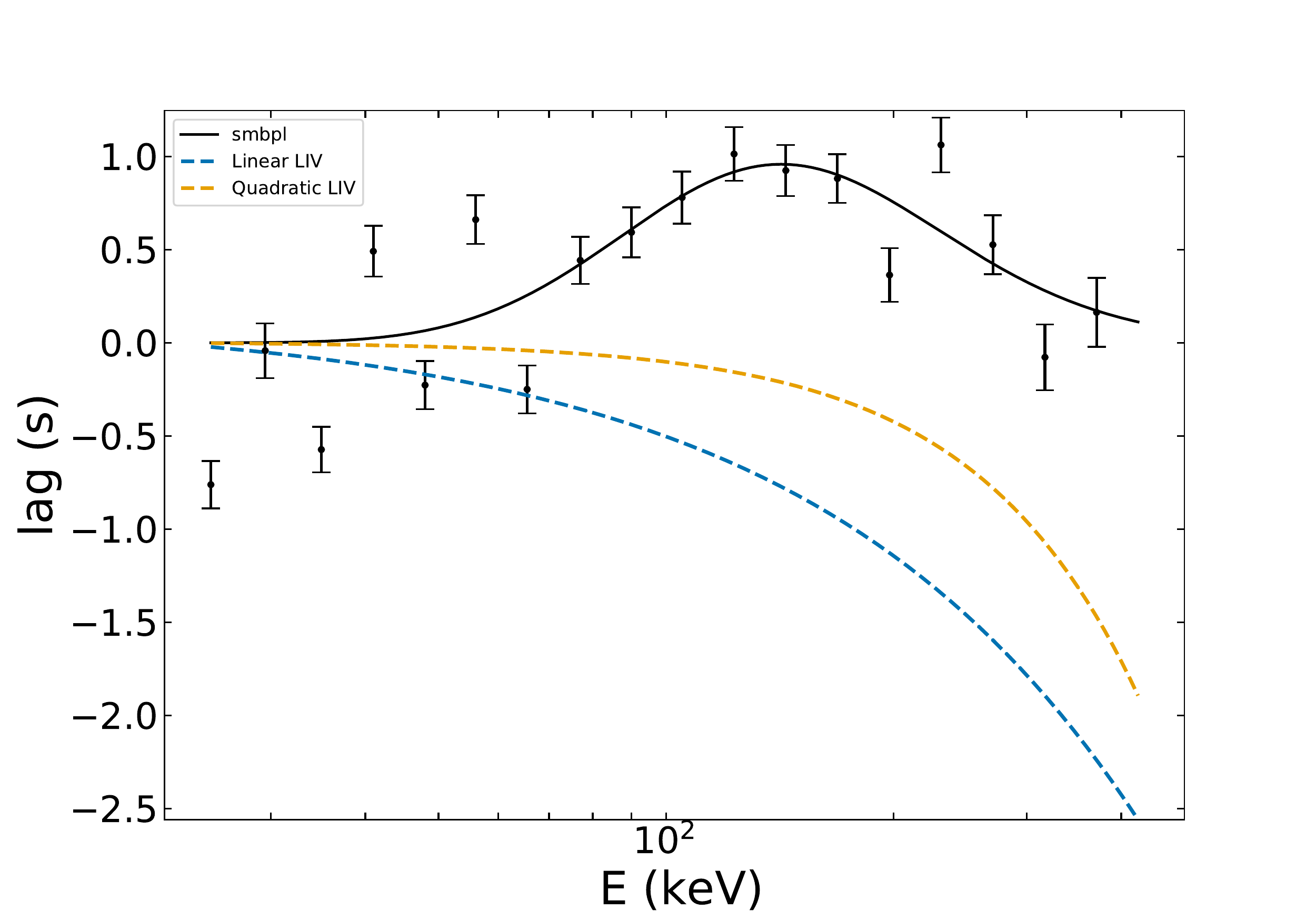}{0.29\textwidth}{GRB 180703A} \hspace{-0.4cm}\fig{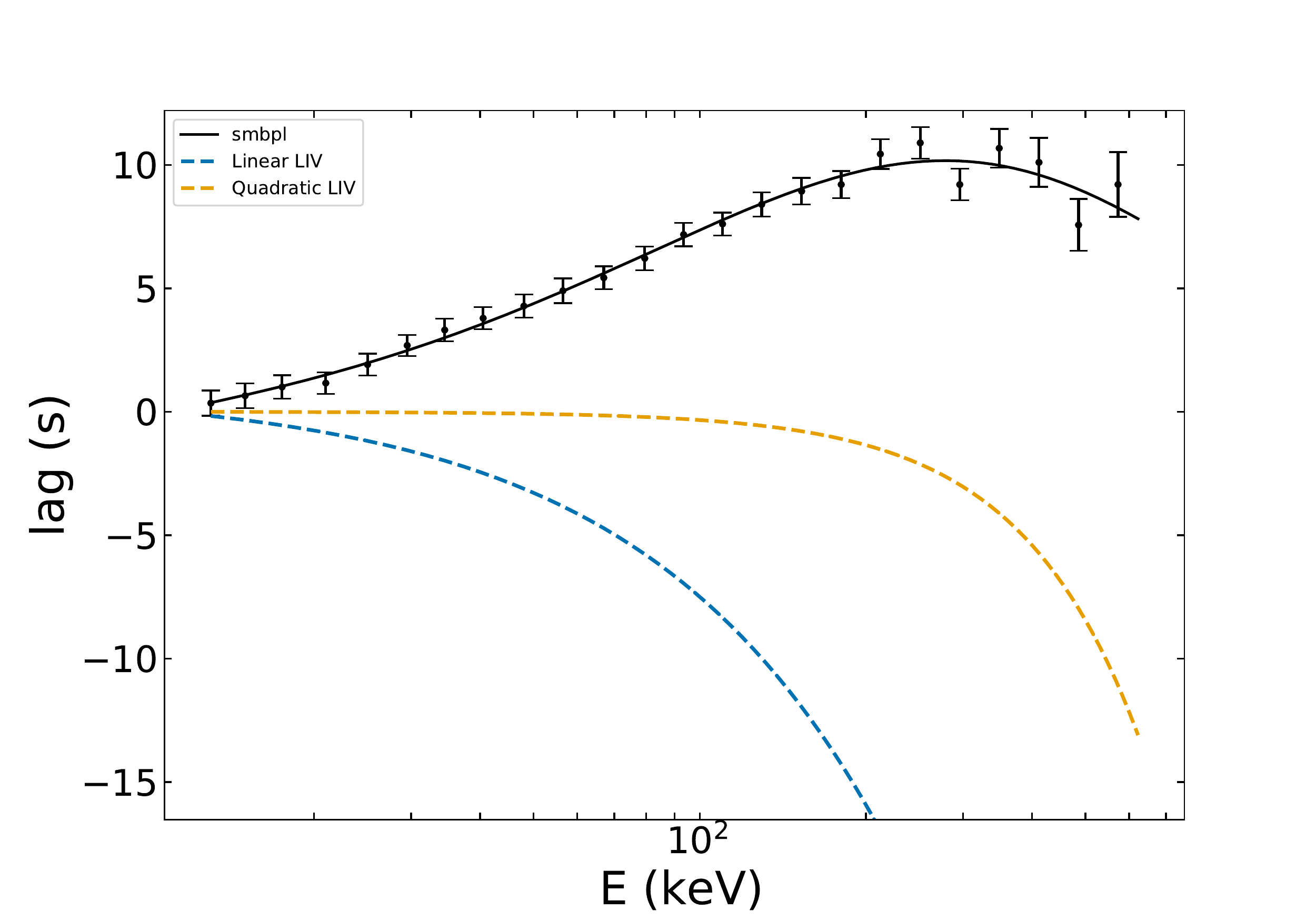}{0.29\textwidth}{GRB 171010A}
 \hspace{-0.4cm}\fig{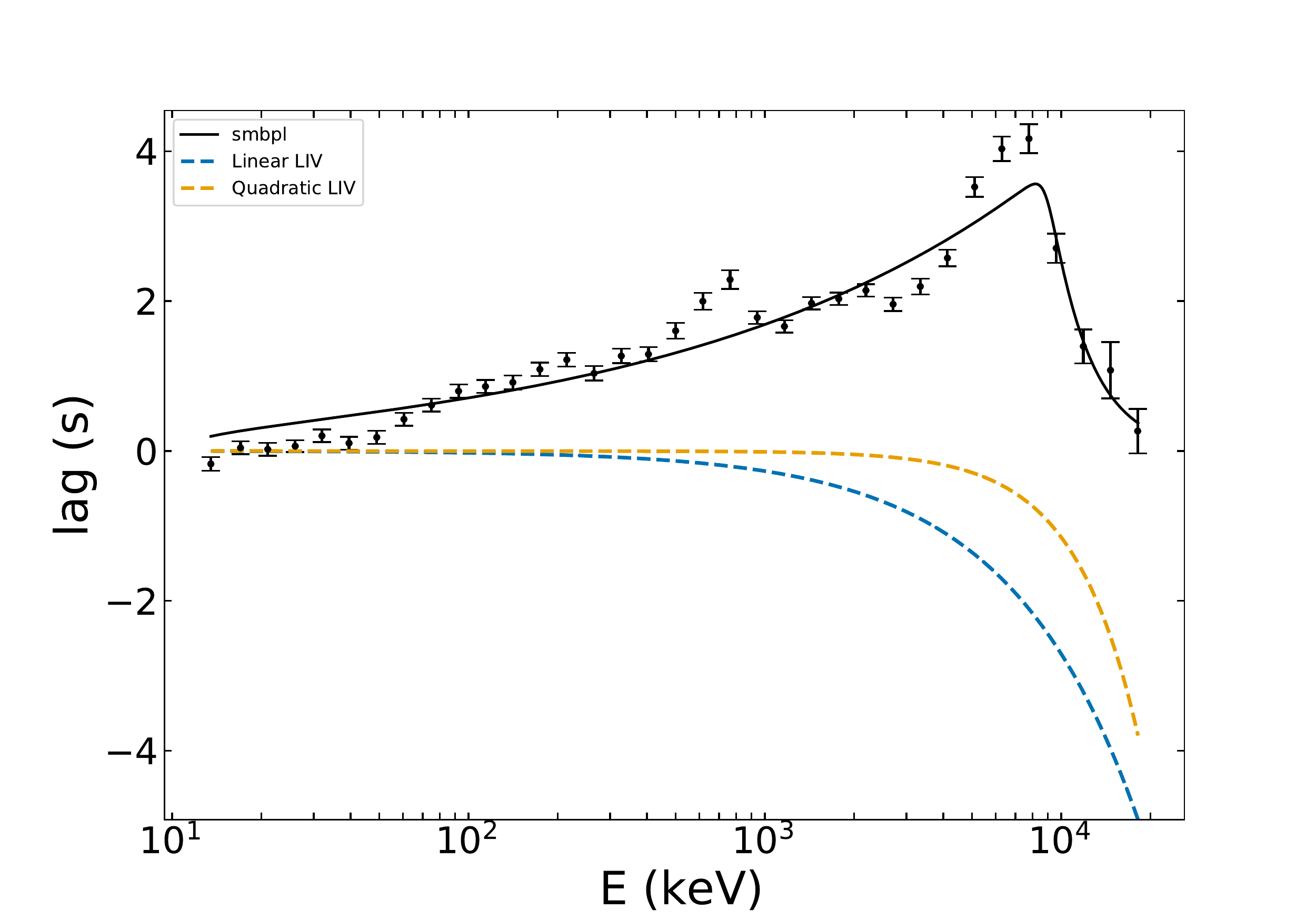}{0.29\textwidth}{GRB 160625B}
 \hspace{-0.4cm}\fig{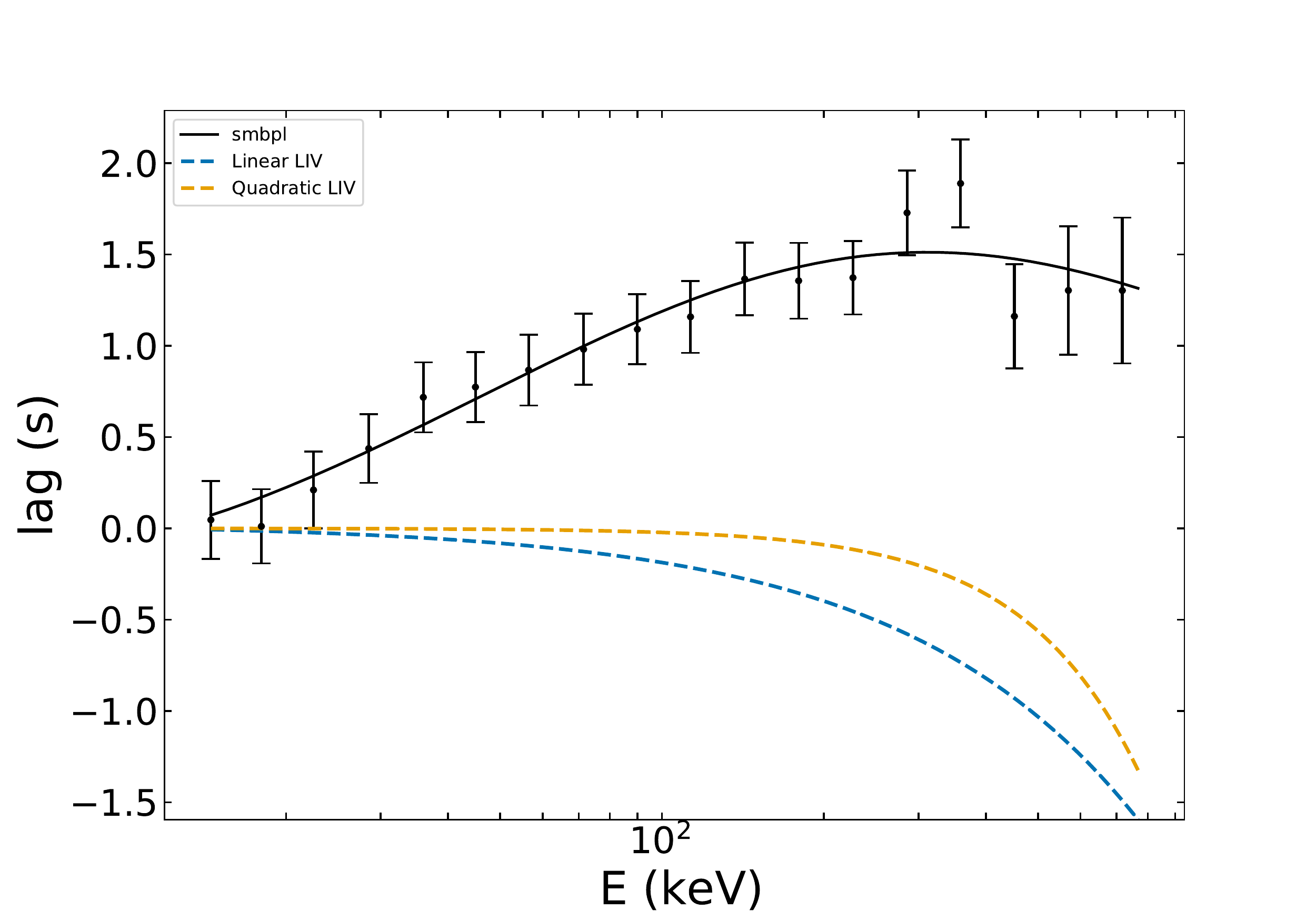}{0.29\textwidth}{GRB 160509A}
 }
\vspace{-0.5cm}
\gridline{\hspace{-1.0cm}\fig{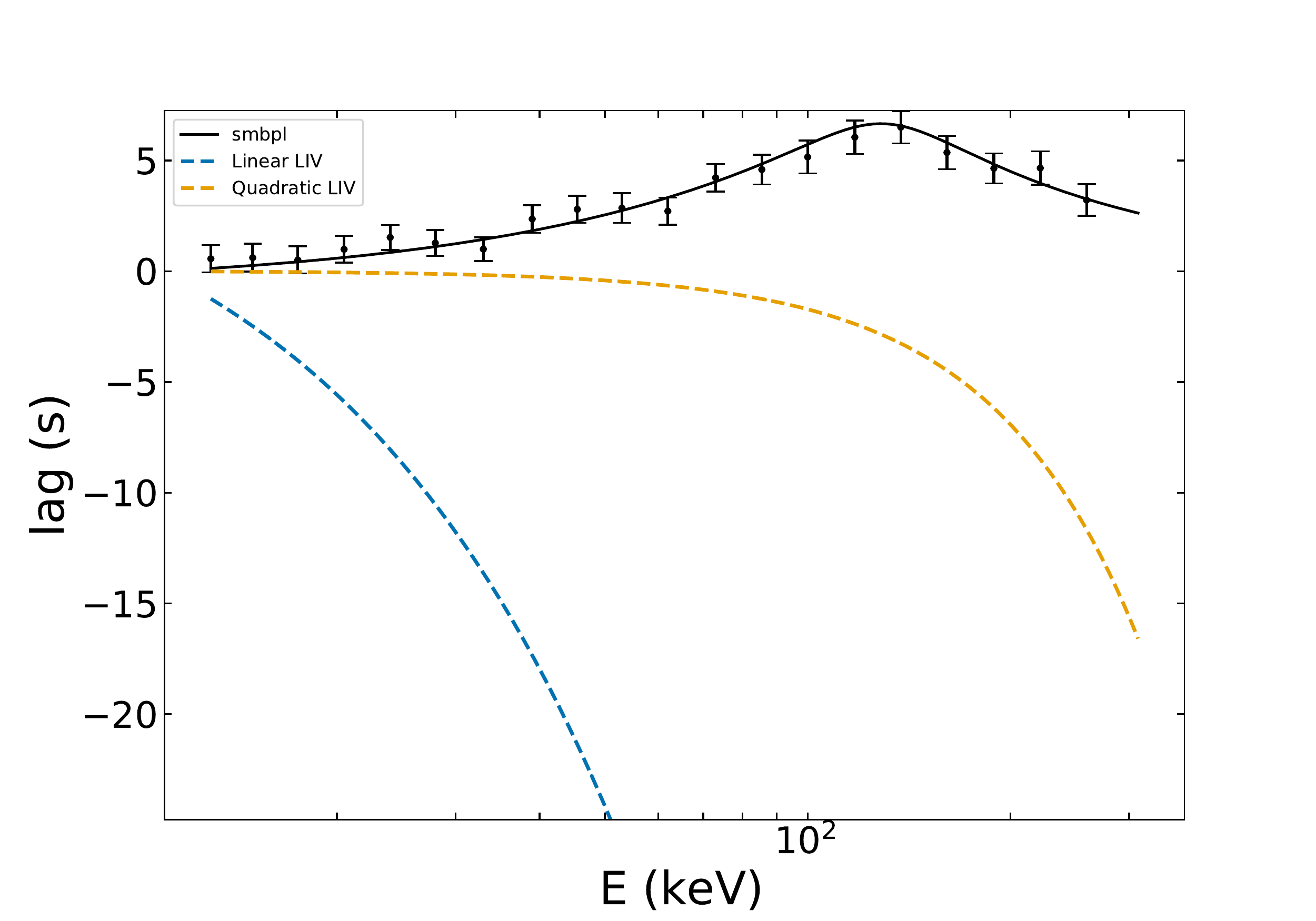}{0.29\textwidth}{GRB 150821A} \hspace{-0.4cm}\fig{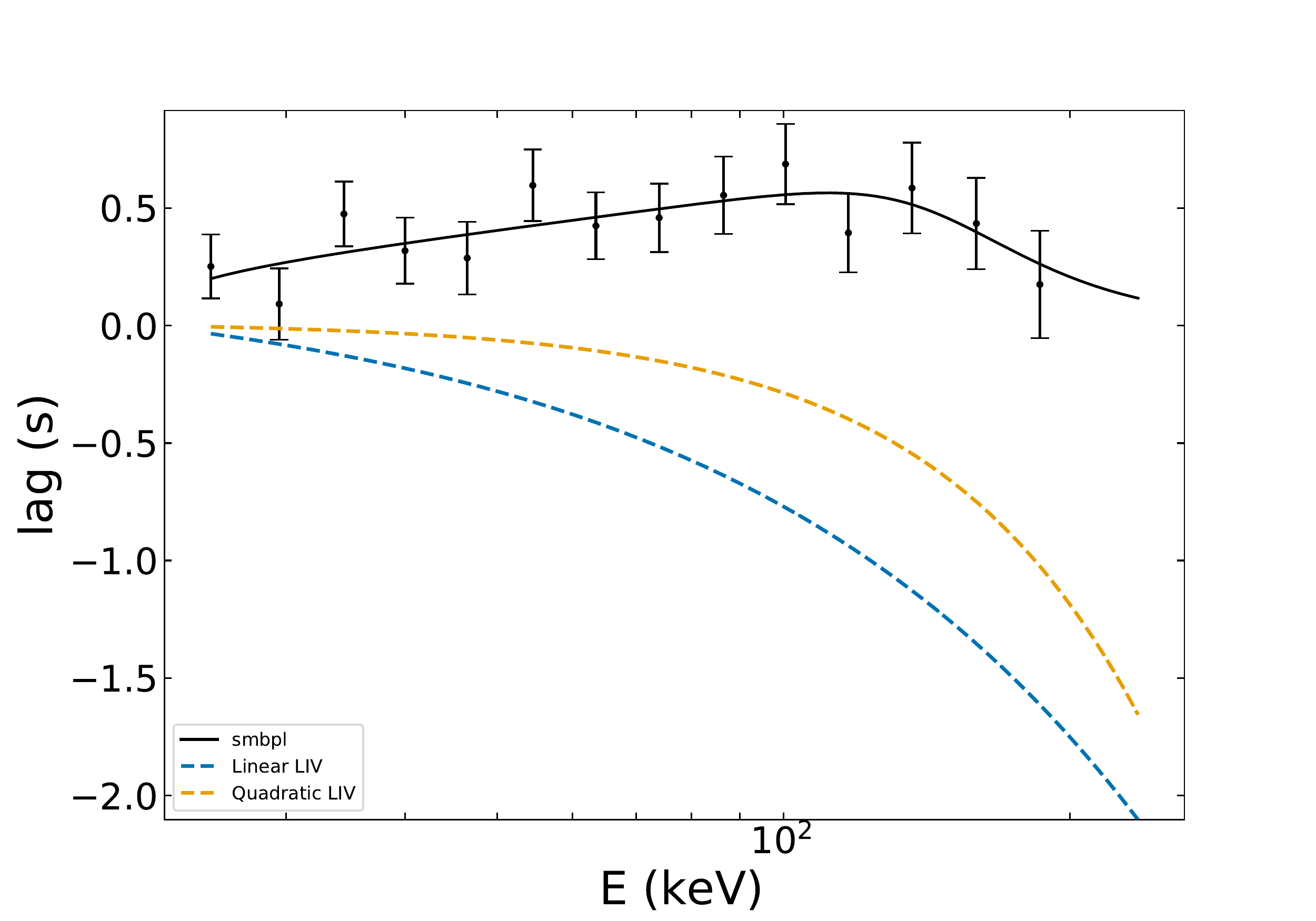}{0.29\textwidth}{GRB 150514A}
 \hspace{-0.4cm} \fig{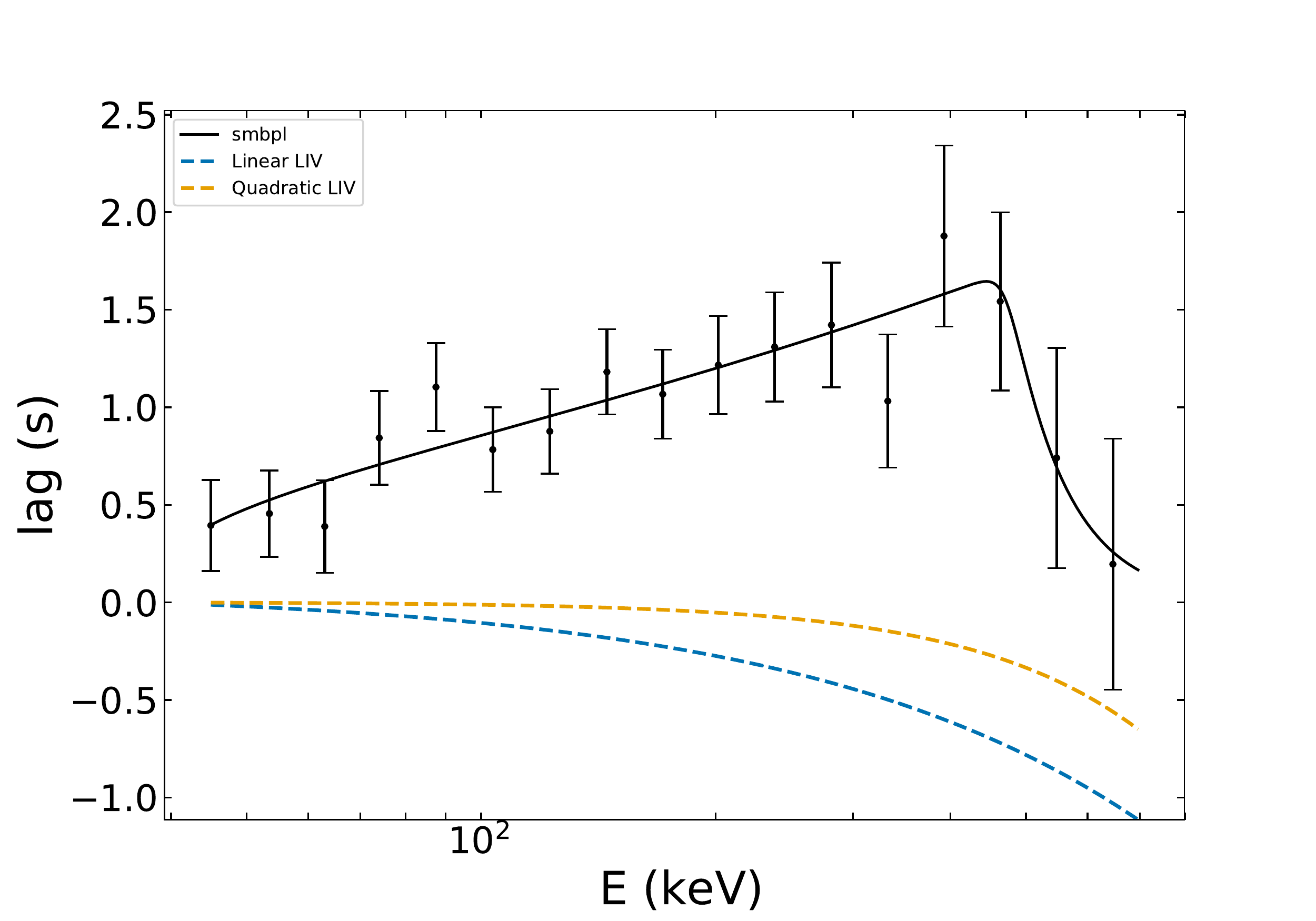}{0.29\textwidth}{GRB 150403A}
 \hspace{-0.4cm} \fig{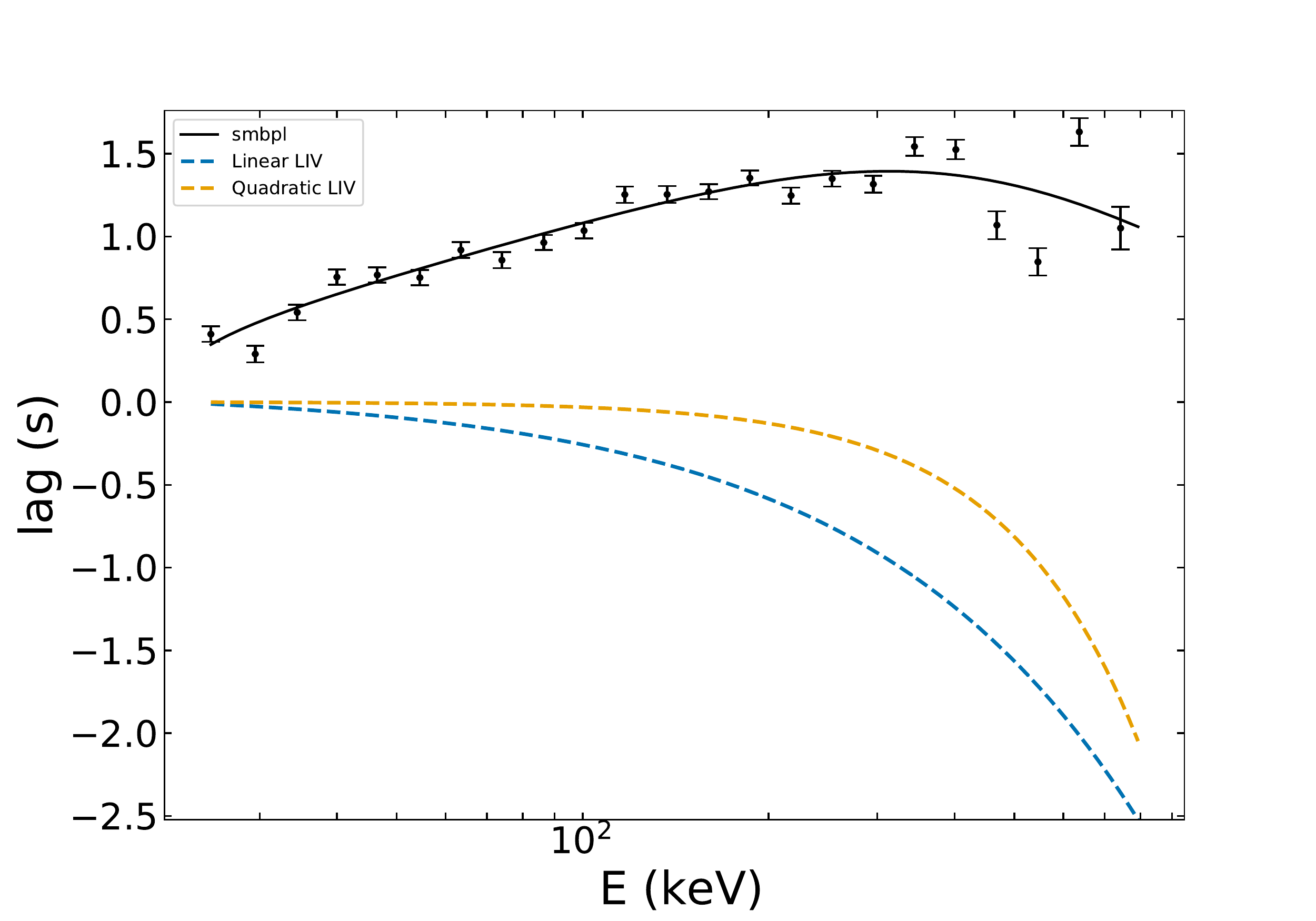}{0.29\textwidth}{GRB 150314A}
 }
\vspace{-0.5cm}
\gridline{\hspace{-1.0cm}\fig{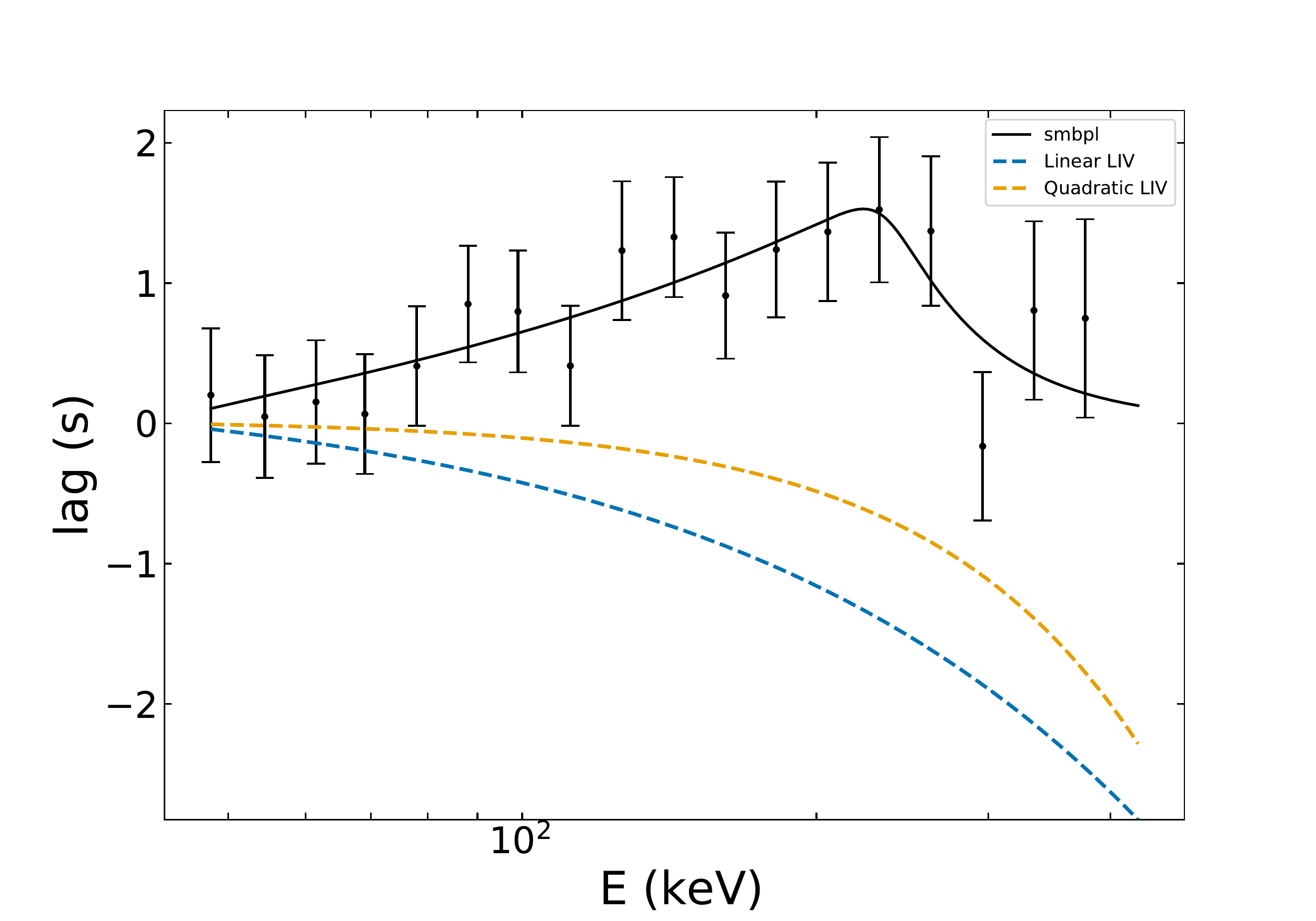}{0.29\textwidth}{GRB 141028A} \hspace{-0.4cm}\fig{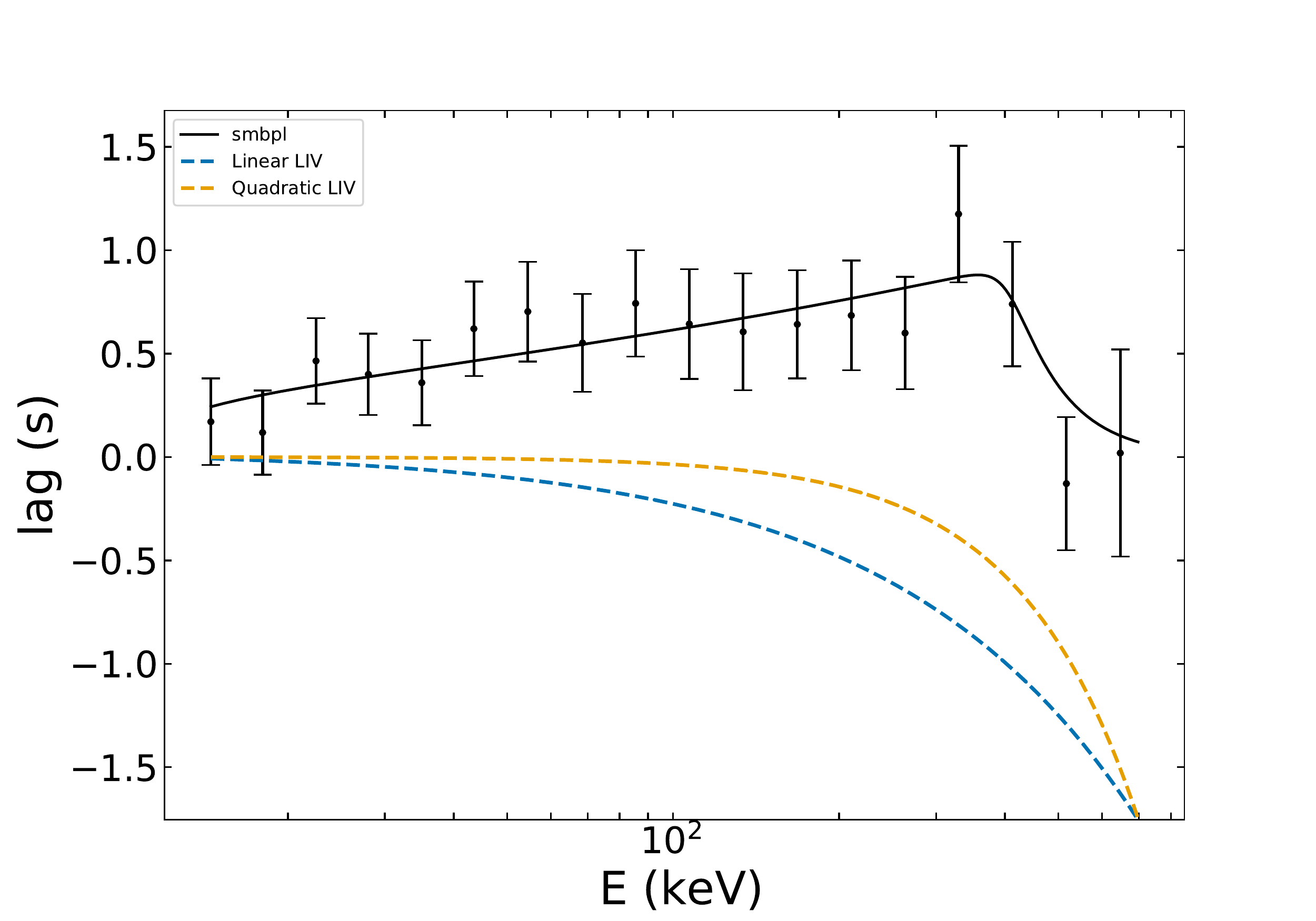}{0.29\textwidth}{GRB 140508A}
 \hspace{-0.4cm} \fig{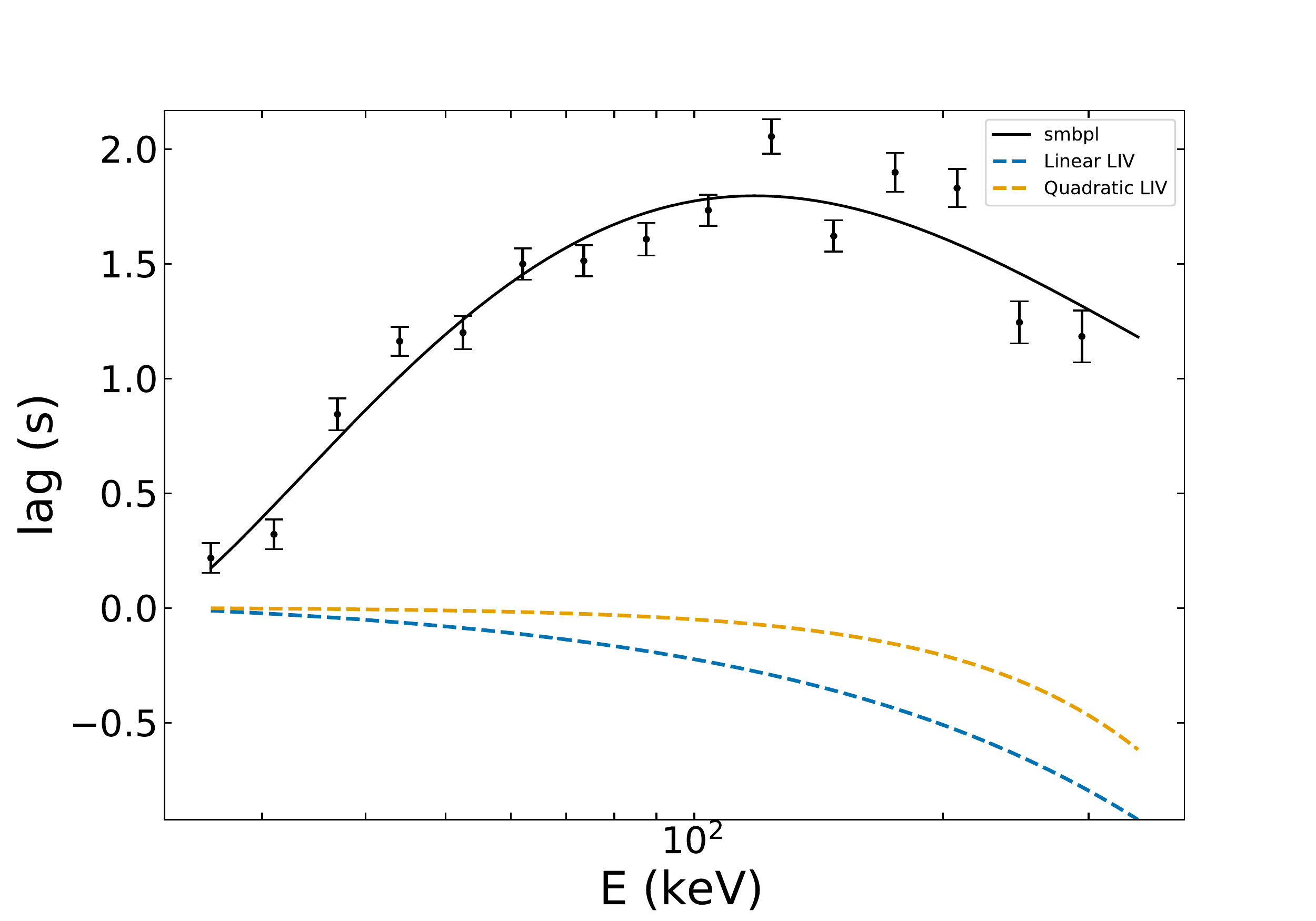}{0.29\textwidth}{GRB 140206A}
 \hspace{-0.4cm} \fig{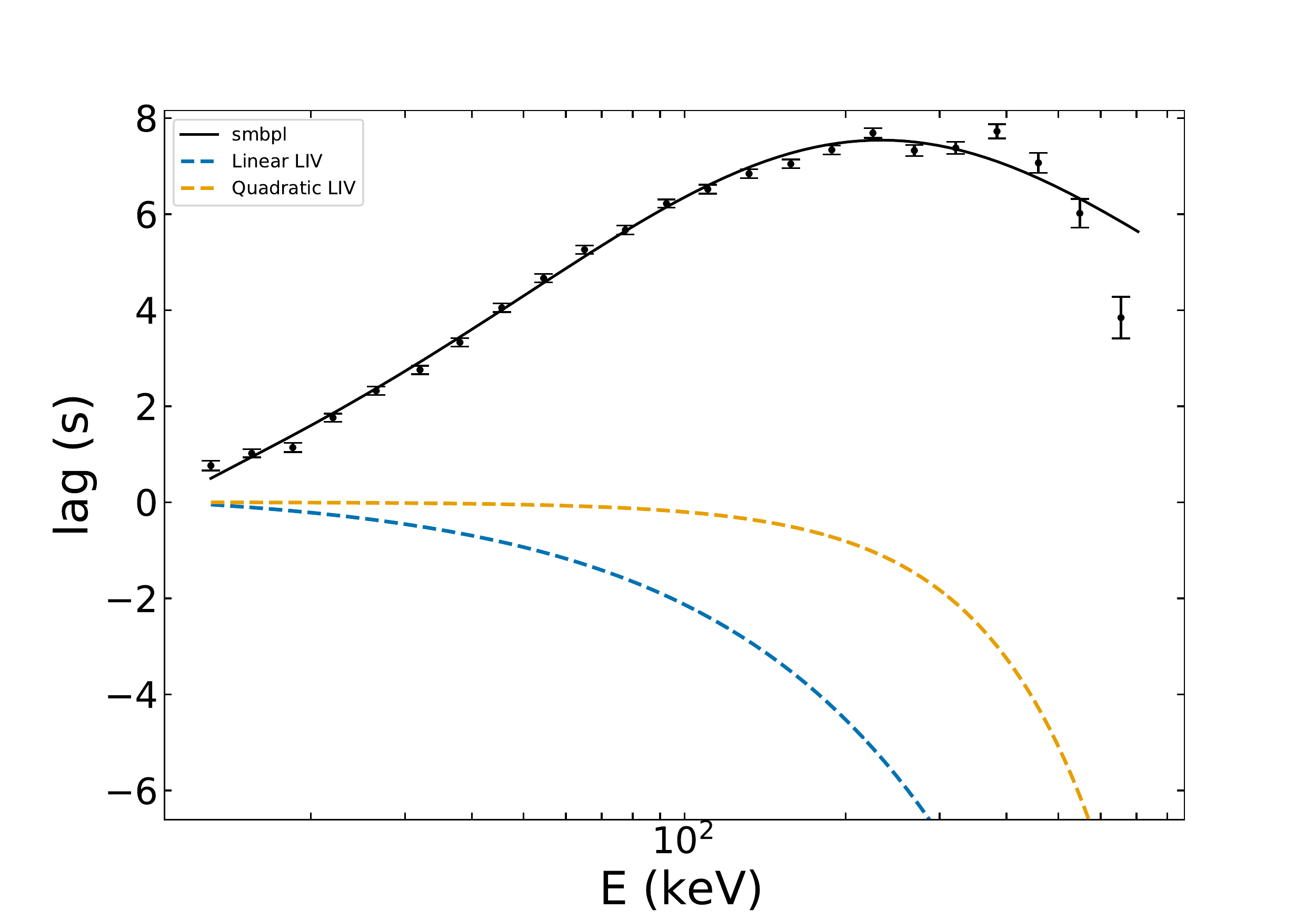}{0.29\textwidth}{GRB 131231A}
 }

\vspace{0cm}
\figcaption{Lag-energy dependence of each GRB in our sample. Solid black lines show the smoothly broken power law (SBPL) fits. Blue and orange dotted lines indicate the maximally allowed LIV-induced lags in linear and quadratic cases, which define the lower limits of the QE energy for each GRB.}
\label{fig1}
\end{figure*}

\label{sec2}
 
To date, more than a thousand GRBs have been observed by \fermi/\gbm \citep{FermiGBMCATELOAG}, of which 135 \textcolor{black}{long-duration} bursts with redshift are measured. These bursts constitute our initial sample which is further screened for the positive-to-negative lag transitions. The final sample for this study, as shown in Table 1, includes 32 GRBs. For each GRB, we extracted its multi-wavelength light curves and calculated its lags using the following steps.

\begin{figure*}[t]
\vspace{-0.5cm}
\gridline{\hspace{-1.0cm}\fig{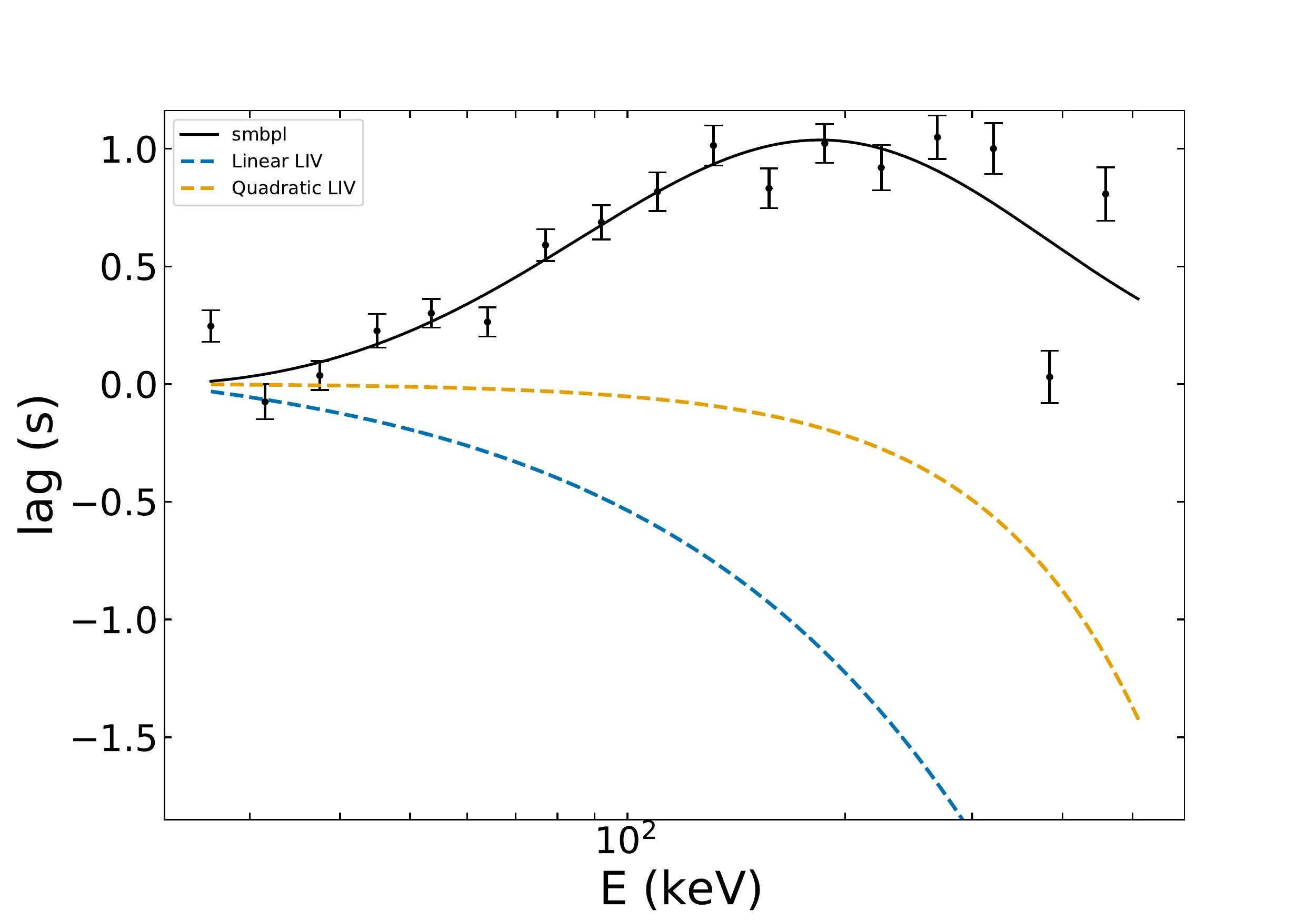}{0.29\textwidth}{GRB 131108A} \hspace{-0.4cm}\fig{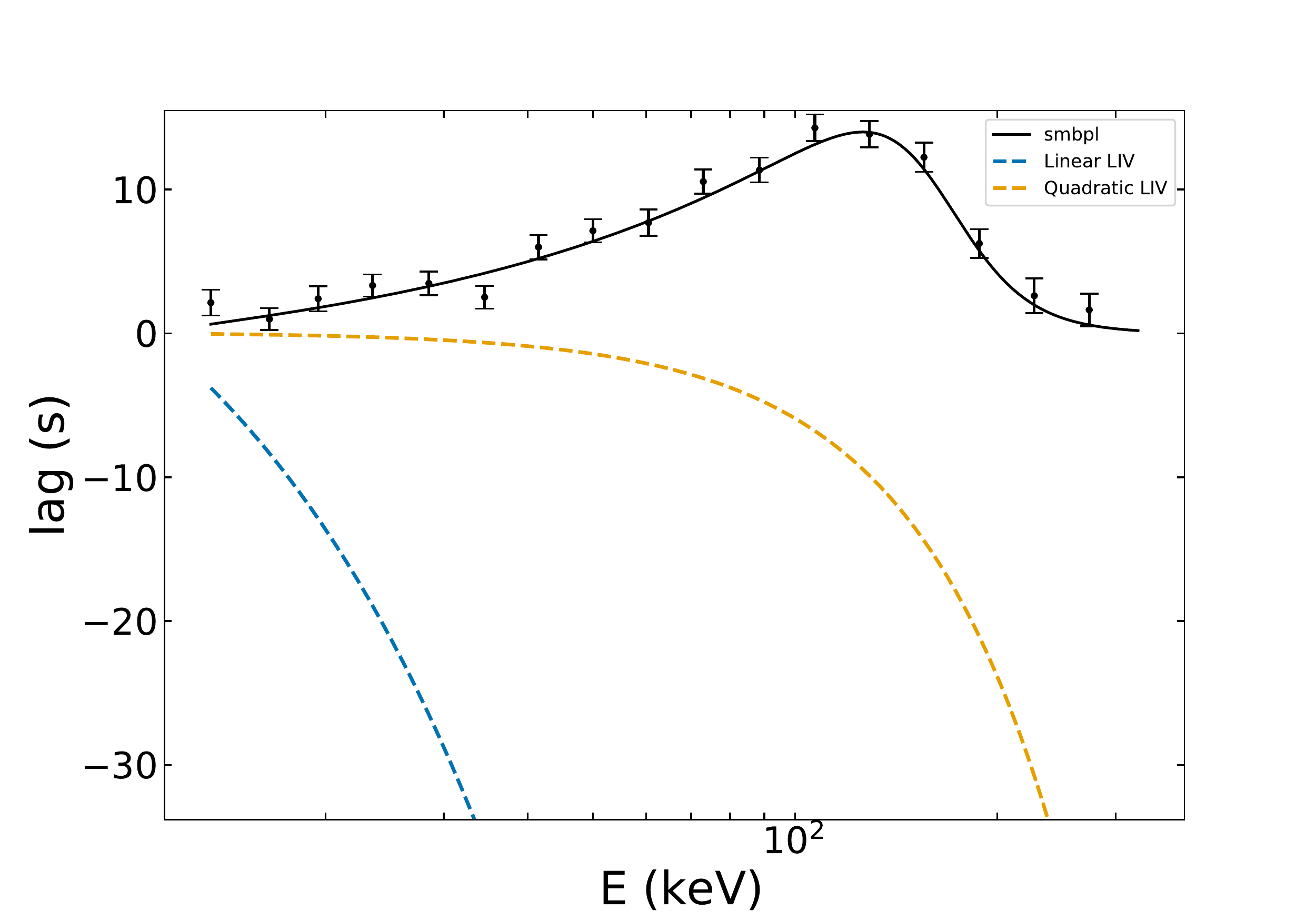}{0.29\textwidth}{GRB 130925A}
 \hspace{-0.4cm}\fig{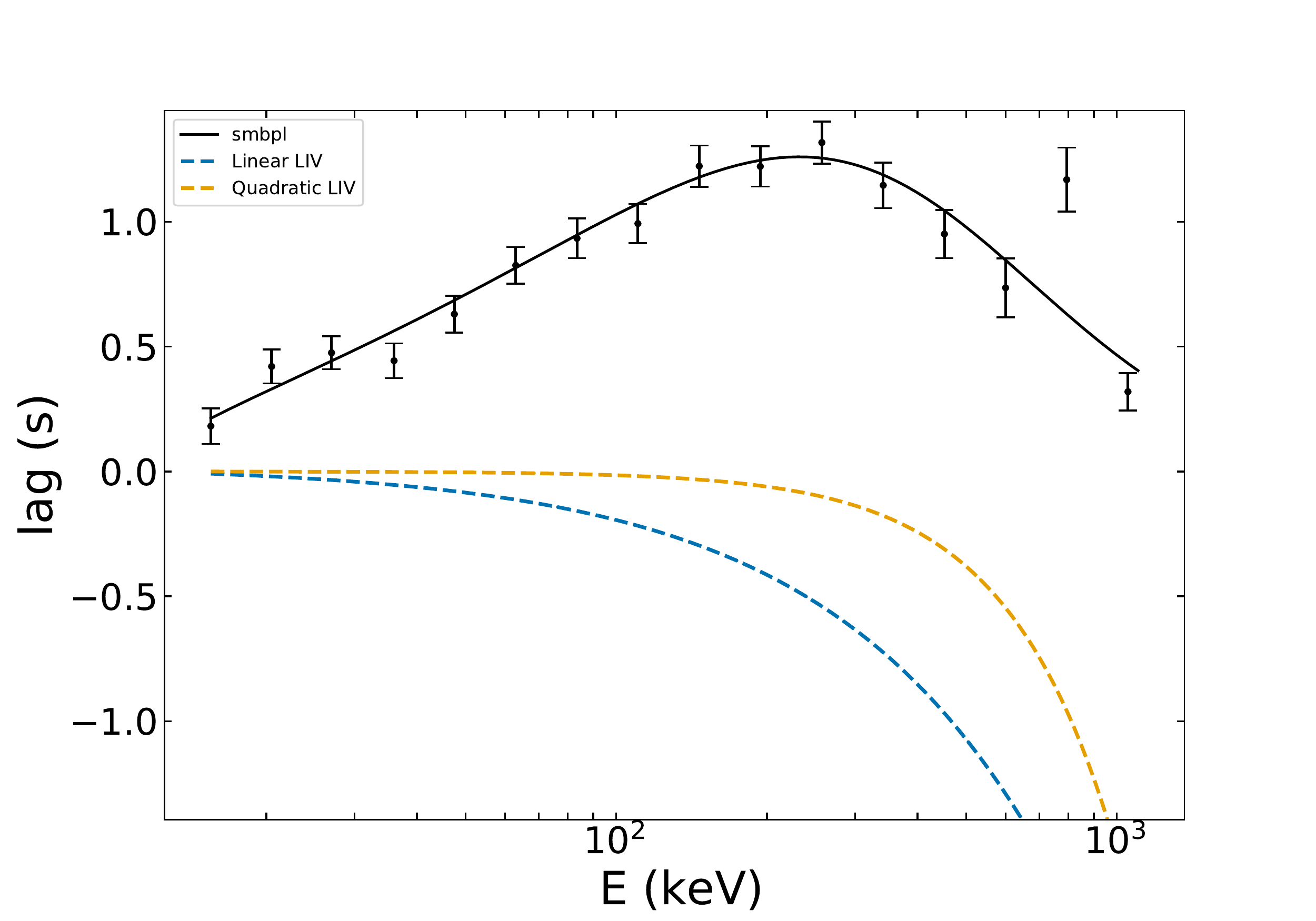}{0.29\textwidth}{GRB 130518A}
 \hspace{-0.4cm}\fig{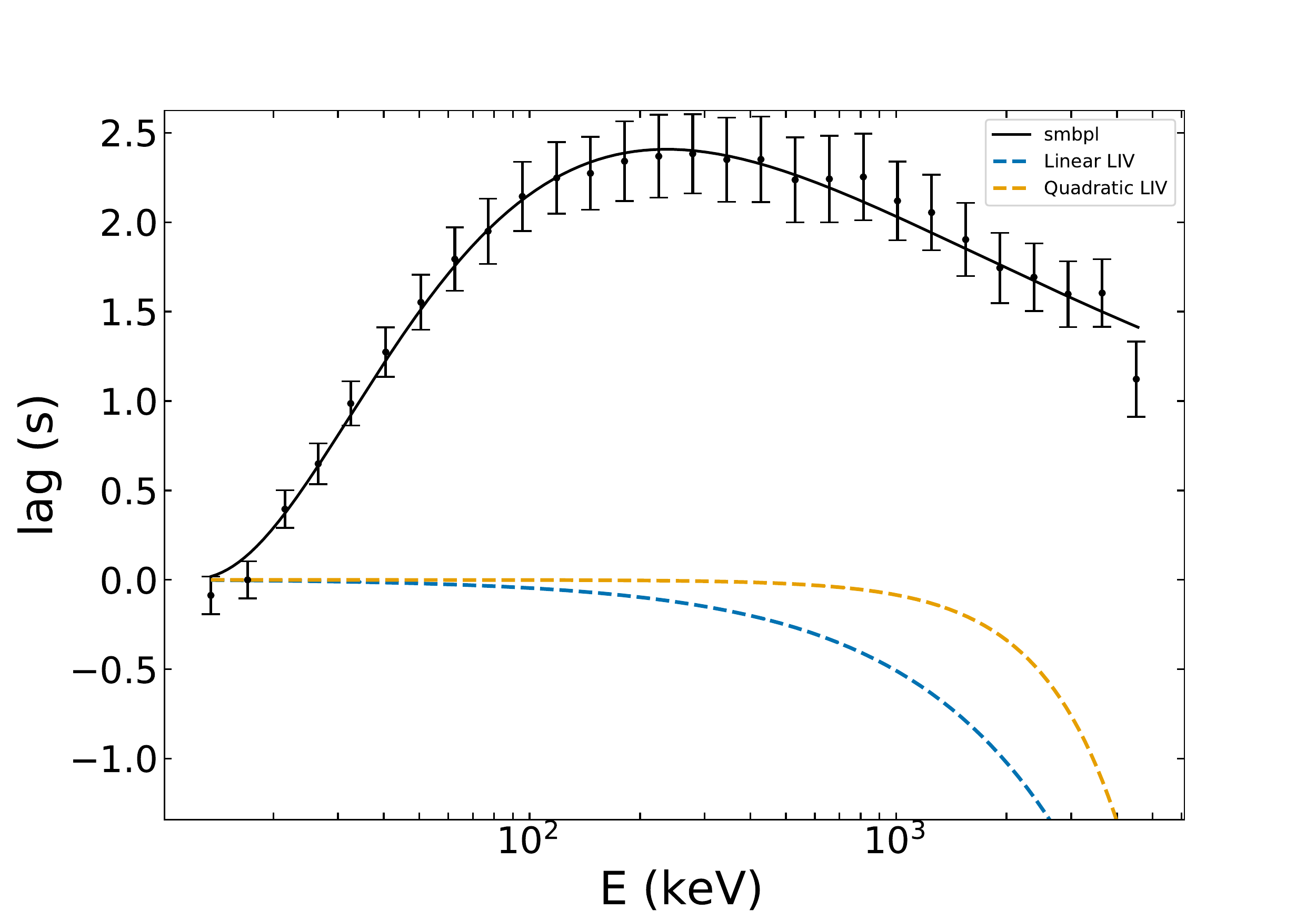}{0.29\textwidth}{GRB 130427A}
 }
\gridline{\hspace{-1.0cm}\fig{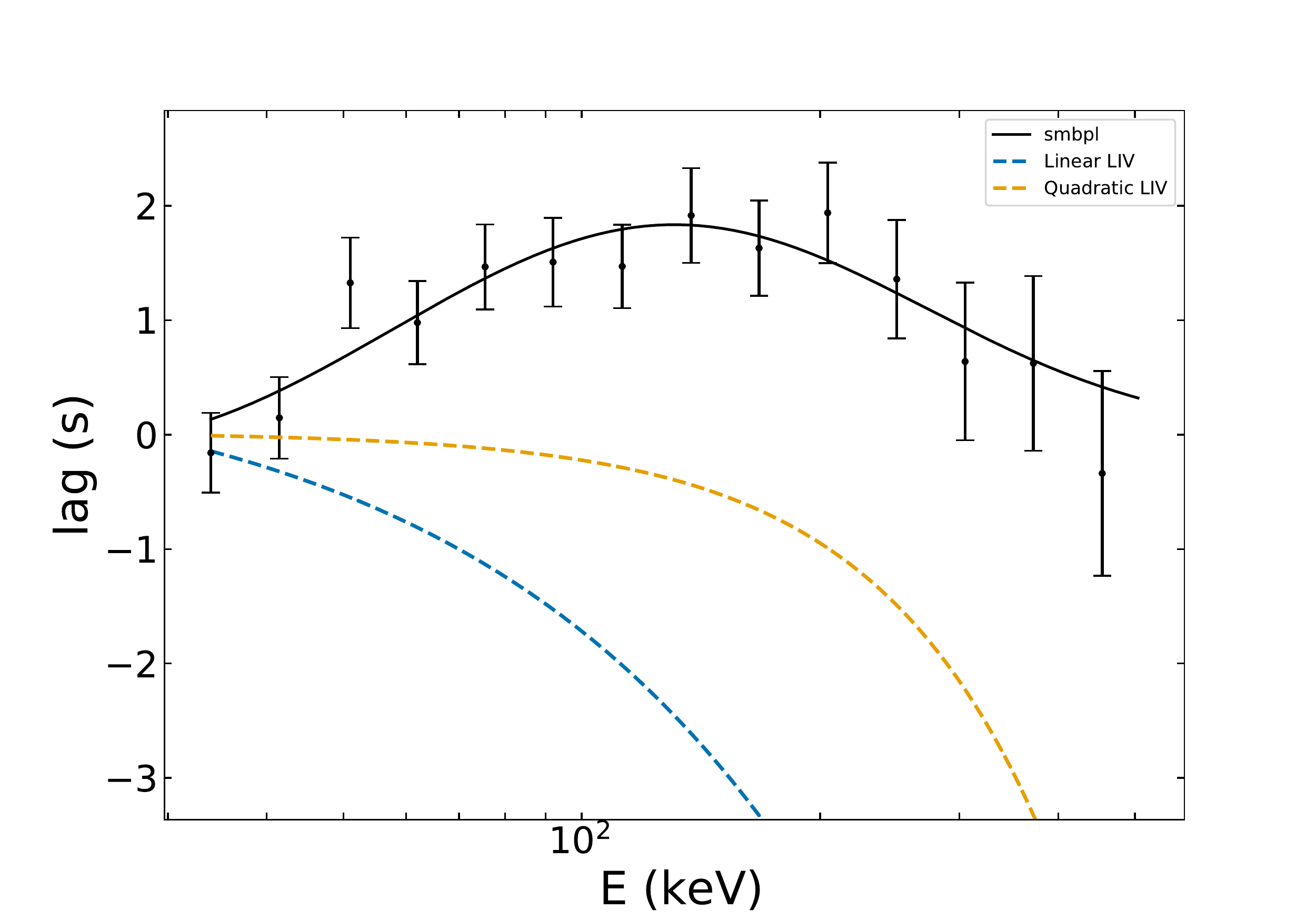}{0.29\textwidth}{GRB 120119A} \hspace{-0.4cm}\fig{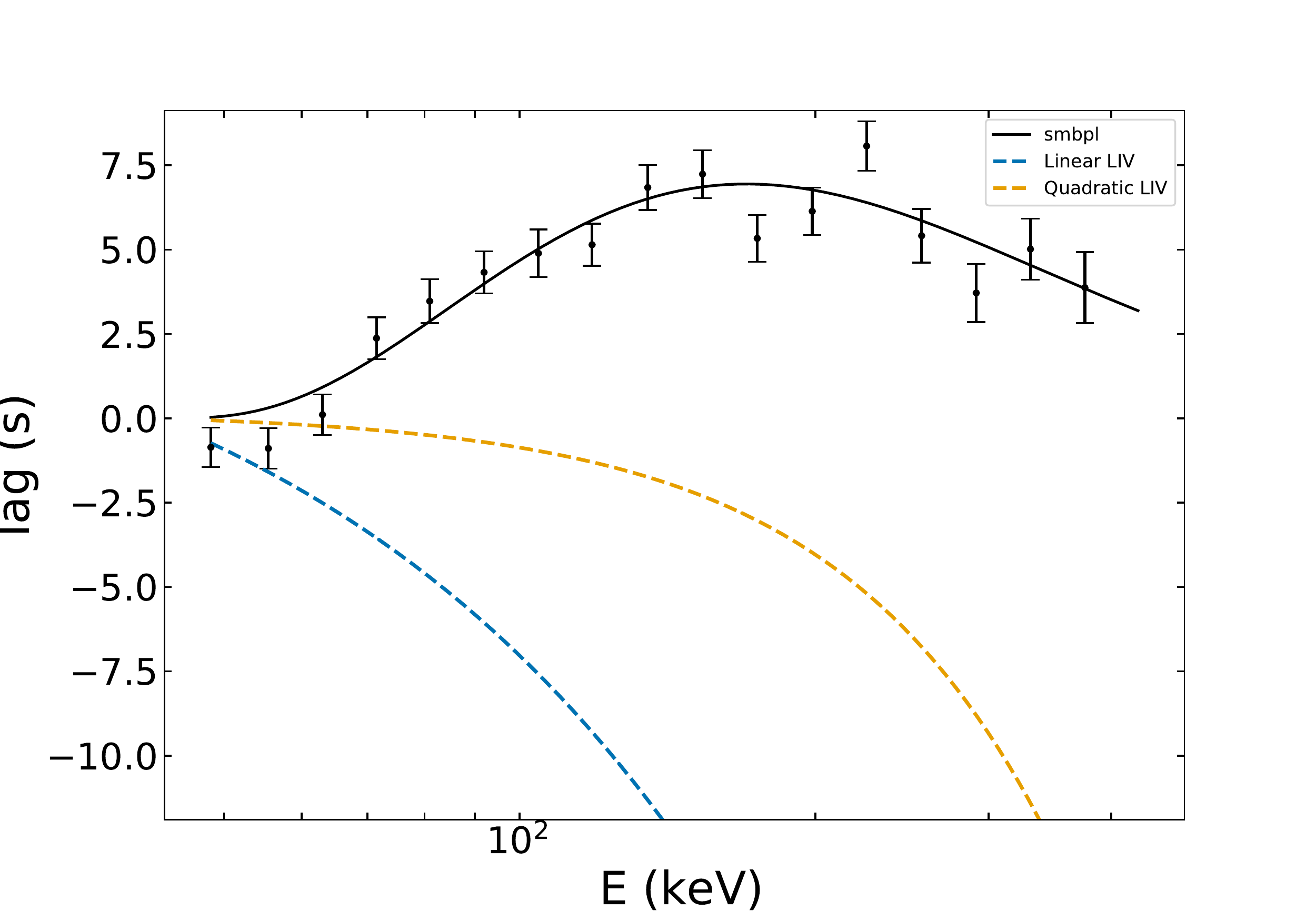}{0.29\textwidth}{GRB 100728A}
 \hspace{-0.4cm} \fig{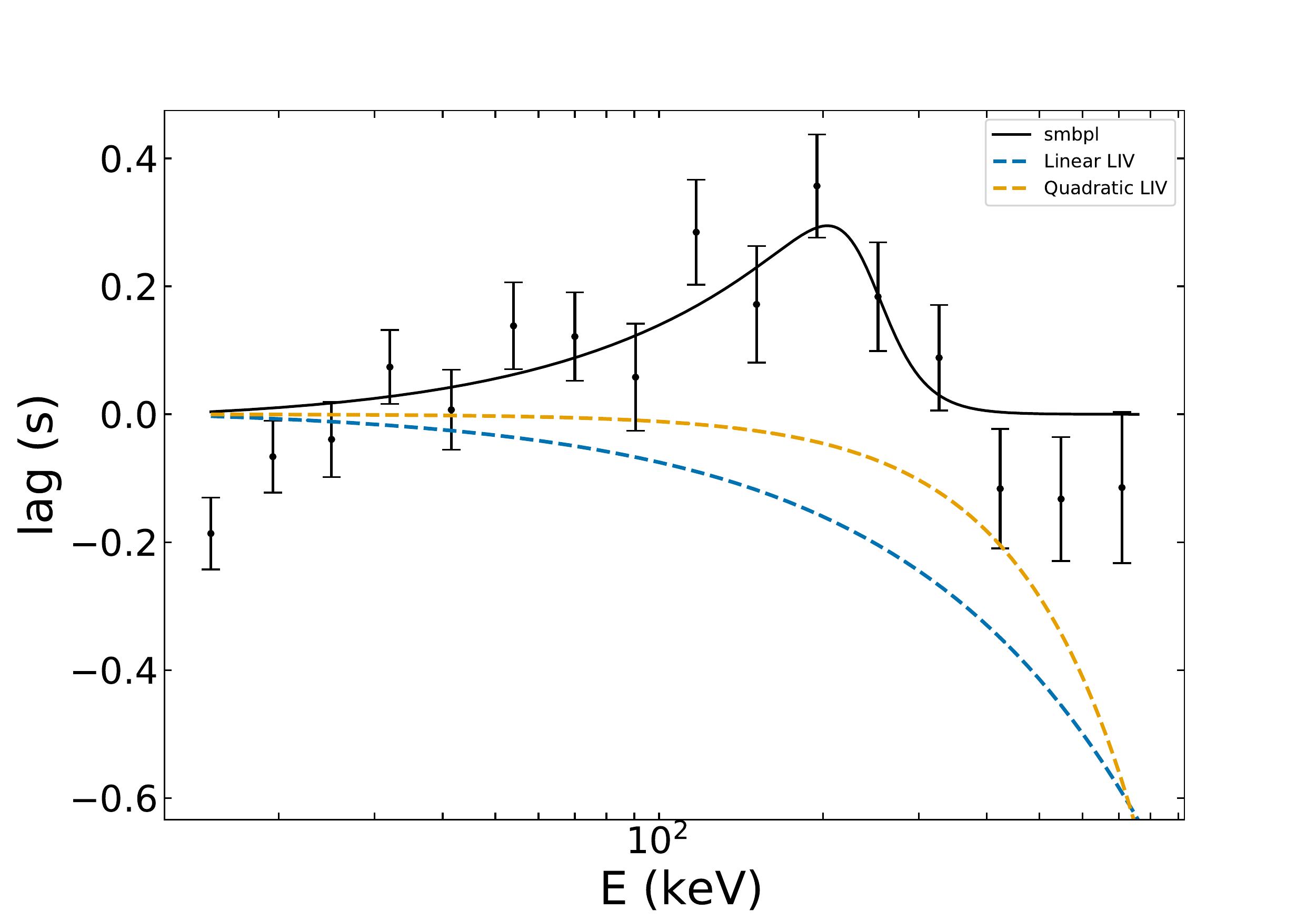}{0.29\textwidth}{GRB 091003A}
 \hspace{-0.4cm} \fig{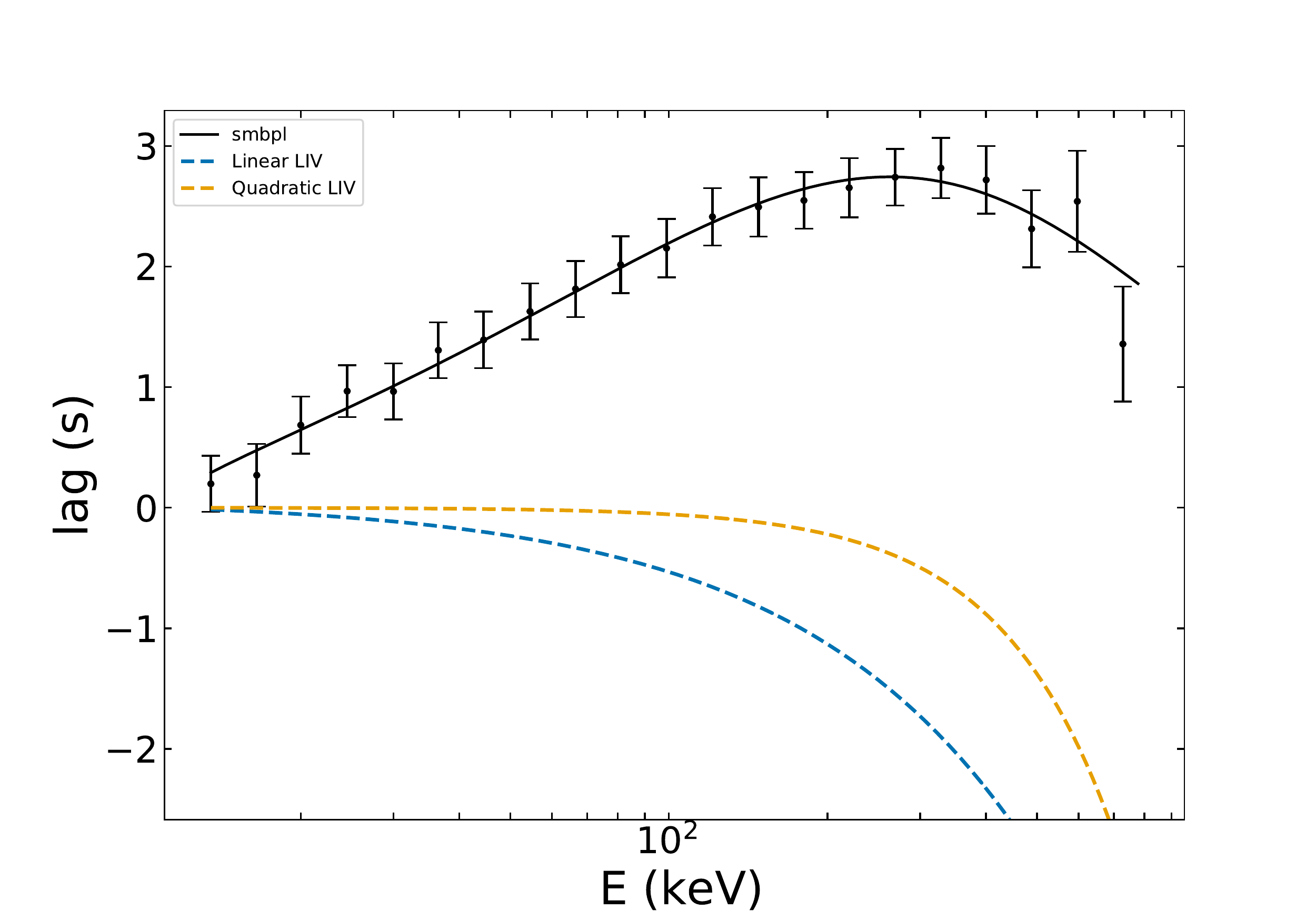}{0.29\textwidth}{GRB 090926A}
 } 
\gridline{\hspace{-1.0cm}\fig{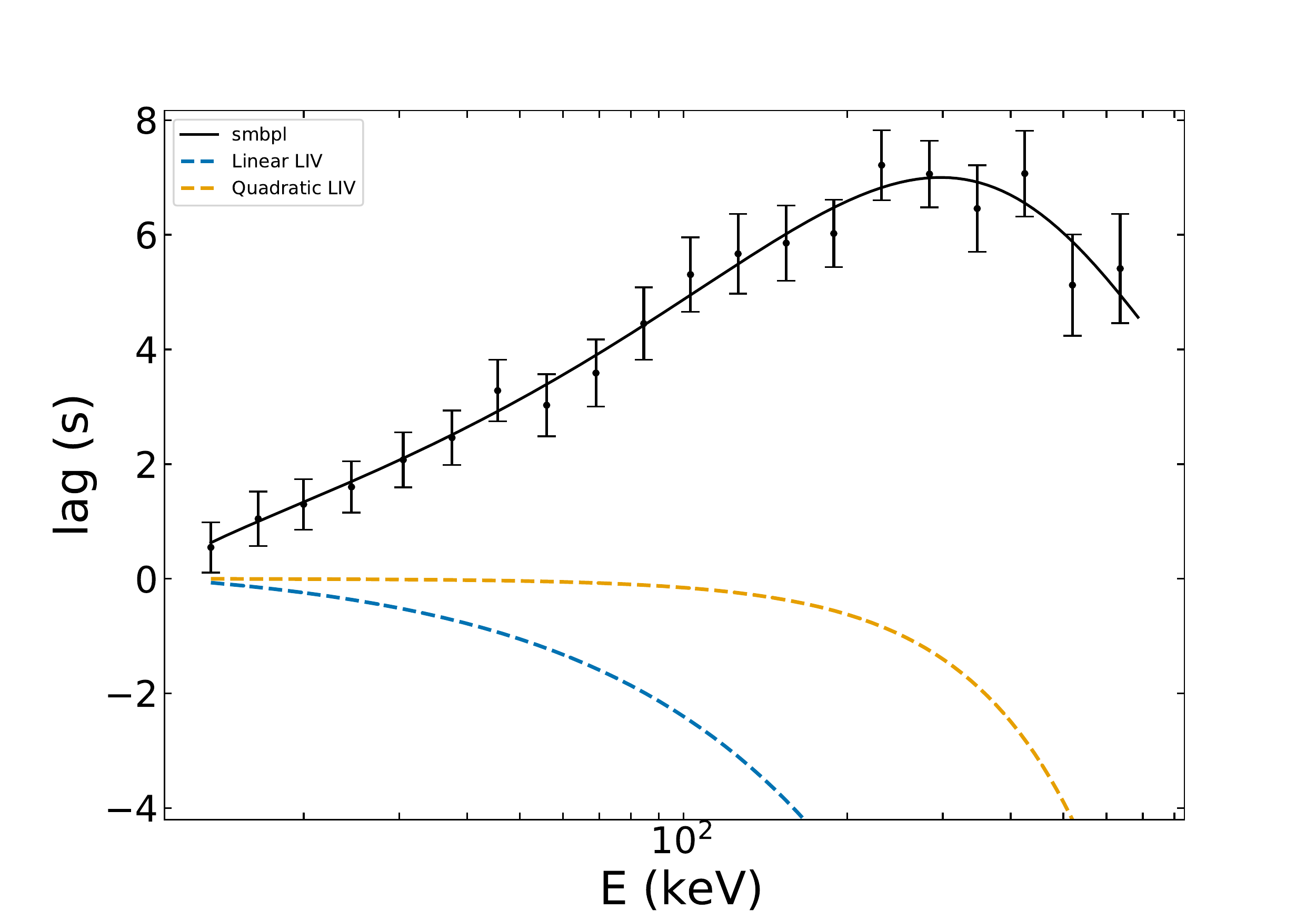}{0.29\textwidth}{GRB 090618} \hspace{-0.4cm}\fig{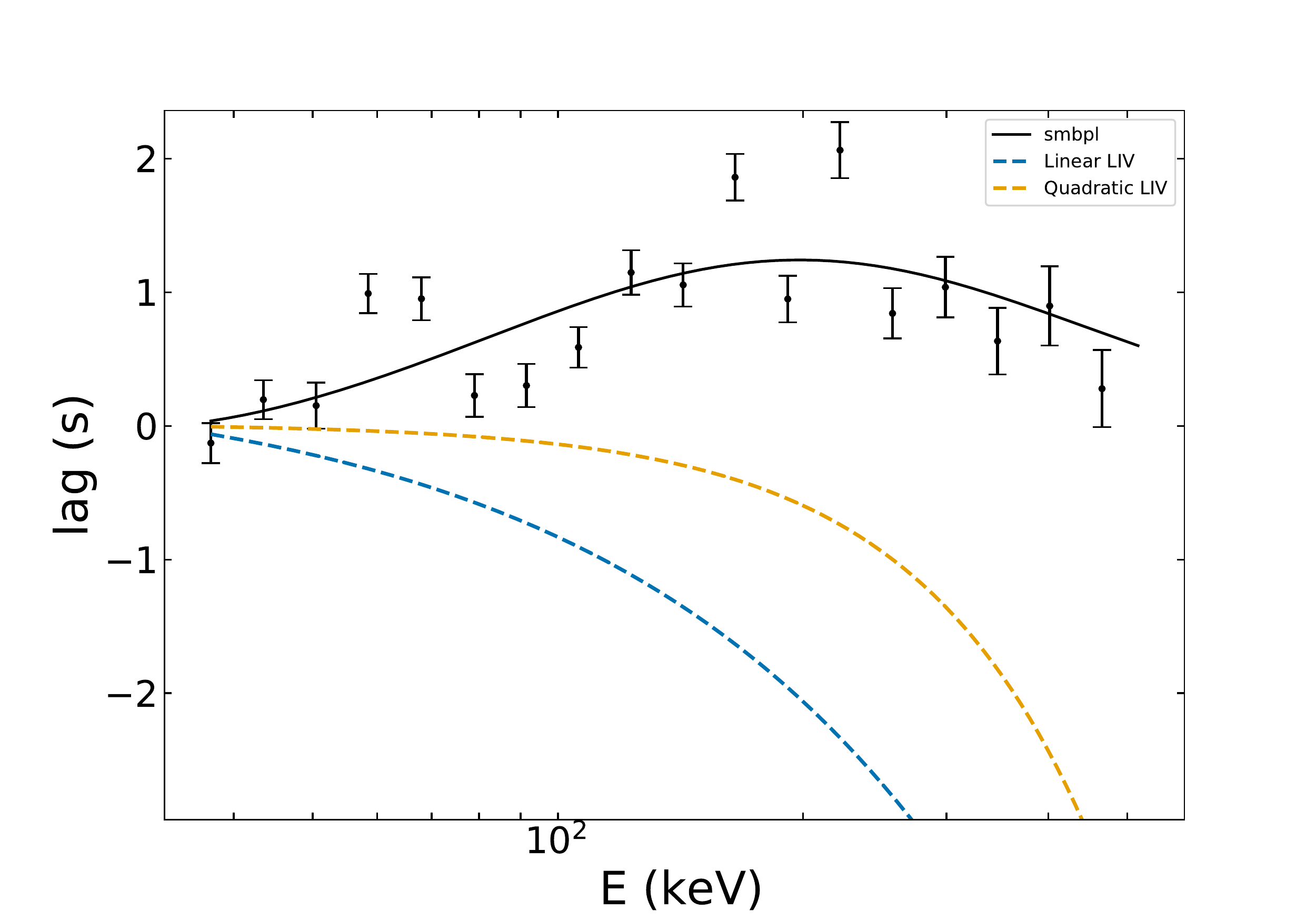}{0.29\textwidth}{GRB 090328}
 \hspace{-0.4cm} \fig{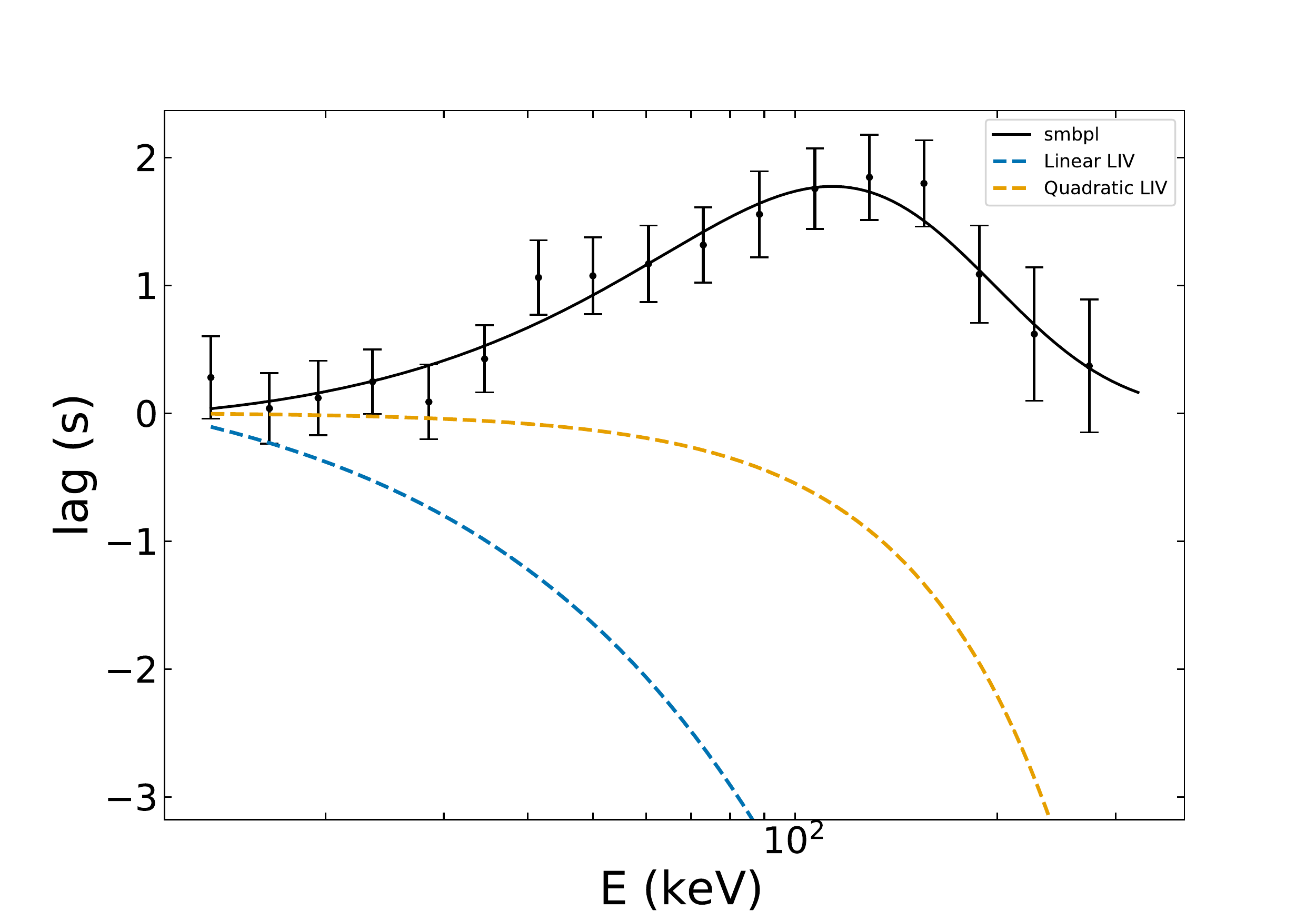}{0.29\textwidth}{GRB 081221}
 \hspace{-0.4cm} \fig{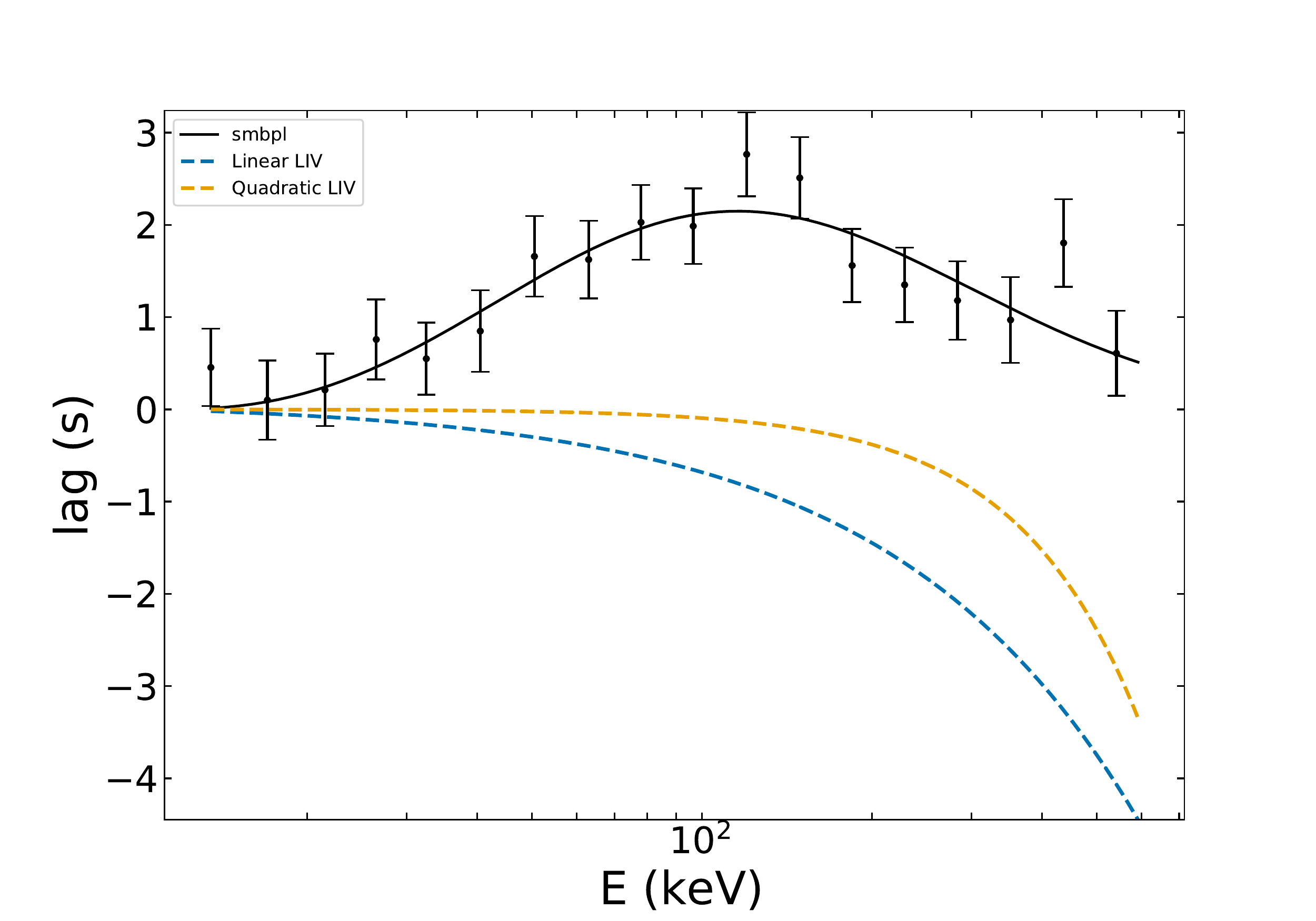}{0.29\textwidth}{GRB 080916C}
 } 
\center{\textbf{Figure 1.}---Continued.}
\end{figure*}

\begin{enumerate}
 \item Light curve extraction. In accordance with \citet{2020ApJ...899..106Y,2021NatAs...5..911Z,2021ApJ...922..237W}, we selected the time-tagged event (TTE) data from the Sodium Iodide (NaI) Scintillation and Bismuth Germanate (BGO) Scintillation detectors onboard Fermi/GBM that had the smallest separation from the location of the burst. Using those data, we then extract the multi-wavelength light curves in a number of {\it N} energy bands, with a bin size of {$\Delta$t } s, between the energy range of [$E_1$, $E_2$]. Here $N$, $\Delta$t, $E_1$, $E_2$ are initially set as free parameters and determined by a series of trials on a burst-by-burst basis so that each of the light curves must have a signal-to-noise ratio $\sigma \geqslant$ 5. The final choices of those parameters are listed in Table 1.

 \item Lag calculation. We calculate spectral lags for any pair of light curves between the lowest energy band and any other band, using the method described in \cite{2012ApJ...748..132Z}. As shown in Figure 1, we plot the lags for each burst as a function of E, the median of the energy boundaries of the higher energy bands. One can see that each of our sample exhibits a positive-to-negative lag transition.

\end{enumerate}
 

\begin{deluxetable*}{cccccc}
\linespread{0.01} 
\tablecaption{\textcolor{black}{List of the 32 GRBs in our sample and the corresponding parameters used for light curve extraction and lag calculation }}
\label{tab1}
\tabletypesize{\small}
\tablewidth{2pt} \tablehead{
\colhead{GRB}&\colhead{Binsize (s)} & \colhead{\textbf{N} (Number of bands) } & \colhead{Energy interval for analysis (keV)}& \colhead{Time interval for analysis (s)}&\colhead{Redshift}}
\startdata
GRB 210619B&0.1&24&[10, 11000]&[-5.0, 13.0]&1.937\tablenotemark{$\small{(1)}$}\\\\
GRB 210610B&0.3&22&[30, 380]&[10.0, 130.0]&1.13\tablenotemark{$(2)$}\\\\
GRB 210204A&0.1&24&[10, 400]&[150.0, 300.0]&0.876\tablenotemark{$(3)$}\\\\
GRB 201216C&0.1&21&[15, 700]&[-5.0, 50.0]&1.10\tablenotemark{$(4)$}\\\\
GRB 200829A&0.1&24&[25, 3500]&[10.0, 35.0]&1.25\tablenotemark{$(5)$}\\\\
GRB 200613A&0.08&22&[30, 300]&[-2.0, 50.0]&1.22\tablenotemark{$(6)$}\\\\
GRB 190114C&0.1&21&[10, 5000]&[-1.0, 14.0]&0.425\tablenotemark{$(7)$}\\\\
GRB 180720B&0.2&17&[25, 17000]&[-5.0, 25.0]&0.654\tablenotemark{$(8)$}\\\\
GRB 180703A&0.1&19&[20, 400]&[-6.0, 35.0]&0.6678\tablenotemark{$(9)$}\\\\
GRB 171010A&0.2&25&[10, 620]&[-8.0, 120.0]&0.3285\tablenotemark{$(10)$}\\\\
GRB 160625B&0.1&29&[10, 20000]&[180.0, 215.7]&1.41\tablenotemark{$(11)$}\\\\
GRB 160509A&0.1&19&[10, 800]&[-3.0, 43.0]&1.17\tablenotemark{$(12)$}\\\\
GRB 150821A&0.6&21&[10, 280]&[-5.0, 140.0]&0.755\tablenotemark{$(13)$}\\\\
GRB 150514A&0.07&15&[20, 200]&[-2.0, 13.0]&0.807\tablenotemark{$(14)$}\\\\
GRB 150403A&0.08&18&[35, 700]&[0.0, 35.0]&2.06\tablenotemark{$(15)$}\\\\
GRB 150314A&0.08&24&[20, 800]&[-3.0, 20.0]&1.758\tablenotemark{$(16)$}\\\\
GRB 141028A&0.1&19&[40, 400]&[-3.0, 45.0]&2.33\tablenotemark{$(17)$}\\\\
GRB 140508A&0.07&19&[10, 720]&[-3.0, 20.0]&1.027\tablenotemark{$(18)$}\\\\
GRB 140206A&0.07&16&[20, 320]&[-1.0, 20.0]&2.73\tablenotemark{$(19)$}\\\\
GRB 131231A&0.1&24&[10, 713]&[-3.0, 63.0]&0.642\tablenotemark{$(20)$}\\\\
GRB 131108A&0.1&18&[20, 500]&[-3.0, 15.0]&2.40\tablenotemark{$(21)$}\\\\
GRB 130925A&0.5&18&[10, 300]&[-10.0, 300.0]&0.347\tablenotemark{$(22)$}\\\\
GRB 130518A&0.1&17&[10, 1200]&[15.0, 42.0]&2.488\tablenotemark{$(23)$}\\\\
GRB 130427A&0.06&29&[10, 5000]&[2.0, 21.0]&0.3399\tablenotemark{$(24)$}\\\\
GRB 120119A&0.1&15&[25, 500]&[-5.0, 55.0]&1.728\tablenotemark{$(25)$}\\\\
GRB 100728A&0.3&18&[40, 400]&[0.0, 225.0]&1.567\tablenotemark{$(26)$}\\\\
GRB 091003A&0.1&17&[10, 800]&[12.0, 25.0]&0.8969\tablenotemark{$(27)$}\\\\
GRB 090926A&0.1&22&[10, 800]&[-2.0, 23.0]&2.1062\tablenotemark{$(28)$}\\\\
GRB 090618&0.1&21&[10, 700]&[40.0, 120.0]&0.54\tablenotemark{$(29)$}\\\\
GRB 090328&0.1&19&[30, 500]&[-8.0, 50.0]&0.736\tablenotemark{$(30)$}\\\\
GRB 081221&0.1&18&[10, 300]&[10.0, 43.0]&2.26\tablenotemark{$(31)$}\\\\
GRB 080916C&0.1&19&[10, 600]&[-3.0, 55.0]&4.35\tablenotemark{$(32)$}\\\\
\enddata
\tablerefs{\scriptsize{(1) GCN Circular 30272 \citep{2021GCN30272}, (2) GCN Circular 30194 \citep{2021GCN.30194....1D}, (3) GCN Circular 29432 \citep{2021GCN.29432....1X}, (4) GCN Circular 29077 \citep{2020GCN.29077....1V}, (5) GCN Circular 28338\citep{2020GCN.28338....1O}, (6) GCN Circular 29320\citep{2020GCN29320}, (7) GCN Circular 23708\citep{2019GCN.23708....1C}, (8) GCN Circular 22996\citep{2018GCN.22996....1V}, (9) GCN Circular 23889\citep{2019GCN.23889....1I}, (10) GCN Circular 22096\citep{2017GCN22096}, (11) GCN Circular 19600\citep{2016GCN.19600....1X}, (12) GCN Circular 18187\citep{2016GCN.19419....1T}, (13) GCN Circular 18187 \citep{2015GCN.18187....1D}, (14) GCN Circular 17822 \citep{2015GCN.17822....1D}, (15) GCN Circular 17672 \citep{2015GCN.17672....1P}, (16) GCN Circular 17583 \citep{2015GCN.17583....1D}, (17) GCN Circular 16983 \citep{2014GCN.16983....1X}, (18) GCN Circular 16231 \citep{2014GCN.16231....1W}, (19) GCN Circular 15800 \citep{2014GCN.15800....1M}, (20) GCN Circular 15645 \citep{2014GCN.15645....1X}, (21) GCN Circular 15470\citep{2013GCN.15470....1D}, (22) GCN Circular 15249\citep{2013GCN.15249....1V}, (23) GCN Circular 14685\citep{2013GCN.14685....1S}, (24) GCN Circular 14491 \citep{2013GCN.14491....1F}, (25) GCN Circular 12867\citep{2012GCN.12867....1M}, (26) GCN Circular 14500\citep{2013GCN.14500....1K}, (27) GCN Circular 10031\citep{2009GCN.10031....1C}, (28) GCN Circular 9942\citep{2009GCN..9942....1M}, (29) GCN Circular 9518 \citep{2009GCN..9518....1C}, (30) GCN Circular 8766\citep{2009GCN..8766....1D}, (31) \cite{2012ApJ...749...68S}, (32) \cite{2013ApJ...774...76A}
}}\end{deluxetable*}

\section{The Model}
\label{sec3}

To model the observed lag-E behaviors in Figure 1, one has to consider the following two components:
\begin{enumerate}
 \item The intrinsic lag, $\Delta t_{\mathrm{int}}$, due to the GRB radiation itself. As pointed out in \S 1, the exact physical process causing spectral lags remains an open question. Nevertheless, some previous studies 
 \citep[e.g.,][]{2017ApJ...834L..13W} assume a simple power law model for $\Delta t_{\mathrm{int}}$. However, such a power law model doesn't account for the negative lags. In this study, we use a smoothly broken power-law function to model the energy-dependent lags, i.e., $\Delta t_{\mathrm{int}}$ is in form of
 
 \begin{equation}
 \begin{aligned}
\hspace{-0.4
cm}\Delta t_{\mathrm{int}}=\zeta\left(\frac{E-E_{0}}{E_{b}}\right)^{\alpha_{1}}\left\{\frac{1}{2}\left[1+\left(\frac{E-E_{0}}{E_{b}}\right)^{1 / \mu}\right]\right\}^{\left(\alpha_{2}-\alpha_{1}\right) \mu},
\end{aligned}
\label{eq2}
\end{equation}
where $\zeta$ is the normalization amplitude, $\alpha_{1}$ and $\alpha_{2}$ are the two slopes before and after the transition energy, $E_{\rm b}$ and $\mu$ measures the smoothness of the transition. We note that when $\alpha_1=\alpha_2$, Eq. (1) becomes a simple power law, as used in \cite{2017ApJ...834L..13W}.

\begin{table}
\caption{Allowed range of the fitting parameters}
\begin{center}
\begin{tabular}{cc}
\hline 
Parameters & Range \\
\hline
$\zeta$ & [0.0, 4.0] \\[4pt]
$E_{\rm b}$&[0.0, 5000.0] \\[4pt]
$\alpha_{1}$ &[-3.0, 10.0] \\[4pt]
$\mu$ (G) &[0.0, 3.0] \\[4pt]
$\alpha_{2}$ (s) & [-10, 3] \\[4pt]
\hline
$E_{\rm QG,1}$ (GeV) & [0, 10$^{20}$] \\[4pt]
$E_{\rm QG, 2}$ (GeV) & [0, 10$^{15}$] \\[4pt]

\hline
 		
\label{tab2}
\end{tabular}%
\end{center}
\end{table}

 \item The LIV lag. In the QG theory, a small-scale structure in spacetime can cause a deformed photon dispersion relation, which can be formulated as an \citep{1998Natur.393..763A,2008JCAP...01..031J}:
\begin{equation}
 \begin{aligned}
c^{2}{p}^{2}=E^{2}\left[1+f\left(E / E_{\mathrm{QG}}\right)\right],
\end{aligned}
\label{eq3}
\end{equation}
where $c$ is the speed of light, $p$ is the photon momentum, $E_{\rm QG}$ is the QG energy scale, $f\left(E / E_{\mathrm{QG}}\right)$ is the model-dependent function of the dimensionless ratio of $E / E_{\mathrm{QG}}$. Noticing $f$ =0 at $E=0$, $f$ can be expanded as Taylor series in a small energy condition($E \ll E_{\mathrm{QG}}$) at $a=0$ as :
\begin{equation}
 f={ \sum _{n=0}^{\infty }{\frac {f^{(n)}(a)}{n!}}(\frac{E}{E_{\rm QG}}-a)^{n}}= \sum _{n=1}^{\infty }{\frac {f^{(n)}(0)}{n!}}(\frac{E}{E_{QG}})^{n}.
\end{equation}
Defining $E_{\rm QG,n}$ as the $n$th-order quantum gravity energy by 
\begin{equation}
E_{QG,n}^{n} =s_\pm{E^n_{\mathrm{QG}}}\frac{n!}{f^{(n)}(0)},
\end{equation}
we can further substitute Eqs. (3) and (4) into Eq. (2), so 
\begin{equation}
 c^{2} {p}^{2}\simeq E^{2}\left[1+\sum_{\textcolor{black}{n=1}}^{\infty} s_{\pm}\left(\frac{E}{E_{\mathrm{QG,n}}}\right)^{n}\right],
\end{equation}

where 
$s_\pm$ represents the sign of the LIV effect. $s_\pm=+1$ ($-1$) corresponds to the subluminal (superluminal) scenario so the high energy photons travel slower (faster) than low energy photons. We only consider the case where $s_\pm=+1$ in this work to account for the negative lags.

In practice, it is convenient to replace the right side of Eq. (5) with its leading term in order n, so 

\begin{equation}
 c^{2} {p}^{2}\simeq E^{2}\left[1+ s_{\pm}\left(\frac{E}{E_{\mathrm{QG,n}}}\right)^{n}\right].
\end{equation}

One can further derive the photon propagating speed as
\begin{equation}
\label{eq5}
 \begin{aligned}
v(E)=\frac{\partial E}{\partial p} \simeq c\left[1- s_{\pm} \frac{n+1}{2}\left(\frac{E}{E_{\mathrm{QG}, n}}\right)^{n}\right].
\end{aligned}
\end{equation}

\begin{figure}
\gridline{\hspace{-1cm}\fig{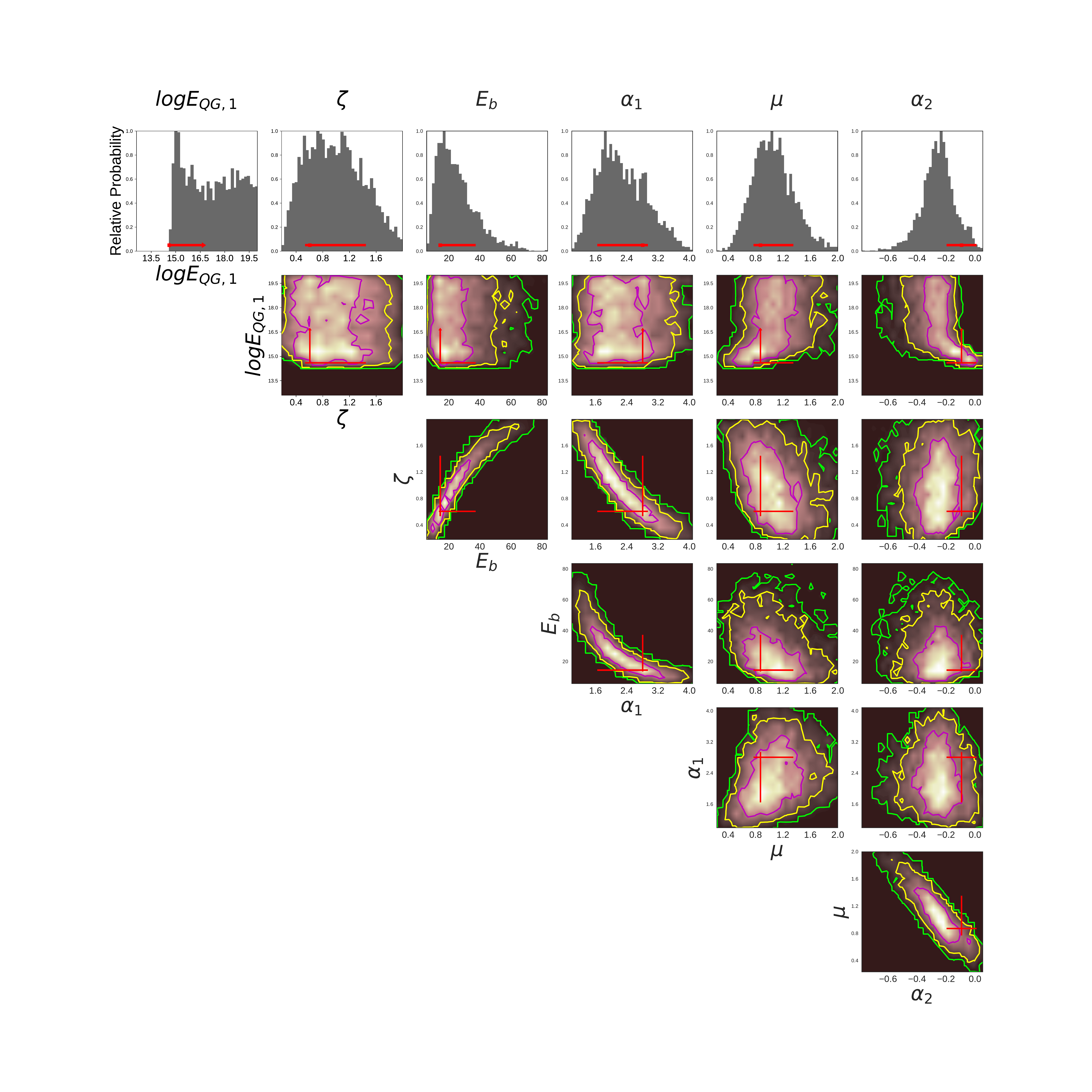}{0.5\textwidth}{}}
\vspace{-2cm}
\gridline{\hspace{-1cm}\fig{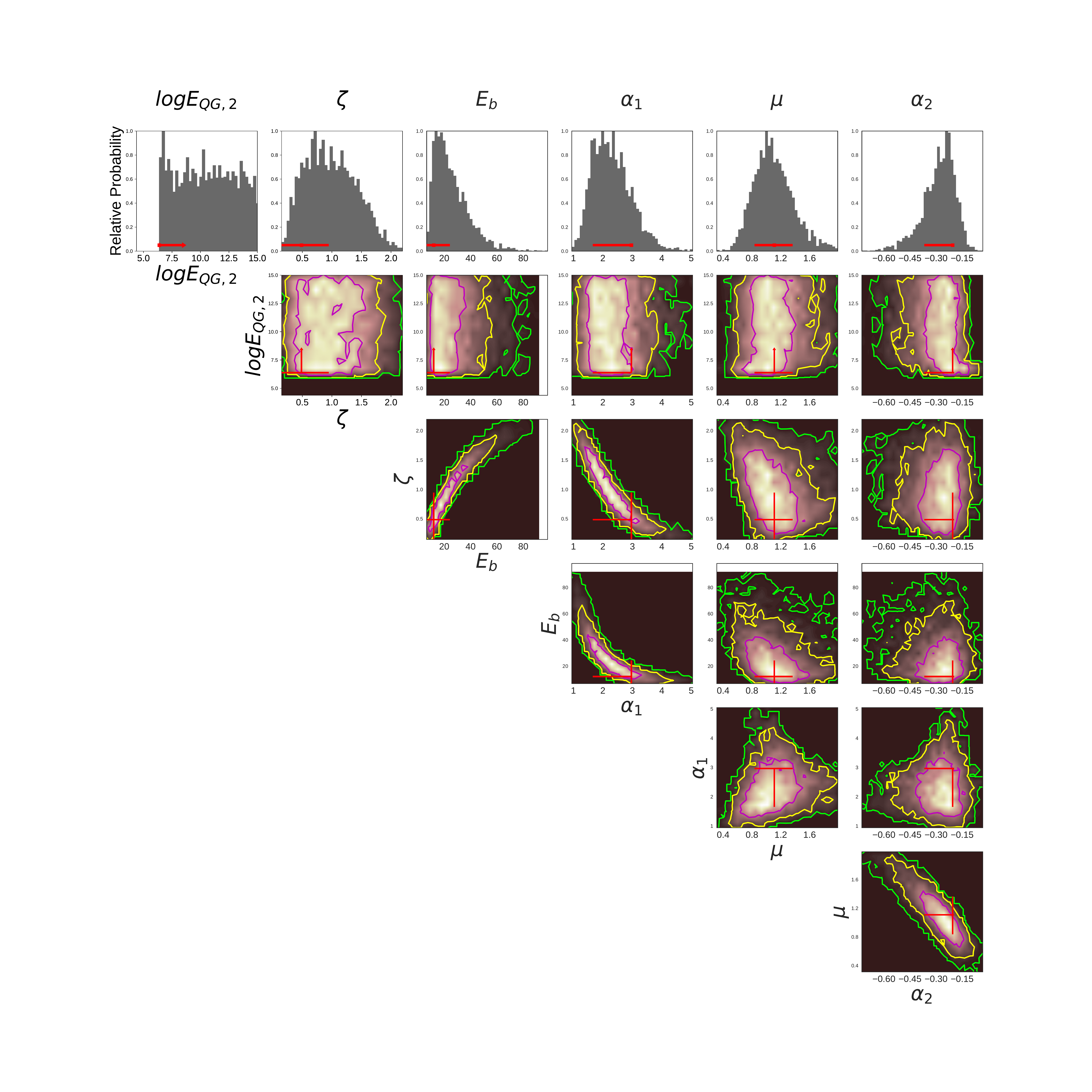}{0.5\textwidth}{}}
\caption{
Distribution of posterior probability of parameters of linear and quadratic model for GRB 130427A. Upper panel: Gray histograms and contours represents the distributions of posterior probability of parameters for linear LIV model in one and two dimensions. Red arrows for parameter $E_{\rm QG}$ indicates its lower limits and red crosses show the best-fit values and their 1-$\sigma$ error ranges. Lower panel: same but for quadratic (n = 2) case.
\label{fig2}}
\end{figure}

Eq. (7) suggests that two photons with different energy arrive at observers with a time delay, even both are emitted from one GRB concurrently. Considering cosmological expansion, the LIV-induced lag can be written as

\begin{equation}
 \begin{aligned}
 \Delta t_{\mathrm{LIV}}=-\frac{1+n}{2 H_{0}} \frac{E^{n}-E_{0}^{n}}{E_{\mathrm{QG}, n}^{n}} \int_{0}^{z} \frac{\left(1+z^{\prime}\right)^{n} d z^{\prime}}{\sqrt{\Omega_{\mathrm{m}}\left(1+z^{\prime}\right)^{3}+\Omega_{\Lambda}}},
\end{aligned}
\label{eq6}
\end{equation}
where the following cosmological parameters are adopted \citep{2020A&A...641A...6P}: $H_0=67.36$~km$\cdot$s$^{-1}\cdot$Mpc$^{-1}$,$\Omega_{\rm m,0}=0.315$, and $\Omega_{\Lambda,0}=1-\Omega_{\rm m,0}$.

In this {\it Letter}, we consider both the linear (n=1) and quadratic (n=2) cases of Eq. (8), which, when fitted the observational data, can provide constraints on the corresponding QG energy scales, $E_{\rm QG,1}$ and $E_{\rm QG,2}$.

\end{enumerate}

\begin{deluxetable*}{cccccc}
\linespread{0.5}
\tablecaption{ Lower limits of Linear and Quadratic of Quantum Gravity Energy and the Best-fit Parameters of the SBPL Fit }
\label{tab3}
\tablehead{\multirow{2}{*}[-6pt]{GRB Name}&\multicolumn{2}{c}{Lower limits of log$E_{\mathrm{QG}}$ (GeV)} & \multicolumn{3}{c}{Best-fit Parameters of SBPL Model}\\[1ex]
\cmidrule(r){2-3} 		\cmidrule(r){4-6}\vspace{-1cm}
&\colhead{Linear LIV} & \colhead{Quadratic LIV} & \colhead{$\alpha_{\rm 1}$}& \colhead{$\alpha_{\rm 2}$}& \colhead{$E_{\rm break}$ (keV)} \vspace{1.cm}}
\tabletypesize{\small}

\startdata
GRB 210619B&$\geqslant 5.52\times10^{15}$&$\geqslant 1.72\times10^{7}$&1.35$_{-0.46}^{+0.17}$&-0.85$_{-0.35}^{+0.33}$&46.13$_{-20.68}^{+29.83}$\\\\
GRB 210610B&$\geqslant 3.23\times10^{13}$&$\geqslant 2.03\times10^{5}$&0.68$_{-0.2}^{+0.49}$&-0.51$_{-0.57}^{+0.28}$&80.1$_{-10.17}^{+39.81}$\\\\
GRB 210204A&$\geqslant 1.02\times10^{13}$&$\geqslant 1.31\times10^{5}$&0.71$_{-0.36}^{+0.45}$&-0.64$_{-0.76}^{+0.62}$&107.89$_{-21.52}^{+66.86}$\\\\
GRB 201216C&$\geqslant 1.9\times10^{14}$&$\geqslant 3.38\times10^{5}$&0.42$_{-0.15}^{+0.21}$&-0.19$_{-0.22}^{+0.19}$&147.81$_{-39.02}^{+37.68}$\\\\
GRB 200829A&$\geqslant 4.28\times10^{14}$&$\geqslant 2.52\times10^{6}$&0.38$_{-0.05}^{+0.09}$&-3.53$_{-0.68}^{+1.0}$&1887.23$_{-57.0}^{+75.43}$\\\\
GRB 200613A&$\geqslant 4.74\times10^{13}$&$\geqslant 1.68\times10^{5}$&1.03$_{-0.48}^{+0.51}$&-3.38$_{-1.03}^{+2.08}$&381.27$_{-23.26}^{+9.85}$\\\\
GRB 190114C&$\geqslant 5.43\times10^{14}$&$\geqslant 2.22\times10^{6}$&0.94$_{-0.11}^{+0.72}$&-3.85$_{-0.36}^{+0.22}$&1754.88$_{-120.86}^{+14.11}$\\\\
GRB 180720B&$\geqslant 9.54\times10^{14}$&$\geqslant 6.46\times10^{6}$&3.21$_{-0.71}^{+0.56}$&-2.57$_{-0.63}^{+0.63}$&839.9$_{-39.86}^{+169.8}$\\\\
GRB 180703A&$\geqslant 3.6\times10^{13}$&$\geqslant 1.19\times10^{5}$&3.64$_{-1.38}^{+2.45}$&-5.21$_{-0.29}^{+0.81}$&155.11$_{-65.16}^{+47.98}$\\\\
GRB 171010A&$\geqslant 5.68\times10^{13}$&$\geqslant 9.96\times10^{4}$&0.9$_{-0.15}^{+0.03}$&-4.15$_{-1.0}^{+0.76}$&1473.3$_{-69.51}^{+26.52}$\\\\
GRB 160625B&$\geqslant 1.35\times10^{15}$&$\geqslant 7.28\times10^{6}$&0.36$_{-0.02}^{+0.04}$&-3.21$_{-1.94}^{+1.14}$&8862.15$_{-668.15}^{+800.83}$\\\\
GRB 160509A&$\geqslant 1.73\times10^{14}$&$\geqslant 5.21\times10^{5}$&1.64$_{-0.61}^{+0.12}$&-1.75$_{-0.24}^{+0.54}$&354.19$_{-36.81}^{+32.02}$\\\\
GRB 150821A&$\geqslant 1.01\times10^{13}$&$\geqslant 1.1\times10^{5}$&1.0$_{-0.11}^{+0.34}$&-1.19$_{-0.82}^{+0.82}$&119.55$_{-2.56}^{+42.72}$\\\\
GRB 150514A&$\geqslant 3.57\times10^{13}$&$\geqslant 1.3\times10^{5}$&0.34$_{-0.12}^{+0.19}$&-3.38$_{-0.27}^{+1.07}$&128.37$_{-27.98}^{+20.68}$\\\\
GRB 150403A&$\geqslant 2.18\times10^{14}$&$\geqslant 5.8\times10^{5}$&0.35$_{-0.06}^{+0.14}$&-5.55$_{-0.45}^{+1.29}$&433.61$_{-51.16}^{+27.42}$\\\\
GRB 150314A&$\geqslant 3.91\times10^{14}$&$\geqslant 8.5\times10^{5}$&0.38$_{-0.06}^{+0.13}$&-1.9$_{-0.52}^{+0.62}$&913.34$_{-13.27}^{+33.66}$\\\\
GRB 141028A&$\geqslant 1.53\times10^{14}$&$\geqslant 2.75\times10^{5}$&0.77$_{-0.22}^{+0.26}$&-3.72$_{-0.36}^{+0.19}$&192.74$_{-16.13}^{+29.62}$\\\\
GRB 140508A&$\geqslant 1.21\times10^{14}$&$\geqslant 4.77\times10^{5}$&0.27$_{-0.24}^{+0.57}$&-4.58$_{-0.42}^{+1.0}$&393.97$_{-125.62}^{+133.49}$\\\\
GRB 140206A&$\geqslant 4.57\times10^{14}$&$\geqslant 9.33\times10^{5}$&1.63$_{-0.68}^{+2.36}$&-2.15$_{-2.35}^{+1.21}$&152.58$_{-84.54}^{+22.67}$\\\\
GRB 131231A&$\geqslant 2.09\times10^{13}$&$\geqslant 1.91\times10^{5}$&0.81$_{-0.09}^{+0.1}$&-1.14$_{-0.13}^{+0.5}$&318.34$_{-140.88}^{+42.66}$\\\\
GRB 131108A&$\geqslant 2.1\times10^{14}$&$\geqslant 6.04\times10^{5}$&1.75$_{-0.56}^{+1.05}$&-6.9$_{-0.41}^{+0.1}$&612.89$_{-227.96}^{+63.75}$\\\\
GRB 130925A&$\geqslant 8.18\times10^{12}$&$\geqslant 6.15\times10^{4}$&0.84$_{-0.11}^{+0.14}$&-6.49$_{-1.62}^{+1.1}$&152.94$_{-10.65}^{+4.67}$\\\\
GRB 130518A&$\geqslant 2.55\times10^{14}$&$\geqslant 1.64\times10^{6}$&0.55$_{-0.16}^{+0.24}$&-2.32$_{-1.02}^{+1.25}$&591.49$_{-307.74}^{+160.03}$\\\\
GRB 130427A&$\geqslant 3.85\times10^{14}$&$\geqslant 2.75\times10^{6}$&3.94$_{-0.63}^{+0.02}$&-0.3$_{-0.1}^{+0.09}$&7.34$_{-2.0}^{+10.01}$\\\\
GRB 120119A&$\geqslant 1.25\times10^{14}$&$\geqslant 2.97\times10^{5}$&1.47$_{-0.55}^{+0.51}$&-6.17$_{-1.12}^{+2.23}$&447.46$_{-34.64}^{+32.16}$\\\\
GRB 100728A&$\geqslant 3.46\times10^{13}$&$\geqslant 2.15\times10^{5}$&7.98$_{-2.17}^{+1.98}$&-12.79$_{-0.58}^{+4.22}$&799.79$_{-198.37}^{+95.55}$\\\\
GRB 091003A&$\geqslant 2.34\times10^{14}$&$\geqslant 6.22\times10^{5}$&1.1$_{-1.1}^{+2.41}$&-8.27$_{-4.95}^{+5.28}$&231.06$_{-120.62}^{+62.0}$\\\\
GRB 090926A&$\geqslant 1.09\times10^{14}$&$\geqslant 8.45\times10^{5}$&0.64$_{-0.11}^{+0.1}$&-2.66$_{-1.09}^{+0.6}$&1048.68$_{-102.28}^{+97.2}$\\\\
GRB 090618&$\geqslant 4.01\times10^{13}$&$\geqslant 1.97\times10^{5}$&0.59$_{-0.07}^{+0.07}$&-4.14$_{-0.88}^{+0.14}$&947.71$_{-30.69}^{+37.42}$\\\\
GRB 090328&$\geqslant 6.17\times10^{13}$&$\geqslant 2.17\times10^{5}$&1.49$_{-0.8}^{+0.8}$&-5.14$_{-1.76}^{+1.45}$&620.23$_{-20.2}^{+66.86}$\\\\
GRB 081221&$\geqslant 7.14\times10^{13}$&$\geqslant 2.41\times10^{5}$&1.2$_{-0.71}^{+0.89}$&-8.7$_{-1.25}^{+1.85}$&252.52$_{-83.12}^{+69.84}$\\\\
GRB 080916C&$\geqslant 6.45\times10^{14}$&$\geqslant 1.22\times10^{6}$&3.34$_{-0.81}^{+0.66}$&-5.81$_{-1.17}^{+1.29}$&368.55$_{-100.22}^{+145.0}$\\\\
\enddata
\end{deluxetable*}

\begin{figure*}

\gridline{\hspace{-1.0cm}\fig{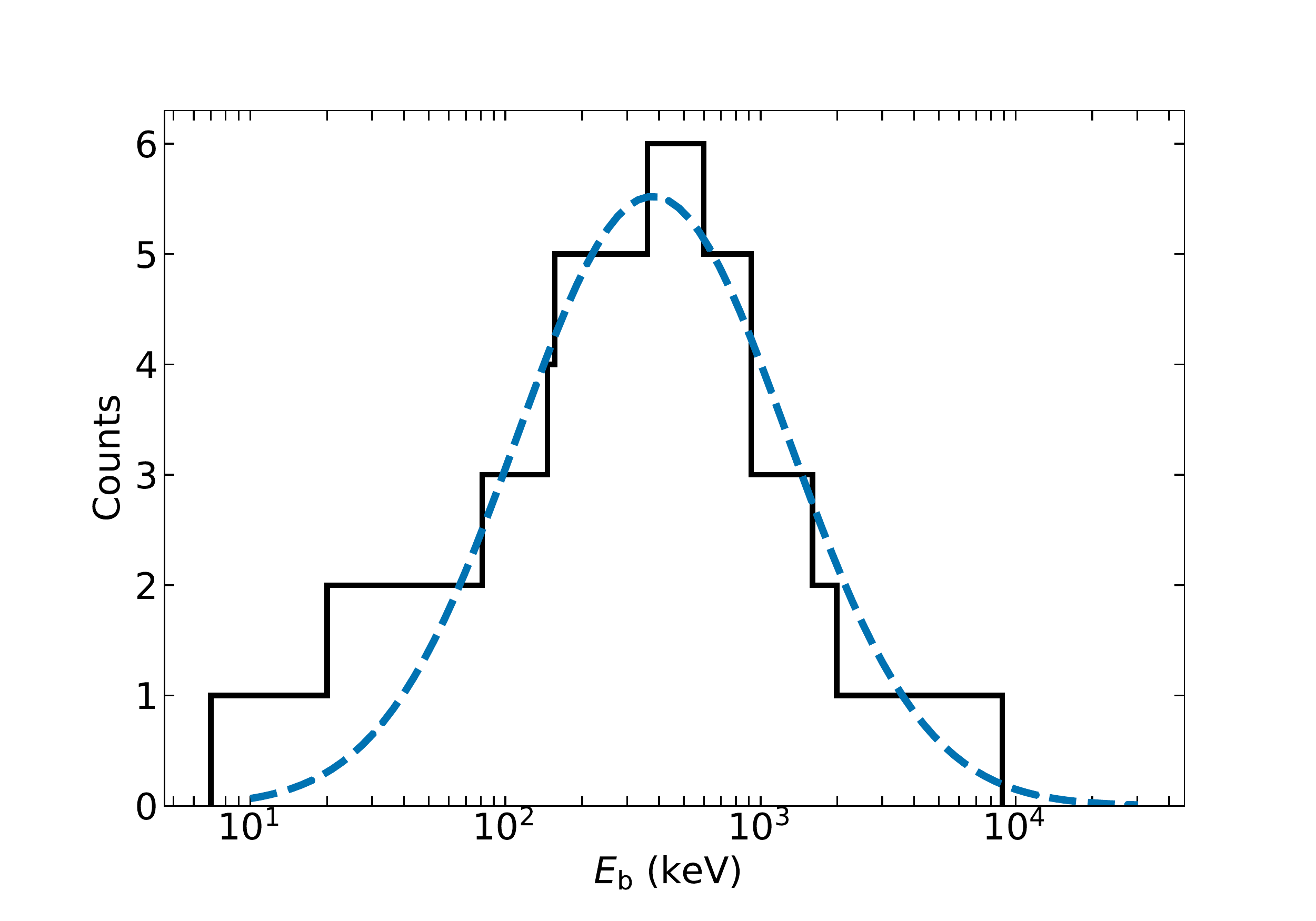}{0.5\textwidth}{a}\fig{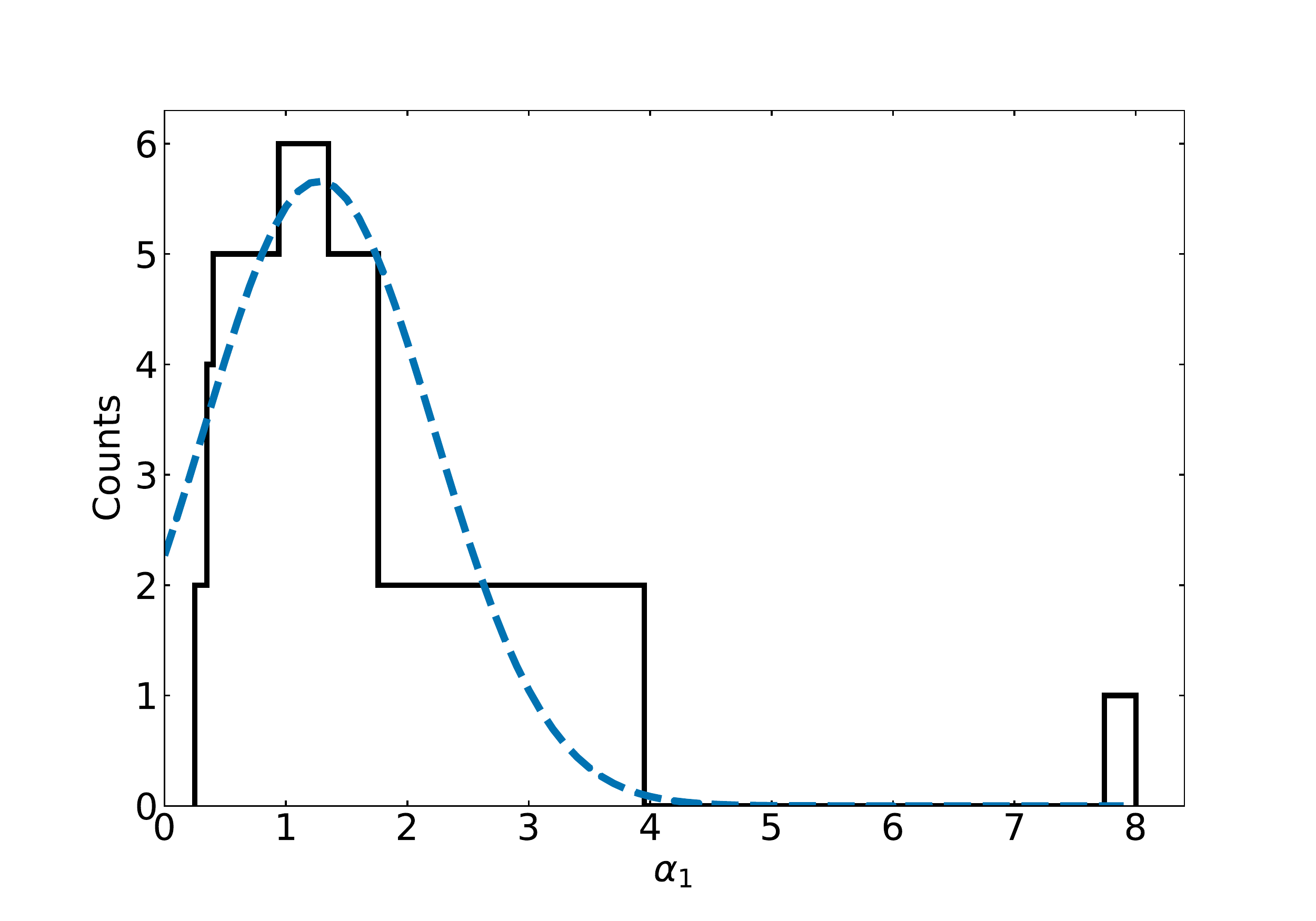}{0.5\textwidth}{b}} 
\gridline{
 \hspace{-1.0cm} \fig{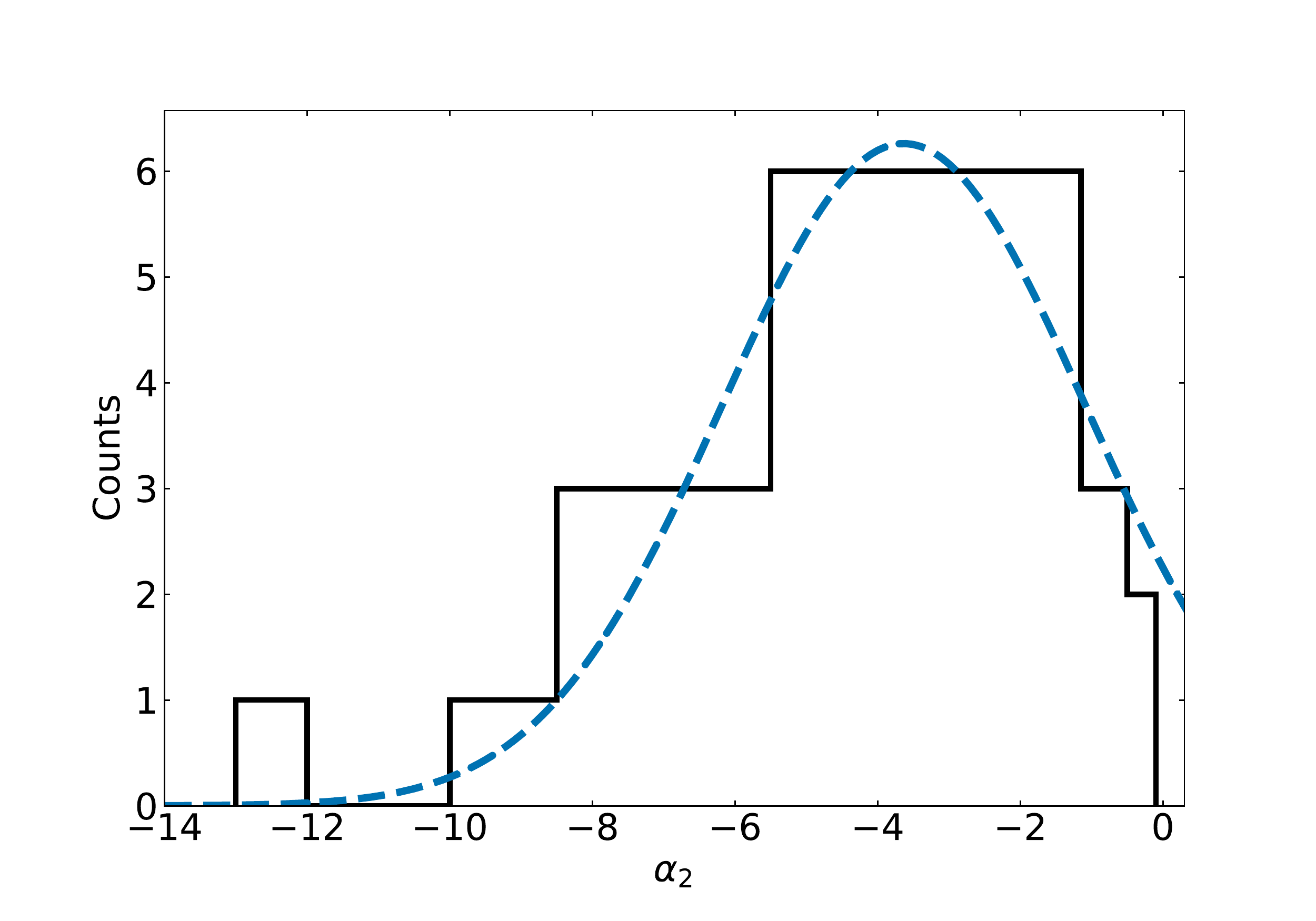}{0.5\textwidth}{c}
 \fig{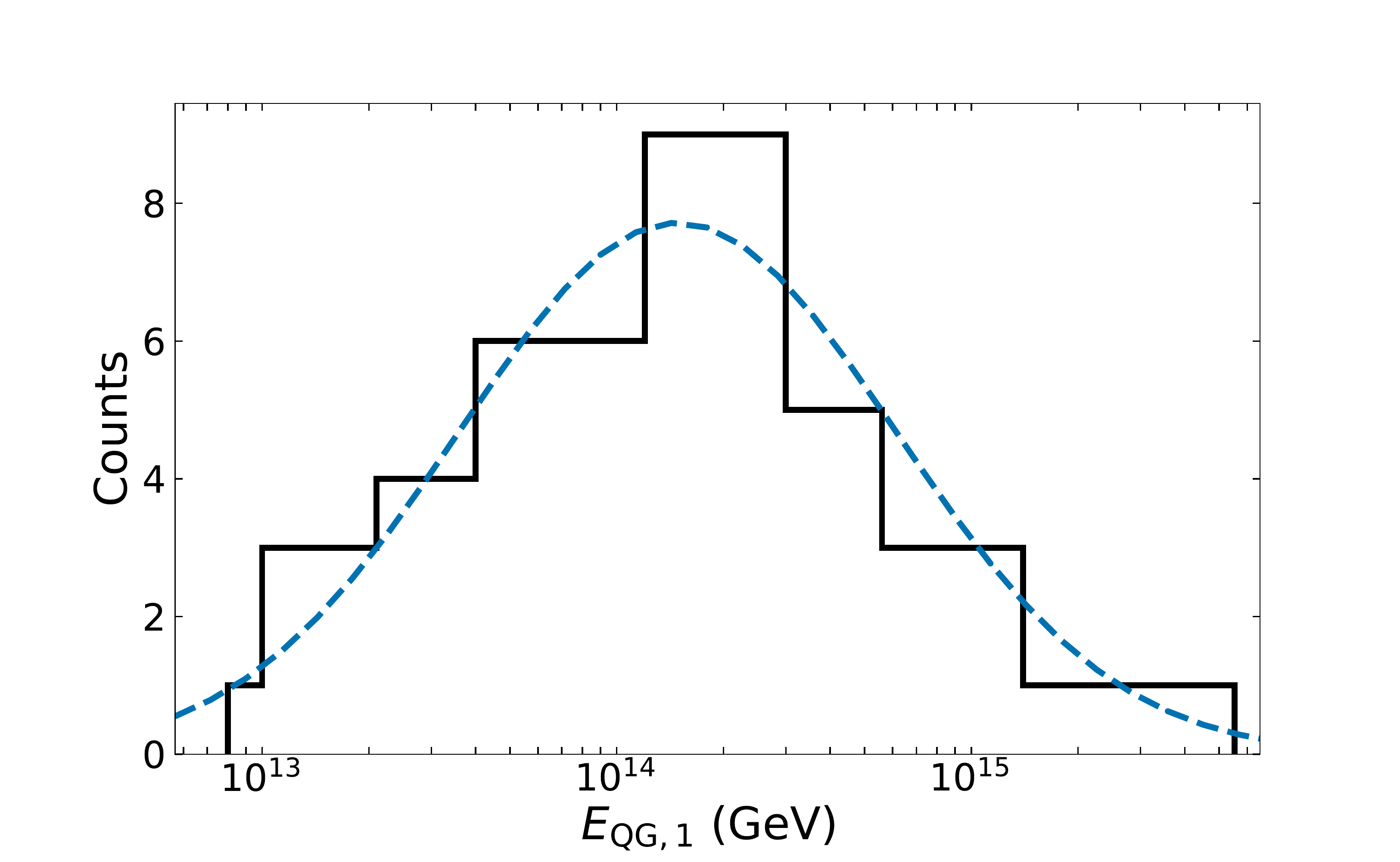}{0.5\textwidth}{d}} 
\gridline{ \hspace{-10.5cm}\fig{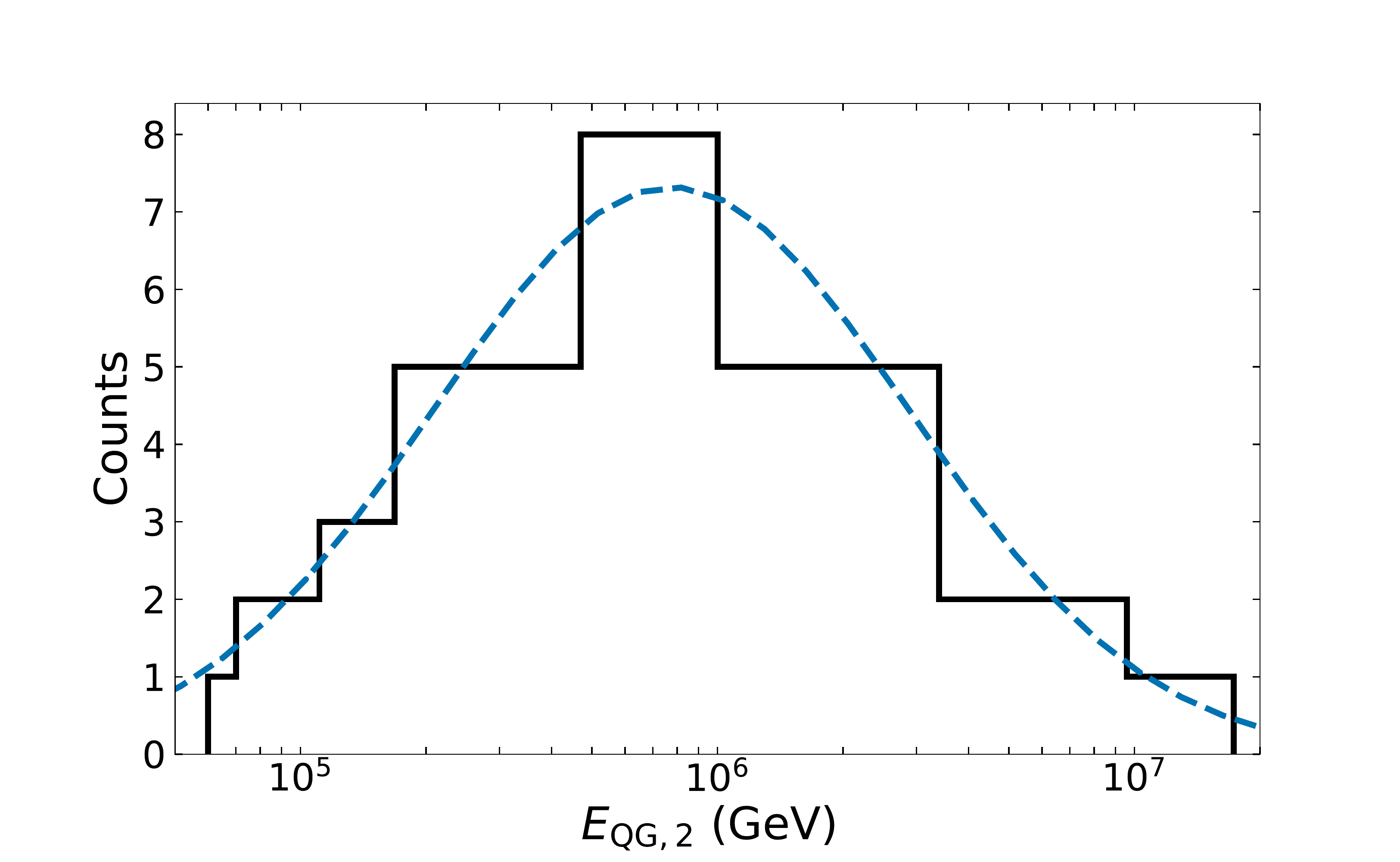}{0.5\textwidth}{e}}

\caption{\textcolor{black}{
The distributions of the best-fit parameters ($E_{\rm b}, \alpha_{1}, \alpha_{2}$) of the SBPL model and the constrained lower limits of the quantum gravity energy $(E_{\rm QG,1}, E_{\rm QG,2}$). Each distribution is fitted by a Gaussian or log-normal function. }}

\label{fig3}
\end{figure*}

Finally, the observed energy-dependent lags 
can be modeled by
\begin{equation}
\begin{aligned}
\tau=\Delta t_{\mathrm{int}}+\Delta t_{\mathrm{LIV}},
\end{aligned}
\end{equation}
which is directly fitted to the observational data in \S 4. 

\section{The Fit}
\label{sec4}

For each burst, we first fit its observed lags using the smoothly broken power-law as shown in Eq. (1). We can use this information to determine various features of the lag behavior, such as the break energy $E_{\rm b}$, and the slopes prior and post $E_{\rm b}$. The fit is performed using the {\it McEasyFit} \citep{2015ApJ...806...15Z} tool, which is a self-developed Bayesian Monte-Carlo fitting package ensuring reliable and realistic best-fit parameters and their uncertainties based on converged Markov chains. The priors of the free parameters are set to uniform distributions in the range listed in Table \ref{tab2}. Our model successfully fit the data. The best-fit parameters as well as their constraints are listed in Table \ref{tab3}. The model-predicted curves using the best-fit parameters are over-plotted as black solid lines in each panel of Figure 1.

Next, we fit the observed data with the LIV-induced model as in Eq. (9). This introduces an additional free parameter, $E_{\rm QG,n}$. The prior of $E_{\rm QG,n}$ is set as a uniform distribution in logarithmic scale in range of [0,$10^{20}$] GeV for n=1, or [0,$10^{15}$] GeV for n=2 (Table 2). In addition, we require the $\Delta t_{\rm LIV}$ term in Eq. (9) not to dominate over $\Delta t_{\rm int}$ so that $\Delta t_{\rm int}$ still shows a negative-to-positive transition in order to be consistent with the observations. Such a requirement can be reflected in the log-likelihood function in {\it McEasyFit} as:
\begin{equation}
\label{eq7}
L(\Theta)= \begin{cases}-\infty & \alpha_{1}<\alpha_{2} \\ -\frac{1}{2} \Sigma\left(\frac{\tau_{\text {obs }}-\tau_{\text {model}}(\Theta)}{\sigma(\tau_{\text {obs })}}\right)^{2} & \alpha_{1} \geqslant \alpha_{2}\end{cases},
\end{equation}
where $\Theta$ represents the free fitting parameters. The above approach allowed us to successfully fit the observed data of each GRB and constrain the lower limits of its linear and quadratic quantum gravity energies. Figure 2 shows an example of the posterior probability distributions of fitting parameters for GRB 130427A. For all of the GRBs in our sample, the $E_{\rm QG}$ lower limits are listed in Table 3 and correspond to the dashed lines in Figure 1.

Upon fitting the entire sample of 32 GRBs, we obtained the following key statistical properties of the lag behavior, as well as the constraints on the quantum gravity energy:
\begin{enumerate}
 \item The distribution of the lag transition energy, $E_{\rm b}$ is a log-normal shape with a median value of $E_{\rm b}$ = 398 keV (Figure 3a); 
 
 \item The slopes for the $\tau-E$ relation prior and post the break are distributed as a Gaussian function. The median values of the $\alpha_1$ and $\alpha_2$ are 1.27 and -3.52 respectively (Figure 3b and Figure 3c);
\item The linear quantum gravity energy lower limits are constrained at a large range from 8.2 $\times$ 10$^{12}$ GeV to 5.5 $\times$ 10$^{15}$ GeV. The distribution of the lower limits is a log-normal shape with a median value of $1.5\times 10^{14}$ GeV (Figure 3d);
 
 \item The quadratic quantum gravity energy lower limits are also constrained at a large range from 6.2 $\times$ 10$^{4}$ GeV to 1.7 $\times$ 10$^{7}$ GeV . The distribution of the lower limits is a log-normal shape with a median value of $8\times 10^{5}$ GeV (Figure 3e).
 
\end{enumerate}

\section{Summary \& Discussions}
In this study, a total of 32 GRBs with positive-to-negative transitions in their spectral lags have been found among the 135 {\it Fermi}/GBM long GRBs with redshift measurement, suggesting lag transitions are not uncommon. We systematically processed and analyzed the lags of these 32 GRBs. The observed lag-E relationship of each burst can successfully be fitted by an empirically smoothly broken power-law function. Such fits yield a typical value of $400$ keV for the transition energy. Our results are further applied to constrain the Lorentz invariance violation. Incorporating the LIV effect into the fit, the lower limits of linear and quadratic quantum gravity energy are derived for each burst. The typical lower limits of $E_{\rm QG,1}\geqslant1.5\times 10^{14}$ GeV and $E_{\rm QG,2}\geqslant8\times 10^{5}$ GeV of our study are consistent with, and sometimes deeper than, those of previous case studies, such as for GRB 160625B \citep{2017ApJ...834L..13W} and GRB 190114C \citep{2021ApJ...906....8D}.

Our findings offer some insight into understanding the lag origins. For instance, some significant negative lags (e.g., $\tau\simeq1$ s is observed in several GRBs) cannot be explained solely by the LIV effect. Thus the SBPL function we proposed in this work appears to be more accurate than the simple power law function \citep{2017ApJ...834L..13W} to describe the intrinsic lag behaviors in our sample. In addition, our study provides a rich sample for investigating the underlying physical processes that result in the lag transition. These theories include, but are not limited to, the hard-to-soft evolution of a curved spectrum \citep{1996Natur.381...49L, 2012ApJ...756..112L} and modified photosphere models \citep{2018ApJ...860...72M,2019ApJ...882...26M,2022MNRAS.509.6047M}.

\section*{acknowledgements}

Z.-K.L. thanks Ken Chen for the helpful discussion on this paper. We acknowledges support by the National Key Research and Development Programs of China (2018YFA0404204), the National Natural Science Foundation of China (Grant Nos. 11833003, U2038105, 12121003), the science research grants from the China Manned Space Project with NO.CMS-CSST-2021-B11, and the Program for Innovative Talents, Entrepreneur in Jiangsu. Y.-Z.M. is supported by the National Postdoctoral Program for Innovative Talents (grant no. BX20200164). We acknowledge the use of public data from the Fermi Science Support Center (FSSC).
%

\begin{thebibliography}{}
\expandafter\ifx\csname natexlab\endcsname\relax\def\natexlab#1{#1}\fi
\providecommand{\url}[1]{\href{#1}{#1}}
\providecommand{\dodoi}[1]{doi:~\href{http://doi.org/#1}{\nolinkurl{#1}}}
\providecommand{\doeprint}[1]{\href{http://ascl.net/#1}{\nolinkurl{http://ascl.net/#1}}}
\providecommand{\doarXiv}[1]{\href{https://arxiv.org/abs/#1}{\nolinkurl{https://arxiv.org/abs/#1}}}

\bibitem[{{Abdo} {et~al.}(2009{\natexlab{a}}){Abdo}, {Ackermann}, {Ajello},
 {Asano}, {Atwood}, {Axelsson}, {Baldini}, {Ballet}, {Barbiellini}, {Baring},
 {Bastieri}, {Bechtol}, {Bellazzini}, {Berenji}, {Bhat}, {Bissaldi}, {Bloom},
 {Bonamente}, {Bonnell}, {Borgland}, {Bouvier}, {Bregeon}, {Brez}, {Briggs},
 {Brigida}, {Bruel}, {Burgess}, {Burnett}, {Caliandro}, {Cameron}, {Caraveo},
 {Casandjian}, {Cecchi}, {{\c{C}}elik}, {Chaplin}, {Charles}, {Cheung},
 {Chiang}, {Ciprini}, {Claus}, {Cohen-Tanugi}, {Cominsky}, {Connaughton},
 {Conrad}, {Cutini}, {Dermer}, {de Angelis}, {de Palma}, {Digel}, {Dingus},
 {Do Couto E Silva}, {Drell}, {Dubois}, {Dumora}, {Farnier}, {Favuzzi},
 {Fegan}, {Finke}, {Fishman}, {Focke}, {Foschini}, {Fukazawa}, {Funk},
 {Fusco}, {Gargano}, {Gasparrini}, {Gehrels}, {Germani}, {Gibby}, {Giebels},
 {Giglietto}, {Giordano}, {Glanzman}, {Godfrey}, {Granot}, {Greiner},
 {Grenier}, {Grondin}, {Grove}, {Grupe}, {Guillemot}, {Guiriec}, {Hanabata},
 {Harding}, {Hayashida}, {Hays}, {Hoversten}, {Hughes}, {J{\'o}hannesson},
 {Johnson}, {Johnson}, {Johnson}, {Kamae}, {Katagiri}, {Kataoka}, {Kawai},
 {Kerr}, {Kippen}, {Kn{\"o}dlseder}, {Kocevski}, {Kouveliotou}, {Kuehn},
 {Kuss}, {Lande}, {Latronico}, {Lemoine-Goumard}, {Longo}, {Loparco}, {Lott},
 {Lovellette}, {Lubrano}, {Madejski}, {Makeev}, {Mazziotta}, {McBreen},
 {McEnery}, {McGlynn}, {M{\'e}sz{\'a}ros}, {Meurer}, {Michelson},
 {Mitthumsiri}, {Mizuno}, {Moiseev}, {Monte}, {Monzani}, {Moretti},
 {Morselli}, {Moskalenko}, {Murgia}, {Nakamori}, {Nolan}, {Norris}, {Nuss},
 {Ohno}, {Ohsugi}, {Omodei}, {Orlando}, {Ormes}, {Ozaki}, {Paciesas},
 {Paneque}, {Panetta}, {Parent}, {Pelassa}, {Pepe}, {Pesce-Rollins},
 {Petrosian}, {Piron}, {Porter}, {Preece}, {Rain{\`o}}, {Ramirez-Ruiz},
 {Rando}, {Razzano}, {Razzaque}, {Reimer}, {Reimer}, {Reposeur}, {Ritz},
 {Rochester}, {Rodriguez}, {Roth}, {Ryde}, {Sadrozinski}, {Sanchez}, {Sander},
 {Saz Parkinson}, {Scargle}, {Schalk}, {Sgr{\`o}}, {Siskind}, {Smith},
 {Smith}, {Spandre}, {Spinelli}, {Stamatikos}, {Stecker}, {Strickman},
 {Suson}, {Tajima}, {Takahashi}, {Takahashi}, {Tanaka}, {Thayer}, {Thayer},
 {Thompson}, {Tibaldo}, {Toma}, {Torres}, {Tosti}, {Troja}, {Uchiyama},
 {Uehara}, {Usher}, {van der Horst}, {Vasileiou}, {Vilchez}, {Vitale}, {von
 Kienlin}, {Waite}, {Wang}, {Wilson-Hodge}, {Winer}, {Wood}, {Wu}, {Yamazaki},
 {Ylinen}, {Ziegler}, \& {Fermi LAT Collaboration}}]{2009Natur.462..331A}
{Abdo}, A.~A., {Ackermann}, M., {Ajello}, M., {et~al.} 2009{\natexlab{a}},
 \nat, 462, 331, \dodoi{10.1038/nature08574}

\bibitem[{{Abdo} {et~al.}(2009{\natexlab{b}}){Abdo}, {Ackermann}, {Arimoto},
 {Asano}, {Atwood}, {Axelsson}, {Baldini}, {Ballet}, {Band}, {Barbiellini},
 {Baring}, {Bastieri}, {Battelino}, {Baughman}, {Bechtol}, {Bellardi},
 {Bellazzini}, {Berenji}, {Bhat}, {Bissaldi}, {Blandford}, {Bloom}, {Bogaert},
 {Bogart}, {Bonamente}, {Bonnell}, {Borgland}, {Bouvier}, {Bregeon}, {Brez},
 {Briggs}, {Brigida}, {Bruel}, {Burnett}, {Burrows}, {Busetto}, {Caliandro},
 {Cameron}, {Caraveo}, {Casandjian}, {Ceccanti}, {Cecchi}, {Celotti},
 {Charles}, {Chekhtman}, {Cheung}, {Chiang}, {Ciprini}, {Claus},
 {Cohen-Tanugi}, {Cominsky}, {Connaughton}, {Conrad}, {Costamante}, {Cutini},
 {DeKlotz}, {Dermer}, {de Angelis}, {de Palma}, {Digel}, {Dingus}, {do Couto e
 Silva}, {Drell}, {Dubois}, {Dumora}, {Edmonds}, {Evans}, {Fabiani},
 {Farnier}, {Favuzzi}, {Finke}, {Fishman}, {Focke}, {Frailis}, {Fukazawa},
 {Funk}, {Fusco}, {Gargano}, {Gasparrini}, {Gehrels}, {Germani}, {Giebels},
 {Giglietto}, {Giommi}, {Giordano}, {Glanzman}, {Godfrey}, {Goldstein},
 {Granot}, {Greiner}, {Grenier}, {Grondin}, {Grove}, {Guillemot}, {Guiriec},
 {Haller}, {Hanabata}, {Harding}, {Hayashida}, {Hays}, {Morata}, {Hoover},
 {Hughes}, {J{\'o}hannesson}, {Johnson}, {Johnson}, {Johnson}, {Johnson},
 {Kamae}, {Katagiri}, {Kataoka}, {Kavelaars}, {Kawai}, {Kelly}, {Kennea},
 {Kerr}, {Kippen}, {Kn{\"o}dlseder}, {Kocevski}, {Kocian}, {Komin},
 {Kouveliotou}, {Kuehn}, {Kuss}, {Lande}, {Landriu}, {Larsson}, {Latronico},
 {Lavalley}, {Lee}, {Lee}, {Lemoine-Goumard}, {Lichti}, {Longo}, {Loparco},
 {Lott}, {Lovellette}, {Lubrano}, {Madejski}, {Makeev}, {Marangelli},
 {Mazziotta}, {McBreen}, {McEnery}, {McGlynn}, {Meegan}, {M{\'e}sz{\'a}ros},
 {Meurer}, {Michelson}, {Minuti}, {Mirizzi}, {Mitthumsiri}, {Mizuno},
 {Moiseev}, {Monte}, {Monzani}, {Moretti}, {Morselli}, {Moskalenko}, {Murgia},
 {Nakamori}, {Nelson}, {Nolan}, {Norris}, {Nuss}, {Ohno}, {Ohsugi}, {Okumura},
 {Omodei}, {Orlando}, {Ormes}, {Ozaki}, {Paciesas}, {Paneque}, {Panetta},
 {Parent}, {Pelassa}, {Pepe}, {Perri}, {Pesce-Rollins}, {Petrosian},
 {Pinchera}, {Piron}, {Porter}, {Preece}, {Rain{\`o}}, {Ramirez-Ruiz},
 {Rando}, {Rapposelli}, {Razzano}, {Razzaque}, {Rea}, {Reimer}, {Reimer},
 {Reposeur}, {Reyes}, {Ritz}, {Rochester}, {Rodriguez}, {Roth}, {Ryde},
 {Sadrozinski}, {Sanchez}, {Sander}, {Parkinson}, {Scargle}, {Schalk},
 {Segal}, {Sgr{\`o}}, {Shimokawabe}, {Siskind}, {Smith}, {Smith}, {Spandre},
 {Spinelli}, {Stamatikos}, {Starck}, {Stecker}, {Steinle}, {Stephens},
 {Strickman}, {Suson}, {Tagliaferri}, {Tajima}, {Takahashi}, {Takahashi},
 {Tanaka}, {Tenze}, {Thayer}, {Thayer}, {Thompson}, {Tibaldo}, {Torres},
 {Tosti}, {Tramacere}, {Turri}, {Tuvi}, {Usher}, {van der Horst}, {Vigiani},
 {Vilchez}, {Vitale}, {von Kienlin}, {Waite}, {Williams}, {Wilson-Hodge},
 {Winer}, {Wood}, {Wu}, {Yamazaki}, {Ylinen}, {Ziegler}, {Fermi LAT
 Collaboration}, \& {Fermi GBM Collaboration}}]{2009Sci...323.1688A}
{Abdo}, A.~A., {Ackermann}, M., {Arimoto}, M., {et~al.} 2009{\natexlab{b}},
 Science, 323, 1688, \dodoi{10.1126/science.1169101}

\bibitem[{{Acciari} {et~al.}(2020){Acciari}, {Ansoldi}, {Antonelli}, {Arbet
 Engels}, {Baack}, {Babi{\'c}}, {Banerjee}, {Barres de Almeida}, {Barrio},
 {Becerra Gonz{\'a}lez}, {Bednarek}, {Bellizzi}, {Bernardini}, {Berti},
 {Besenrieder}, {Bhattacharyya}, {Bigongiari}, {Biland}, {Blanch}, {Bonnoli},
 {Bo{\v{s}}njak}, {Busetto}, {Carosi}, {Ceribella}, {Cerruti}, {Chai},
 {Chilingarian}, {Cikota}, {Colak}, {Colin}, {Colombo}, {Contreras},
 {Cortina}, {Covino}, {D'Amico}, {D'Elia}, {da Vela}, {Dazzi}, {de Angelis},
 {de Lotto}, {Delfino}, {Delgado}, {Depaoli}, {di Pierro}, {di Venere}, {Do
 Souto Espi{\~n}eira}, {Dominis Prester}, {Donini}, {Dorner}, {Doro},
 {Elsaesser}, {Fallah Ramazani}, {Fattorini}, {Ferrara}, {Foffano}, {Fonseca},
 {Font}, {Fruck}, {Fukami}, {Garc{\'\i}a L{\'o}pez}, {Garczarczyk},
 {Gasparyan}, {Gaug}, {Giglietto}, {Giordano}, {Gliwny}, {Godinovi{\'c}},
 {Green}, {Hadasch}, {Hahn}, {Herrera}, {Hoang}, {Hrupec}, {H{\"u}tten},
 {Inada}, {Inoue}, {Ishio}, {Iwamura}, {Jouvin}, {Kajiwara}, {Karjalainen},
 {Kerszberg}, {Kobayashi}, {Kubo}, {Kushida}, {Lamastra}, {Lelas}, {Leone},
 {Lindfors}, {Lombardi}, {Longo}, {L{\'o}pez}, {L{\'o}pez-Coto},
 {L{\'o}pez-Oramas}, {Loporchio}, {Machado de Oliveira Fraga}, {Maggio},
 {Majumdar}, {Makariev}, {Mallamaci}, {Maneva}, {Manganaro}, {Mannheim},
 {Maraschi}, {Mariotti}, {Mart{\'\i}nez}, {Mazin}, {Mender},
 {Mi{\'c}anovi{\'c}}, {Miceli}, {Miener}, {Minev}, {Miranda}, {Mirzoyan},
 {Molina}, {Moralejo}, {Morcuende}, {Moreno}, {Moretti}, {Munar-Adrover},
 {Neustroev}, {Nigro}, {Nilsson}, {Ninci}, {Nishijima}, {Noda}, {Nogu{\'e}s},
 {Nozaki}, {Ohtani}, {Oka}, {Otero-Santos}, {Palatiello}, {Paneque},
 {Paoletti}, {Paredes}, {Pavleti{\'c}}, {Pe{\~n}il}, {Perennes}, {Peresano},
 {Persic}, {Prada Moroni}, {Prandini}, {Puljak}, {Rhode}, {Rib{\'o}}, {Rico},
 {Righi}, {Rugliancich}, {Saha}, {Sahakyan}, {Saito}, {Sakurai}, {Satalecka},
 {Schleicher}, {Schmidt}, {Schweizer}, {Sitarek}, {{\v{S}}nidari{\'c}},
 {Sobczynska}, {Spolon}, {Stamerra}, {Strom}, {Strzys}, {Suda}, {Suri{\'c}},
 {Takahashi}, {Tavecchio}, {Temnikov}, {Terzi{\'c}}, {Teshima},
 {Torres-Alb{\`a}}, {Tosti}, {van Scherpenberg}, {Vanzo}, {Vazquez Acosta},
 {Ventura}, {Verguilov}, {Vigorito}, {Vitale}, {Vovk}, {Will}, {Zari{\'c}},
 {Nava}, \& {MAGIC Collaboration}}]{2020PhRvL.125b1301A}
{Acciari}, V.~A., {Ansoldi}, S., {Antonelli}, L.~A., {et~al.} 2020, \prl, 125,
 021301, \dodoi{10.1103/PhysRevLett.125.021301}

\bibitem[{{Amelino-Camelia}(2013)}]{2013LRR....16....5A}
{Amelino-Camelia}, G. 2013, Living Reviews in Relativity, 16, 5,
 \dodoi{10.12942/lrr-2013-5}

\bibitem[{{Amelino-Camelia} {et~al.}(1998){Amelino-Camelia}, {Ellis},
 {Mavromatos}, {Nanopoulos}, \& {Sarkar}}]{1998Natur.393..763A}
{Amelino-Camelia}, G., {Ellis}, J., {Mavromatos}, N.~E., {Nanopoulos}, D.~V.,
 \& {Sarkar}, S. 1998, \nat, 393, 763, \dodoi{10.1038/31647}

\bibitem[{{Atwood} {et~al.}(2013){Atwood}, {Baldini}, {Bregeon}, {Bruel},
 {Chekhtman}, {Cohen-Tanugi}, {Drlica-Wagner}, {Granot}, {Longo}, {Omodei},
 {Pesce-Rollins}, {Razzaque}, {Rochester}, {Sgr{\`o}}, {Tinivella}, {Usher},
 \& {Zimmer}}]{2013ApJ...774...76A}
{Atwood}, W.~B., {Baldini}, L., {Bregeon}, J., {et~al.} 2013, \apj, 774, 76,
 \dodoi{10.1088/0004-637X/774/1/76}

\bibitem[{{Castro-Tirado} {et~al.}(2019){Castro-Tirado}, {Hu},
 {Fernandez-Garcia}, {Valeev}, {Sokolov}, {Guziy}, {Oates}, {Jeong}, {Pandey},
 {Carrasco}, \& {Reverte-Paya}}]{2019GCN.23708....1C}
{Castro-Tirado}, A.~J., {Hu}, Y., {Fernandez-Garcia}, E., {et~al.} 2019, GRB
 Coordinates Network, 23708, 1

\bibitem[{{Cenko} {et~al.}(2009){Cenko}, {Perley}, {Junkkarinen}, {Burbidge},
 {Diego}, \& {Miller}}]{2009GCN..9518....1C}
{Cenko}, S.~B., {Perley}, D.~A., {Junkkarinen}, V., {et~al.} 2009, GRB
 Coordinates Network, 9518, 1

\bibitem[{{Cucchiara} {et~al.}(2009){Cucchiara}, {Fox}, {Cenko}, {Tanvir}, \&
 {Berger}}]{2009GCN.10031....1C}
{Cucchiara}, A., {Fox}, D.~B., {Cenko}, S.~B., {Tanvir}, N., \& {Berger}, E.
 2009, GRB Coordinates Network, 1031, 1

\bibitem[{{de Ugarte Postigo}(2020)}]{2020GCN29320}
{de Ugarte Postigo}, A. 2020, GRB Coordinates Network, 29320, 1

\bibitem[{{de Ugarte Postigo} {et~al.}(2009){de Ugarte Postigo}, {Jakobsson},
 {Malesani}, {Fynbo}, {Simpson}, \& {Barros}}]{2009GCN..8766....1D}
{de Ugarte Postigo}, A., {Jakobsson}, P., {Malesani}, D., {et~al.} 2009, GRB
 Coordinates Network, 8766, 1

\bibitem[{{de Ugarte Postigo} {et~al.}(2017){de Ugarte Postigo}, {Selsing},
 {Malesani}, {Xu}, {Izzo }, {Kouveliotou}, {D'Elia}, \&
 {Tanvir}}]{2017GCN22096}
{de Ugarte Postigo}, A., {Selsing}, J., {Malesani}, D., {et~al.} 2017, GRB
 Coordinates Network, 22096, 1

\bibitem[{{de Ugarte Postigo} {et~al.}(2021{\natexlab{a}}){de Ugarte Postigo},
 {Thoene}, {Agui Fernandez}, {Blazek}, {Kann}, {Fynbo}, {Izzo}, \& {Garcia
 Alvarez}}]{2021GCN.30194....1D}
{de Ugarte Postigo}, A., {Thoene}, C., {Agui Fernandez}, J.~F., {et~al.}
 2021{\natexlab{a}}, GRB Coordinates Network, 30194, 1

\bibitem[{{de Ugarte Postigo} {et~al.}(2021{\natexlab{b}}){de Ugarte Postigo},
 {Thoene}, {Agui Fernandez}, {Izzo}, {Tanvir}, \& {Fynbo}}]{2021GCN30272}
---. 2021{\natexlab{b}}, GRB Coordinates Network, 30272, 1

\bibitem[{{de Ugarte Postigo} {et~al.}(2013){de Ugarte Postigo}, {Thoene},
 {Gorosabel}, {Sanchez-Ramirez}, {Fynbo}, {Tanvir}, {Cabrera-Lavers}, \&
 {Garcia}}]{2013GCN.15470....1D}
{de Ugarte Postigo}, A., {Thoene}, C.~C., {Gorosabel}, J., {et~al.} 2013, GRB
 Coordinates Network, 15470, 1

\bibitem[{{de Ugarte Postigo} {et~al.}(2015{\natexlab{a}}){de Ugarte Postigo},
 {Xu}, {Malesani}, \& {Tanvir}}]{2015GCN.17822....1D}
{de Ugarte Postigo}, A., {Xu}, D., {Malesani}, D., \& {Tanvir}, N.~R.
 2015{\natexlab{a}}, GRB Coordinates Network, 17822, 1

\bibitem[{{de Ugarte Postigo} {et~al.}(2015{\natexlab{b}}){de Ugarte Postigo},
 {Fynbo}, {Thoene}, {Tanvir}, {Sanchez-Ramirez}, {Gorosabel}, {Pessev},
 {Alvarez-Iglesias}, \& {Rivero}}]{2015GCN.17583....1D}
{de Ugarte Postigo}, A., {Fynbo}, J.~P.~U., {Thoene}, C., {et~al.}
 2015{\natexlab{b}}, GRB Coordinates Network, 17583, 1

\bibitem[{{D'Elia} {et~al.}(2015){D'Elia}, {Kruehler}, {Wiersema}, {Tanvir},
 {Japelj}, {Pugliese}, {Malesani}, {Milvang-Jensen}, \&
 {Fynbo}}]{2015GCN.18187....1D}
{D'Elia}, V., {Kruehler}, T., {Wiersema}, K., {et~al.} 2015, GRB Coordinates
 Network, 18187, 1

\bibitem[{{Du} {et~al.}(2021){Du}, {Lan}, {Wei}, {Zhou}, {Gao}, {Jiang},
 {Zhang}, {Liu}, {Wu}, {Liang}, \& {Zhu}}]{2021ApJ...906....8D}
{Du}, S.-S., {Lan}, L., {Wei}, J.-J., {et~al.} 2021, \apj, 906, 8,
 \dodoi{10.3847/1538-4357/abc624}

\bibitem[{{Flores} {et~al.}(2013){Flores}, {Covino}, {Xu}, {Kruehler}, {Fynbo},
 {Milvang-Jensen}, {de Ugarte Postigo}, {Kaper}, \&
 {Wiersema}}]{2013GCN.14491....1F}
{Flores}, H., {Covino}, S., {Xu}, D., {et~al.} 2013, GRB Coordinates Network,
 14491, 1

\bibitem[{{Ioka} \& {Nakamura}(2001)}]{2001ApJ...554L.163I}
{Ioka}, K., \& {Nakamura}, T. 2001, \apjl, 554, L163, \dodoi{10.1086/321717}

\bibitem[{{Izzo} {et~al.}(2019){Izzo}, {de Ugarte Postigo}, {Schady},
 {Malesani}, {Kann}, {Kouveliotou}, {D'Elia}, \&
 {Tanvir}}]{2019GCN.23889....1I}
{Izzo}, L., {de Ugarte Postigo}, A., {Schady}, P., {et~al.} 2019, GRB
 Coordinates Network, 23889, 1

\bibitem[{{Jacob} \& {Piran}(2008)}]{2008JCAP...01..031J}
{Jacob}, U., \& {Piran}, T. 2008, \jcap, 2008, 031,
 \dodoi{10.1088/1475-7516/2008/01/031}

\bibitem[{{Kruehler} {et~al.}(2013){Kruehler}, {Greiner}, \&
 {Kann}}]{2013GCN.14500....1K}
{Kruehler}, T., {Greiner}, J., \& {Kann}, D.~A. 2013, GRB Coordinates Network,
 14500, 1

\bibitem[{{Li}(2010)}]{2010ApJ...709..525L}
{Li}, Z. 2010, \apj, 709, 525, \dodoi{10.1088/0004-637X/709/1/525}

\bibitem[{{Liang} \& {Kargatis}(1996)}]{1996Natur.381...49L}
{Liang}, E., \& {Kargatis}, V. 1996, \nat, 381, 49, \dodoi{10.1038/381049a0}

\bibitem[{{Lu} {et~al.}(2012){Lu}, {Wei}, {Liang}, {Zhang}, {L{\"u}}, {L{\"u}},
 {Lei}, \& {Zhang}}]{2012ApJ...756..112L}
{Lu}, R.-J., {Wei}, J.-J., {Liang}, E.-W., {et~al.} 2012, \apj, 756, 112,
 \dodoi{10.1088/0004-637X/756/2/112}

\bibitem[{{Malesani} {et~al.}(2009){Malesani}, {Goldoni}, {Fynbo}, {D'Elia},
 {Covino}, {Flores}, {Levan}, {Vergani}, \& {Wiersema}}]{2009GCN..9942....1M}
{Malesani}, D., {Goldoni}, P., {Fynbo}, J.~P.~U., {et~al.} 2009, GRB
 Coordinates Network, 9942, 1

\bibitem[{{Malesani} {et~al.}(2014){Malesani}, {Xu}, {Fynbo}, {de Ugarte
 Postigo}, {Schulze}, {Finoguenov}, {Jakobsson}, {Melandri}, \&
 {Cucchiara}}]{2014GCN.15800....1M}
{Malesani}, D., {Xu}, D., {Fynbo}, J.~P.~U., {et~al.} 2014, GRB Coordinates
 Network, 15800, 1

\bibitem[{{Mattingly}(2005)}]{2005LRR.....8....5M}
{Mattingly}, D. 2005, Living Reviews in Relativity, 8, 5,
 \dodoi{10.12942/lrr-2005-5}

\bibitem[{{Meng} {et~al.}(2022){Meng}, {Geng}, \& {Wu}}]{2022MNRAS.509.6047M}
{Meng}, Y.-Z., {Geng}, J.-J., \& {Wu}, X.-F. 2022, \mnras, 509, 6047,
 \dodoi{10.1093/mnras/stab3132}

\bibitem[{{Meng} {et~al.}(2019){Meng}, {Liu}, {Wei}, {Wu}, \&
 {Zhang}}]{2019ApJ...882...26M}
{Meng}, Y.-Z., {Liu}, L.-D., {Wei}, J.-J., {Wu}, X.-F., \& {Zhang}, B.-B. 2019,
 \apj, 882, 26, \dodoi{10.3847/1538-4357/ab30c7}

\bibitem[{{Meng} {et~al.}(2018){Meng}, {Geng}, {Zhang}, {Wei}, {Xiao}, {Liu},
 {Gao}, {Wu}, {Liang}, {Huang}, {Dai}, \& {Zhang}}]{2018ApJ...860...72M}
{Meng}, Y.-Z., {Geng}, J.-J., {Zhang}, B.-B., {et~al.} 2018, \apj, 860, 72,
 \dodoi{10.3847/1538-4357/aac2d9}

\bibitem[{{Milisavljevic} {et~al.}(2012){Milisavljevic}, {Drout}, \&
 {Berger}}]{2012GCN.12867....1M}
{Milisavljevic}, D., {Drout}, M., \& {Berger}, E. 2012, GRB Coordinates
 Network, 12867, 1

\bibitem[{{Norris} \& {Bonnell}(2006)}]{2006ApJ...643..266N}
{Norris}, J.~P., \& {Bonnell}, J.~T. 2006, \apj, 643, 266,
 \dodoi{10.1086/502796}

\bibitem[{{Norris} {et~al.}(2000){Norris}, {Marani}, \&
 {Bonnell}}]{2000ApJ...534..248N}
{Norris}, J.~P., {Marani}, G.~F., \& {Bonnell}, J.~T. 2000, \apj, 534, 248,
 \dodoi{10.1086/308725}

\bibitem[{{Norris} {et~al.}(1996){Norris}, {Nemiroff}, {Bonnell}, {Scargle},
 {Kouveliotou}, {Paciesas}, {Meegan}, \& {Fishman}}]{1996ApJ...459..393N}
{Norris}, J.~P., {Nemiroff}, R.~J., {Bonnell}, J.~T., {et~al.} 1996, \apj, 459,
 393, \dodoi{10.1086/176902}

\bibitem[{{Oates} {et~al.}(2020){Oates}, {Kuin}, {De Pasquale}, {Campana},
 {Tohuvavohu}, {Siegel}, \& {Neil Gehrels Swift Observatory
 Team}}]{2020GCN.28338....1O}
{Oates}, S.~R., {Kuin}, N.~P.~M., {De Pasquale}, M., {et~al.} 2020, GRB
 Coordinates Network, 28338, 1

\bibitem[{{Planck Collaboration} {et~al.}(2020){Planck Collaboration},
 {Aghanim}, {Akrami}, {Ashdown}, {Aumont}, {Baccigalupi}, {Ballardini},
 {Banday}, {Barreiro}, {Bartolo}, {Basak}, {Battye}, {Benabed}, {Bernard},
 {Bersanelli}, {Bielewicz}, {Bock}, {Bond}, {Borrill}, {Bouchet}, {Boulanger},
 {Bucher}, {Burigana}, {Butler}, {Calabrese}, {Cardoso}, {Carron},
 {Challinor}, {Chiang}, {Chluba}, {Colombo}, {Combet}, {Contreras}, {Crill},
 {Cuttaia}, {de Bernardis}, {de Zotti}, {Delabrouille}, {Delouis}, {Di
 Valentino}, {Diego}, {Dor{\'e}}, {Douspis}, {Ducout}, {Dupac}, {Dusini},
 {Efstathiou}, {Elsner}, {En{\ss}lin}, {Eriksen}, {Fantaye}, {Farhang},
 {Fergusson}, {Fernandez-Cobos}, {Finelli}, {Forastieri}, {Frailis},
 {Fraisse}, {Franceschi}, {Frolov}, {Galeotta}, {Galli}, {Ganga},
 {G{\'e}nova-Santos}, {Gerbino}, {Ghosh}, {Gonz{\'a}lez-Nuevo}, {G{\'o}rski},
 {Gratton}, {Gruppuso}, {Gudmundsson}, {Hamann}, {Handley}, {Hansen},
 {Herranz}, {Hildebrandt}, {Hivon}, {Huang}, {Jaffe}, {Jones}, {Karakci},
 {Keih{\"a}nen}, {Keskitalo}, {Kiiveri}, {Kim}, {Kisner}, {Knox},
 {Krachmalnicoff}, {Kunz}, {Kurki-Suonio}, {Lagache}, {Lamarre}, {Lasenby},
 {Lattanzi}, {Lawrence}, {Le Jeune}, {Lemos}, {Lesgourgues}, {Levrier},
 {Lewis}, {Liguori}, {Lilje}, {Lilley}, {Lindholm}, {L{\'o}pez-Caniego},
 {Lubin}, {Ma}, {Mac{\'\i}as-P{\'e}rez}, {Maggio}, {Maino}, {Mandolesi},
 {Mangilli}, {Marcos-Caballero}, {Maris}, {Martin}, {Martinelli},
 {Mart{\'\i}nez-Gonz{\'a}lez}, {Matarrese}, {Mauri}, {McEwen}, {Meinhold},
 {Melchiorri}, {Mennella}, {Migliaccio}, {Millea}, {Mitra},
 {Miville-Desch{\^e}nes}, {Molinari}, {Montier}, {Morgante}, {Moss}, {Natoli},
 {N{\o}rgaard-Nielsen}, {Pagano}, {Paoletti}, {Partridge}, {Patanchon},
 {Peiris}, {Perrotta}, {Pettorino}, {Piacentini}, {Polastri}, {Polenta},
 {Puget}, {Rachen}, {Reinecke}, {Remazeilles}, {Renzi}, {Rocha}, {Rosset},
 {Roudier}, {Rubi{\~n}o-Mart{\'\i}n}, {Ruiz-Granados}, {Salvati}, {Sandri},
 {Savelainen}, {Scott}, {Shellard}, {Sirignano}, {Sirri}, {Spencer},
 {Sunyaev}, {Suur-Uski}, {Tauber}, {Tavagnacco}, {Tenti}, {Toffolatti},
 {Tomasi}, {Trombetti}, {Valenziano}, {Valiviita}, {Van Tent}, {Vibert},
 {Vielva}, {Villa}, {Vittorio}, {Wandelt}, {Wehus}, {White}, {White},
 {Zacchei}, \& {Zonca}}]{2020A&A...641A...6P}
{Planck Collaboration}, {Aghanim}, N., {Akrami}, Y., {et~al.} 2020, \aap, 641,
 A6, \dodoi{10.1051/0004-6361/201833910}

\bibitem[{{Pugliese} {et~al.}(2015){Pugliese}, {Xu}, {Tanvir}, {Wiersema},
 {Fynbo}, {Milvang-Jensen}, \& {D'Elia}}]{2015GCN.17672....1P}
{Pugliese}, V., {Xu}, D., {Tanvir}, N.~R., {et~al.} 2015, GRB Coordinates
 Network, 17672, 1

\bibitem[{{Salmonson}(2000)}]{2000ApJ...544L.115S}
{Salmonson}, J.~D. 2000, \apjl, 544, L115, \dodoi{10.1086/317305}

\bibitem[{{Salvaterra} {et~al.}(2012){Salvaterra}, {Campana}, {Vergani},
 {Covino}, {D'Avanzo}, {Fugazza}, {Ghirlanda}, {Ghisellini}, {Melandri},
 {Nava}, {Sbarufatti}, {Flores}, {Piranomonte}, \&
 {Tagliaferri}}]{2012ApJ...749...68S}
{Salvaterra}, R., {Campana}, S., {Vergani}, S.~D., {et~al.} 2012, \apj, 749,
 68, \dodoi{10.1088/0004-637X/749/1/68}

\bibitem[{{Sanchez-Ramirez} {et~al.}(2013){Sanchez-Ramirez}, {Gorosabel},
 {Castro-Tirado}, {Cepa}, \& {Gomez-Velarde}}]{2013GCN.14685....1S}
{Sanchez-Ramirez}, R., {Gorosabel}, J., {Castro-Tirado}, A.~J., {Cepa}, J., \&
 {Gomez-Velarde}, G. 2013, GRB Coordinates Network, 14685, 1

\bibitem[{{Schaefer}(2004)}]{2004ApJ...602..306S}
{Schaefer}, B.~E. 2004, \apj, 602, 306, \dodoi{10.1086/380898}

\bibitem[{{Shen} {et~al.}(2005){Shen}, {Song}, \& {Li}}]{2005MNRAS.362...59S}
{Shen}, R.-F., {Song}, L.-M., \& {Li}, Z. 2005, \mnras, 362, 59,
 \dodoi{10.1111/j.1365-2966.2005.09163.x}

\bibitem[{{Tanvir} {et~al.}(2016){Tanvir}, {Levan}, {Cenko}, {Perley},
 {Cucchiara}, {Roth}, {Wiersema}, {Fruchter}, \&
 {Laskar}}]{2016GCN.19419....1T}
{Tanvir}, N.~R., {Levan}, A.~J., {Cenko}, S.~B., {et~al.} 2016, GRB Coordinates
 Network, 19419, 1

\bibitem[{{Toma} {et~al.}(2009){Toma}, {Wu}, \&
 {M{\'e}sz{\'a}ros}}]{2009ApJ...707.1404T}
{Toma}, K., {Wu}, X.-F., \& {M{\'e}sz{\'a}ros}, P. 2009, \apj, 707, 1404,
 \dodoi{10.1088/0004-637X/707/2/1404}

\bibitem[{{Uhm} \& {Zhang}(2016)}]{2016ApJ...825...97U}
{Uhm}, Z.~L., \& {Zhang}, B. 2016, \apj, 825, 97,
 \dodoi{10.3847/0004-637X/825/2/97}

\bibitem[{{Vielfaure} {et~al.}(2020){Vielfaure}, {Izzo}, {Xu}, {Vergani},
 {Malesani}, {de Ugarte Postigo}, {D'Elia}, {Fynbo}, {Kann}, {Levan},
 {Pugliese}, {Tanvir}, {Burgarella}, {Rossi}, \& {Stargate
 Consortium}}]{2020GCN.29077....1V}
{Vielfaure}, J.~B., {Izzo}, L., {Xu}, D., {et~al.} 2020, GRB Coordinates
 Network, 29077, 1

\bibitem[{{von Kienlin} {et~al.}(2020){von Kienlin}, {Meegan}, {Paciesas},
 {Bhat}, {Bissaldi}, {Briggs}, {Burns}, {Cleveland}, {Gibby}, {Giles},
 {Goldstein}, {Hamburg}, {Hui}, {Kocevski}, {Mailyan}, {Malacaria},
 {Poolakkil}, {Preece}, {Roberts}, {Veres}, \&
 {Wilson-Hodge}}]{FermiGBMCATELOAG}
{von Kienlin}, A., {Meegan}, C.~A., {Paciesas}, W.~S., {et~al.} 2020, \apj,
 893, 46, \dodoi{10.3847/1538-4357/ab7a18}

\bibitem[{{Vreeswijk} {et~al.}(2013){Vreeswijk}, {Malesani}, {Fynbo}, {De Cia},
 \& {Ledoux}}]{2013GCN.15249....1V}
{Vreeswijk}, P.~M., {Malesani}, D., {Fynbo}, J.~P.~U., {De Cia}, A., \&
 {Ledoux}, C. 2013, GRB Coordinates Network, 15249, 1

\bibitem[{{Vreeswijk} {et~al.}(2018){Vreeswijk}, {Kann}, {Heintz}, {de Ugarte
 Postigo}, {Milvang-Jensen}, {Malesani}, {Covino}, {Levan}, \&
 {Pugliese}}]{2018GCN.22996....1V}
{Vreeswijk}, P.~M., {Kann}, D.~A., {Heintz}, K.~E., {et~al.} 2018, GRB
 Coordinates Network, 22996, 1

\bibitem[{{Wang} {et~al.}(2021){Wang}, {Zheng}, {Xiao}, {Yang}, {Liu}, {Yang},
 {Zou}, {Zhang}, {Zeng}, {Xiong}, {Feng}, {Song}, {Wen}, {Xu}, {Chen}, {Ni},
 {Zhang}, {Wu}, {Cai}, {Cang}, {Deng}, {Gao}, {Kong}, {Huang}, {Li}, {Li},
 {Li}, {Liang}, {Lin}, {Liu}, {Long}, {Lu}, {Luo}, {Ma}, {Meng}, {Peng},
 {Qiao}, {Song}, {Tian}, {Wang}, {Wang}, {Wang}, {Xu}, {Yang}, {Yin}, {Zeng},
 {Zeng}, {Zhang}, {Zhang}, {Zhang}, \& {Zhang}}]{2021ApJ...922..237W}
{Wang}, X.~I., {Zheng}, X., {Xiao}, S., {et~al.} 2021, \apj, 922, 237,
 \dodoi{10.3847/1538-4357/ac29bd}

\bibitem[{{Wei} {et~al.}(2017){Wei}, {Zhang}, {Shao}, {Wu}, \&
 {M{\'e}sz{\'a}ros}}]{2017ApJ...834L..13W}
{Wei}, J.-J., {Zhang}, B.-B., {Shao}, L., {Wu}, X.-F., \& {M{\'e}sz{\'a}ros},
 P. 2017, \apjl, 834, L13, \dodoi{10.3847/2041-8213/834/2/L13}

\bibitem[{{Wiersema} {et~al.}(2014){Wiersema}, {Tanvir}, {Levan}, \&
 {Karjalainen}}]{2014GCN.16231....1W}
{Wiersema}, K., {Tanvir}, N., {Levan}, A., \& {Karjalainen}, R. 2014, GRB
 Coordinates Network, 16231, 1

\bibitem[{{Xu} {et~al.}(2014{\natexlab{a}}){Xu}, {Levan}, {Fynbo}, {Tanvir},
 {D'Elia}, \& {Malesani}}]{2014GCN.16983....1X}
{Xu}, D., {Levan}, A.~J., {Fynbo}, J.~P.~U., {et~al.} 2014{\natexlab{a}}, GRB
 Coordinates Network, 16983, 1

\bibitem[{{Xu} {et~al.}(2016){Xu}, {Malesani}, {Fynbo}, {Tanvir}, {Levan}, \&
 {Perley}}]{2016GCN.19600....1X}
{Xu}, D., {Malesani}, D., {Fynbo}, J.~P.~U., {et~al.} 2016, GRB Coordinates
 Network, 19600, 1

\bibitem[{{Xu} {et~al.}(2014{\natexlab{b}}){Xu}, {Malesani}, {Tanvir},
 {D'Elia}, {de Ugarte Postigo}, {Kruehler}, \& {Fynbo}}]{2014GCN.15645....1X}
{Xu}, D., {Malesani}, D., {Tanvir}, N.~R., {et~al.} 2014{\natexlab{b}}, GRB
 Coordinates Network, 15645, 1

\bibitem[{{Xu} {et~al.}(2021){Xu}, {Izzo}, {Fynbo}, {Kann}, {Vergani},
 {Malesani}, {Arabsalmani}, {Rossi}, {Pugliese}, {Vielfaure}, \& {Stargate
 Consortium}}]{2021GCN.29432....1X}
{Xu}, D., {Izzo}, L., {Fynbo}, J.~P.~U., {et~al.} 2021, GRB Coordinates
 Network, 29432, 1

\bibitem[{{Yang} {et~al.}(2020){Yang}, {Chand}, {Zhang}, {Yang}, {Zou}, {Yang},
 {Zhao}, {Shao}, {Xiong}, {Luo}, {Li}, {Xiao}, {Li}, {Liu}, {Joshi}, {Sharma},
 {Chakraborty}, {Li}, \& {Zhang}}]{2020ApJ...899..106Y}
{Yang}, J., {Chand}, V., {Zhang}, B.-B., {et~al.} 2020, \apj, 899, 106,
 \dodoi{10.3847/1538-4357/aba745}

\bibitem[{{Yi} {et~al.}(2006){Yi}, {Liang}, {Qin}, \&
 {Lu}}]{2006MNRAS.367.1751Y}
{Yi}, T., {Liang}, E., {Qin}, Y., \& {Lu}, R. 2006, \mnras, 367, 1751,
 \dodoi{10.1111/j.1365-2966.2006.10083.x}

\bibitem[{{Zhang} {et~al.}(2015){Zhang}, {van Eerten}, {Burrows}, {Ryan},
 {Evans}, {Racusin}, {Troja}, \& {MacFadyen}}]{2015ApJ...806...15Z}
{Zhang}, B.-B., {van Eerten}, H., {Burrows}, D.~N., {et~al.} 2015, \apj, 806,
 15, \dodoi{10.1088/0004-637X/806/1/15}

\bibitem[{{Zhang} {et~al.}(2011){Zhang}, {Zhang}, {Liang}, {Fan}, {Wu},
 {Pe'er}, {Maxham}, {Gao}, \& {Dong}}]{2011ApJ...730..141Z}
{Zhang}, B.-B., {Zhang}, B., {Liang}, E.-W., {et~al.} 2011, \apj, 730, 141,
 \dodoi{10.1088/0004-637X/730/2/141}

\bibitem[{{Zhang} {et~al.}(2012){Zhang}, {Burrows}, {Zhang},
 {M{\'e}sz{\'a}ros}, {Wang}, {Stratta}, {D'Elia}, {Frederiks}, {Golenetskii},
 {Cummings}, {Norris}, {Falcone}, {Barthelmy}, \&
 {Gehrels}}]{2012ApJ...748..132Z}
{Zhang}, B.-B., {Burrows}, D.~N., {Zhang}, B., {et~al.} 2012, \apj, 748, 132,
 \dodoi{10.1088/0004-637X/748/2/132}

\bibitem[{{Zhang} {et~al.}(2021){Zhang}, {Liu}, {Peng}, {Li}, {L{\"u}}, {Yang},
 {Yang}, {Yang}, {Meng}, {Zou}, {Ye}, {Wang}, {Mao}, {Zhao}, {Bai},
 {Castro-Tirado}, {Hu}, {Dai}, {Liang}, \& {Zhang}}]{2021NatAs...5..911Z}
{Zhang}, B.~B., {Liu}, Z.~K., {Peng}, Z.~K., {et~al.} 2021, Nature Astronomy,
 5, 911, \dodoi{10.1038/s41550-021-01395-z}

\end{thebibliography}

\end{document}